\documentclass[a4paper,12pt,openright]{thesis}
\usepackage{amssymb}
\usepackage{amsmath}
\usepackage[comma,numbers,sort&compress,square]{natbib}
\usepackage[british]{babel}
\usepackage[latin1]{inputenc}
\usepackage[dvips]{graphicx}
\usepackage{psfrag}
\usepackage{epsfig}

\newcommand{\fig}[1]{Fig.\ \ref{#1}}

\newcommand{\Eq}[1]{Eq.\ (\ref{#1})}
\newcommand{\Eqs}[2]{Eqs.\ (\ref{#1})--(\ref{#2})}

\newcommand{\h}[1]{\hat{#1}}
\renewcommand{\t}[1]{\tilde{#1}}

\newcommand{\SFCS}{SFCS}
\newcommand{\LOLA}{LOLA}
\newcommand{\PFIDO}{PFIDO}
\newcommand{\gaii}{\sqrt{\gamma_{i\,i}}}
\newcommand{\Svb}{\overline{\Sigma v}}
\renewcommand{\th}{\h{t}}
\newcommand{\xh}{\h{x}}

\newcommand{\vb}{\overline{v}}
\newcommand{\bJ}{\overline{J}}
\newcommand{\brho}{\overline{\rho}_e}

\newcommand{\ga}{\gamma}

\newcommand{\B}{\mathcal{B}}

\newcommand{\E}{\mathcal{E}}

\newcommand{\J}{\mathcal{J}}

\newcommand{\M}{\mathcal{M}}

\renewcommand{\P}{\mathcal{P}}

\renewcommand{\chaptermark}[1]%
        {\markboth{#1}{}}
\renewcommand{\sectionmark}[1]%
        {\markright{\thesection \ #1}}

\begin{document}

\frontmatter
\thispagestyle{empty}
\includegraphics[width=\textwidth]{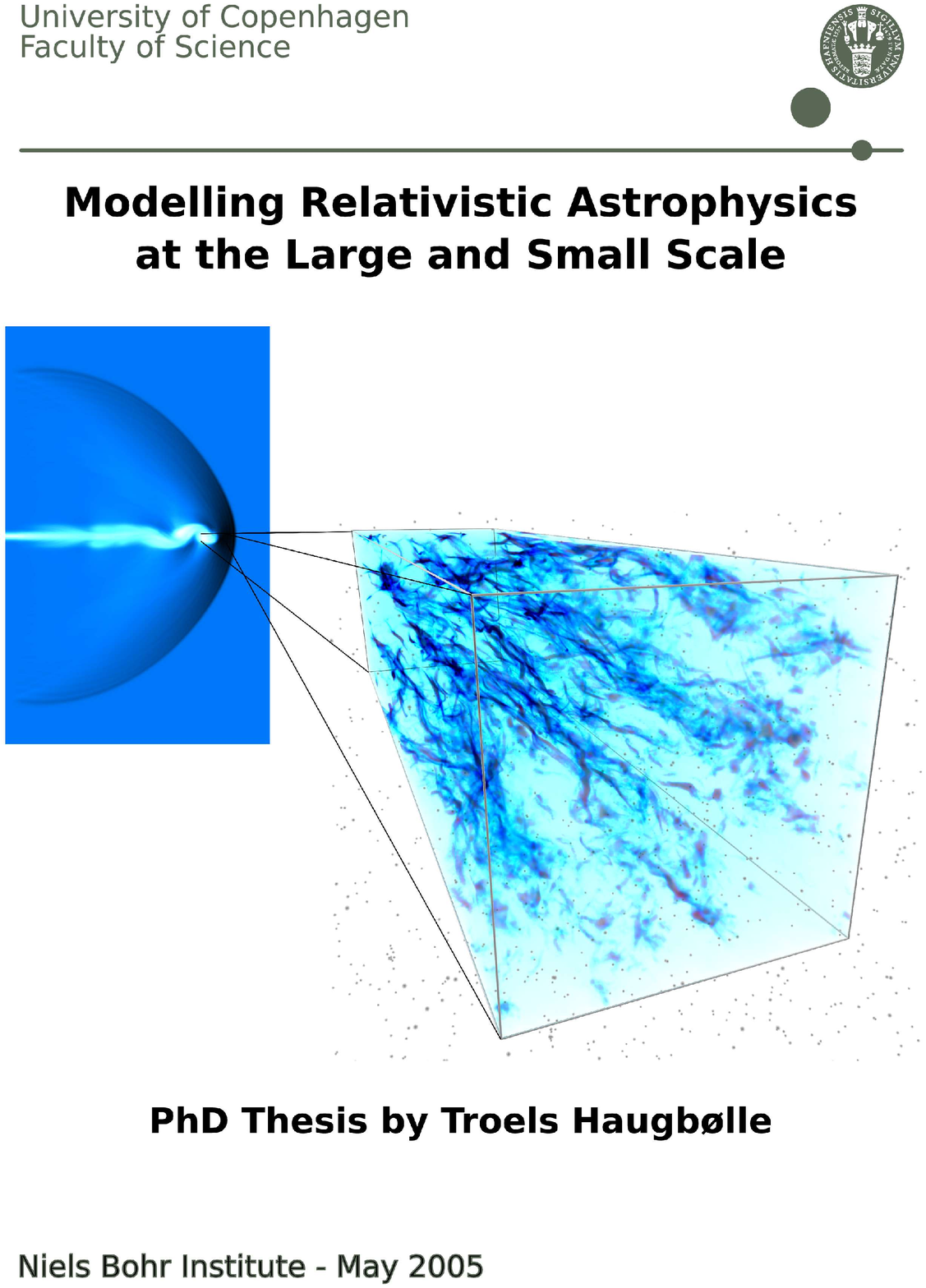}
\thispagestyle{empty}
\section*{\textsc{ACKNOWLEDGEMENTS}}
I would like to thank the people whom I have been
working with the last couple of years, and who have made this thesis
not only possible but also a joy to complete.
The biggest thanks goes to my supervisor {\AA}ke Nordlund;
without him there would be no thesis, and he has always been
willing to answer my questions at all times and guide me
through the labyrinth of numerical astrophysics, not least
in this last stressing month. But fortunately
I have not been left alone on this quest.
It has been a
privilege to be part of the plasma gang;
had it not been for Christian and Trier, my PhD would not
have been the same. They deserve thanks for the
countless discussions over memorable Friday beers, I
owe much of my thesis to them, not least for the many
discussions about pair plasmas and the global
structure of collisionless shocks,
and even though some
of us may shift subjects in the future I will remember this
as the best cooperation I have ever had; I even believe that
in the end they have managed to teach me a bit of that pesky
plasma physics.
The world is not only made of plasma though, nor Friday beers,
and I am grateful to Jakob Hunsballe for the CVS wars
we have waged over the fluid code lately.

The GrMHD code was initiated during a stay in Kyoto in 2003-2004, where
I had the pleasure of working with the group around Shibata-san.
My Japanese may have passed away since then, but I do
remember the good and warm atmosphere and the excellent hospitality
I received there. I learnt a lot about relativistic fluid dynamics,
but also about Japanese culture, the art of eating with chopsticks,
and how everything can be so different and yet you feel at home;
to all of you in Kyoto, specially Shibata-san, Hiro and Mizuno-san:
Thank you for a fantastic half a year in Kyoto, I hope to return soon
again.

Last but not least I am grateful to my proof readers; in spite of
their limited interest in the details of astrophysics, they made
excellent suggestions and enlightened the language at times where
I myself was too tired to do so. Not only I, but also the readers
should thank them:
Ana, Gemma and Sune in this last stressing month, you have made
all the difference. If errors remain, be it linguistic or in the physics,
I surely must have introduced them in the last minute!
\newpage
\thispagestyle{empty}
\section*{\textsc{ABSTRACT}}
In this thesis different numerical methods, as well as applications
of the methods to a number of current problems in relativistic astrophysics,
are presented.

In the first part the theoretical foundation and numerical
implementation of a new general relativistic
magnetohydrodynamics code is discussed. A new form of
the equations of motion using global coordinates, but evolving the dynamical
variables from the point of view of a local observer is presented.
No assumptions are made about the background metric and the design is
ready to be coupled with methods solving the full Einstein equations.

In the second part of the thesis important results concerning the understanding
of collisionless shocks, obtained from experiments with a relativistic
charged particle code, are presented.
Relativistic collisionless shocks are important in a range of 
astrophysical objects;
in particular in gamma ray burst afterglows and other relativistic jets. 
It is shown that a strong small scale, fluctuating, and predominantly
transversal magnetic field
is unavoidably generated by a two-stream instability.
The magnetic energy density  reaches a few percent of equipartition.

A new acceleration mechanism for electrons in
ion-electron collisionless shocks is proposed.
The mechanism is capable of creating a powerlaw 
electron distribution in a collisionless shocked region. 
The non--thermal acceleration of the electrons is directly related to the 
ion current channels generated by the two-stream instability and
is local in nature. 
Thus the observed radiation field may 
be tied directly to the local conditions of the plasma and could be a strong 
handle on the physical processes.

Experiments of colliding pair plasmas are presented and
the formation of a macrophysical shock structure is observed.
A comparable relativistic fluid simulation
 is performed and good agreement is found,
implying that the full structure of the shock has been resolved.
The extent of the shock transition region in a pair plasma is estimated to
50-100 electron skin depths. 

In the third part of the thesis a new particle-in-cell
code is discussed. It solves the full Maxwell equations,
together with direct microphysical particle-particle interactions,
such as relativistic scattering, pair production,
decay, and annihilation of particles.
The inclusion of such relativistic
interaction processes makes it possible to extract
self consistent synthetic photon spectra directly from
the numerical experiments, thereby 
gaining the ability to directly compare models with observations.

\tableofcontents
\mainmatter
\chapter{Introduction \& Overview}
During the last decade we have seen fundamental advances in the
observation of compact objects, active galactic nuclei,
gamma ray bursts, and other objects characterised
by their extreme physical conditions and emittence of
light over the full electromagnetic spectrum. This branch
of astrophysics has aptly been named ``extreme astrophysics'',
and advances in the field are driven by the technical
development and launch of new satellites, such as 
Beppo/Sax, Chandra, XMM and Swift, and the construction of
powerful ground based facilities, such as the HESS telescope and
the Auger observatory to measure X-rays and gamma rays.
Moreover the technique of combining  radio telescopes 
to perform interferometric observations with synthetic dishes
comparable to the entire globe has played an important role for
resolving the inner engines of active galactic nuclei (AGN).

In 1997 the first afterglow
from a gamma ray burst (GRB) was observed, placing GRBs firmly
out of reach at cosmological distances and earning them the
title as the most violent explosions in the Universe.
Very high energy gamma rays have also been observed from
AGNs, and with the increasing resolution of high frequency radio
interferometers, we will soon be able to resolve the
launching region of the jets associated with AGNs, only a
few Schwarzchild radii from the central supermassive
black hole. 

In the decade to come we can foresee that two entirely
new windows to the Universe will be opened to complement the
observations of electromagnetic radiation and cosmic
rays that are made today: On the south pole the IceCube
project will detect cosmic neutrinos generated in the core
of supernovae and possibly in GRBs and other cataclysmic events,
while laser interferometers on the ground, such as LIGO, VIRGO
and GEO 600, together with the space interferometer LISA, will
have reached levels of sensitivity where the gravitational
waves from the coalescence of compact objects and super massive black holes
may be detected.

A decade ago cosmology was still the branch of astrophysics where
one could get along with back-of-the-envelope calculations, since
fundamental parameters such as the Hubble expansion rate, the age
and the matter content of the Universe were all quoted with error bars
as large as the numbers themselves. This is all in the past now.
The Hubble space telescope has finally determined the expansion rate.
Observations of supernovae of type Ia at moderate and high redshifts have
led to the surprising conclusion that the Universe is in fact
accelerating in its expansion. The Boomerang and Maxima balloon
missions and later the WMAP telescope have nailed down fluctuations in
the cosmic microwave background radiation (CMBR) with high precision and
determined the overall geometry of the Universe to be\ldots flat!
Euclid was right. The pieces in the cosmological puzzle are slowly falling
into place. Current and future dedicated facilities to observe
the CMBR, together with large scale galaxy redshift surveys such as the SLOAN
digital sky survey and the 2DF survey, will give strong limits on the
distribution of matter and fields in the early Universe.

It is thus fair to say that both extreme astrophysics
and cosmology, together known as relativistic
astrophysics, are in a golden age and are slowly but firmly entering the
realm of ``messy astrophysics'', where
predictions cannot be based on sketchy ideas anymore but instead
detailed physical models must be worked out, tested, and validated or falsified
by observations.

Parallel to the development in observational relativistic
astrophysics, there has been a revolution in the tools employed by theoretical
astrophysicists. The computational power has for decades been rising
exponentially, doubling every 18 months in accordance with Moores law.
At the end of the nineties three-dimensional computer models of
astrophysical objects became affordable, and for some time computer
modelling has been indispensable in understanding the Universe.

In order to interpret observations, we have to develop theories that in
simple terms grasp the central physical mechanisms and let us
understand how fundamental parameters are related.
As the observations become more complicated, and the quality
of the data improves, so must the theories to be successful in
explaining these new details. Astronomy is different from other natural sciences,
in that we cannot perform experiments in the laboratory,
and in most cases the timescales are so long that we cannot even wait and watch
them unfold in the Universe.

In compact objects and in the early Universe many different physical processes
play important roles to shape the final picture, ranging from the large scale
fluid
dynamics, the curvature of space, the interactions in the plasma between
matter and electromagnetic fields, all the way to the microphysical
generation and scattering of the light, which, ultimately, is
observed on Earth. The computer gives us, as a complement to observations,
the ability to create models, and in contrast to the real Universe,
we can spin our models around and visualise the data in three dimensions,
instead of the projected, two-dimensional view which is the only one that
the real Universe offers.
In this sense computer models have become the virtual laboratory of 
the astrophysicist.
The physical insights gained from these models are essential,
and often the complexity of the phenomena leaves us at loss without
access to such models.

\section{A Swiss Army Knife for Relativistic Astrophysics}
In this thesis I present the application, development and implementation of
several computer codes which may be used to model relativistic astrophysics.
They span a range of scales and interactions. The GrMHD code, presented
in Chapter \ref{chap:GrMHD}, may be used to describe the
flow of matter from cosmological scales down to the scales of black holes.
The charged particle code, used in Chapters \ref{chap:field}--\ref{chap:global},
is applied to understanding the
small scale structure in collisionless shocks. Finally, the photon plasma code,
presented in Chapter \ref{chap:photonplasma}, will
enable us to study a fuller range of plasma
physics, including microphysical interactions, scatterings and the detailed
propagation of radiation.

\section{General Relativistic Magneto-Hydrodynamics}
In Chapter \ref{chap:GrMHD} I present a reformulation of the equations of
motion for general relativistic magnetohydrodynamics (GrMHD) that is
well suited for numerical purposes, and the
implementation in a three-dimensional numerical code that solves the
equations. Before
starting the implementation of the code, I carefully considered the approaches
employed in the handful of existing codes worldwide. My main idea has been
to make a conscious split between the reference frame in which we measure our
coordinates, and the reference frame in which we measure our physical variables.

The coordinate system, naturally, has to cover the whole physical domain. In the
case of compact objects, it is normal to use a coordinate system connected
to observers at infinity. But there is no a priori reason why we
should measure physical variables, such as density, velocity, and internal energy,
as seen by observers at infinity. If one measures them in a locally
defined frame, which is related to a local inertial frame, then
the physics, by the equivalence principle, becomes almost like the physics
of special relativity, for \emph{arbitrary} background space times.

All equations have been derived without placing any constraints on the metric
tensor. It is important to keep everything completely general, to allow
the code in the future to be enhanced with procedures that solve
the Einstein equations and evolve the metric tensor.

The code is based on finite difference techniques, and to handle discontinuities
we have to include some form of artificial viscosity to enhance the entropy
across shock fronts.  I have chosen to employ the correct physical description
of shear viscosity, to enforce energy and momentum conservation. The full
shear viscosity is very complicated, and in the general case of an arbitrary
space time it would be impractical if we did not use variables
tied to a local frame of reference.

\begin{figure}[!t]
\begin{center}
\includegraphics[width=\textwidth]{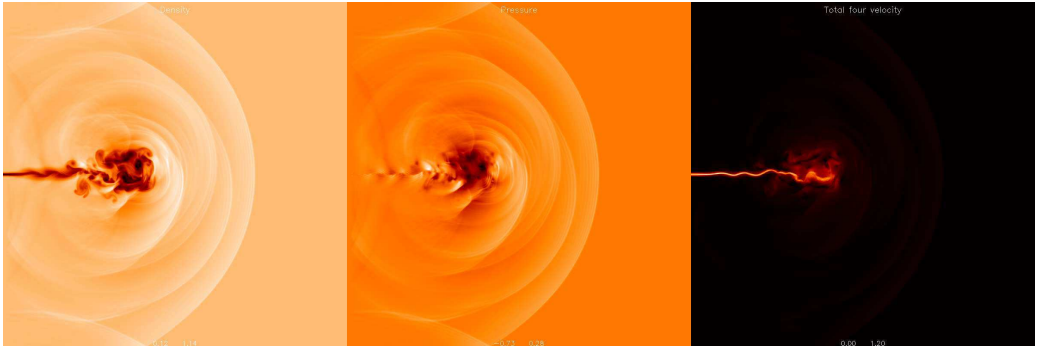}
\caption{A relativistic jet. From left to right: The density, the pressure
and the total four velocity. The jet head has partly destabilised 
and is creating a complex bubble of vortices in the cocoon of the jet.}
\label{fig:jet-intro}
\end{center}
\end{figure}

\begin{figure}[!t]
\begin{center}
\epsfig{figure=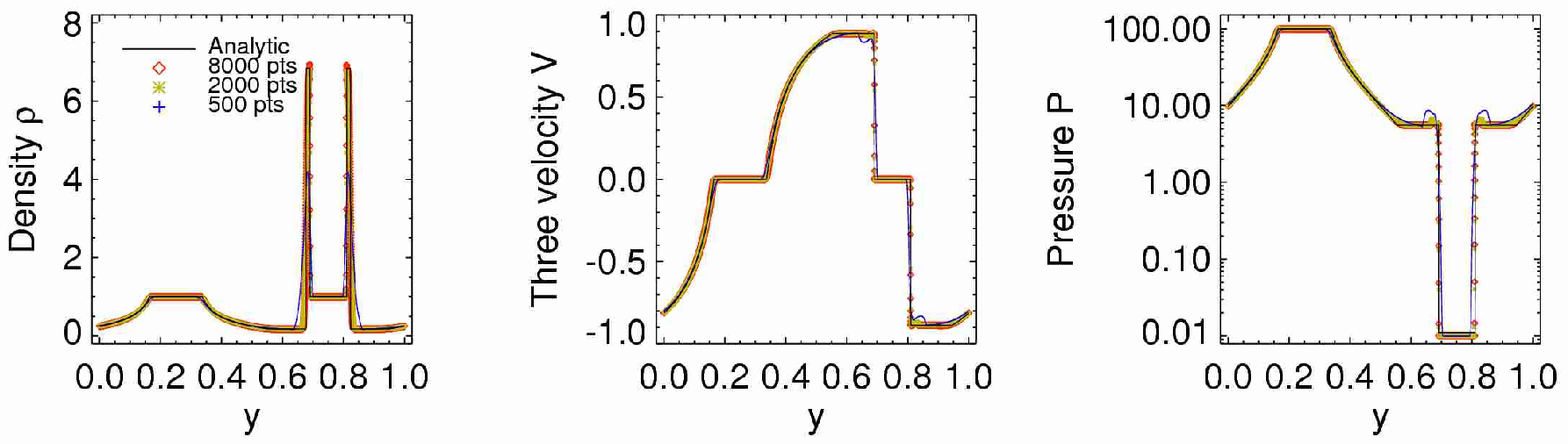,width=\textwidth}
\caption{A highly relativistic shock tube. The solution is shown at $t=0.2$,
just before the two shock waves collide.}
\label{fig:tube}
\end{center} 
\end{figure}

The code has been subjected to an extensive testbed of hydrodynamic and
magnetohydrodynamic problems, and it is demonstrated that it can
handle large gradients in density, pressure and the magnetic field.
As an example Fig.~\ref{fig:tube} shows a highly relativistic
shock tube problem, with two shock waves travelling toward each other,
involving a relative pressure jump of $10^4$. The solution is shown
at different resolutions with the analytic solution overplotted.
This test is described further as problem III in Chapter \ref{chap:GrMHD}.
Moreover, as an example of the capabilities the code, I use it in two
relevant astrophysical applications, one of them being the injection 
of a thin and hot relativistic jet into an ambient medium
shown in Fig.~\ref{fig:jet-intro}.

I have implemented a fully three-dimensional version of the code. The code
is parallelised, has been tested on several supercomputers, and has been shown
to yield excellent performance on hundreds of CPUs.

\section{Magnetic Field Generation in Collisionless Shocks}
\begin{figure}[!t]
\begin{center}
\epsfig{figure=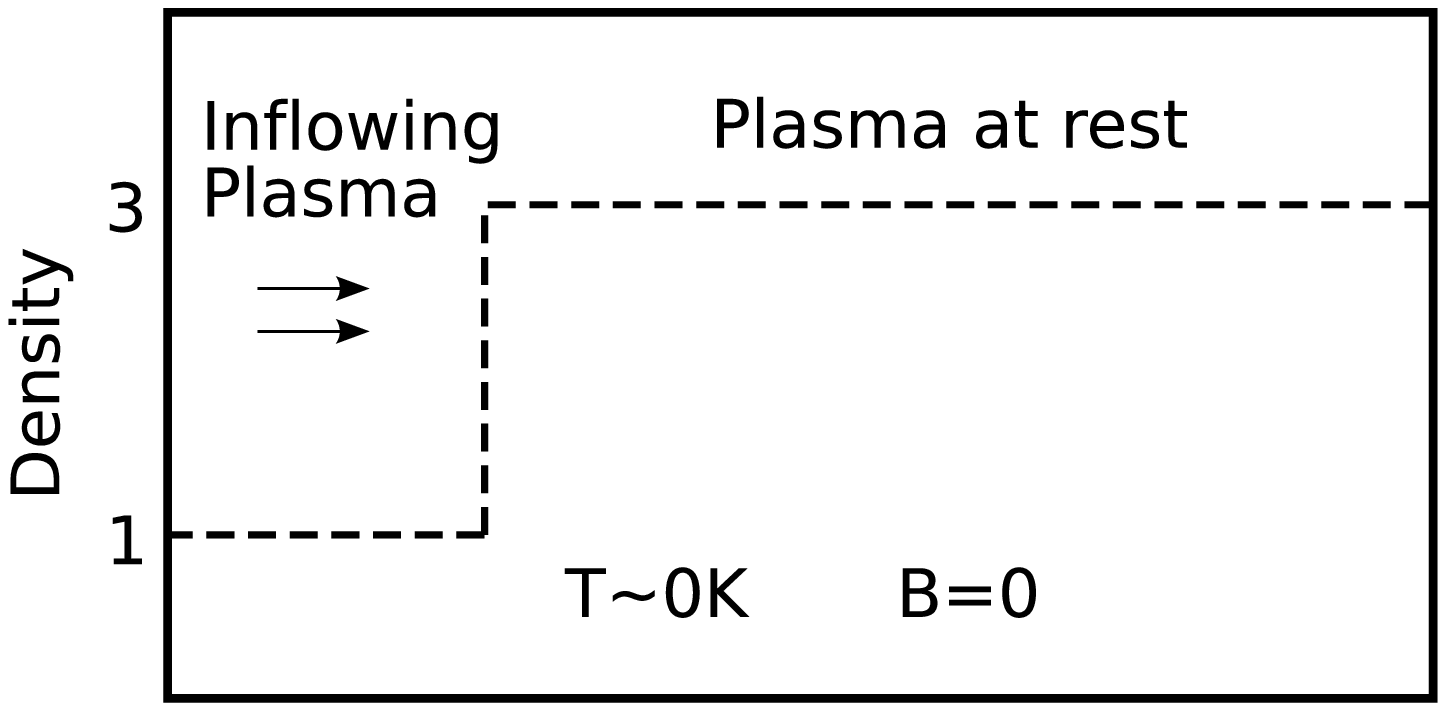,width=0.49\textwidth}
\epsfig{figure=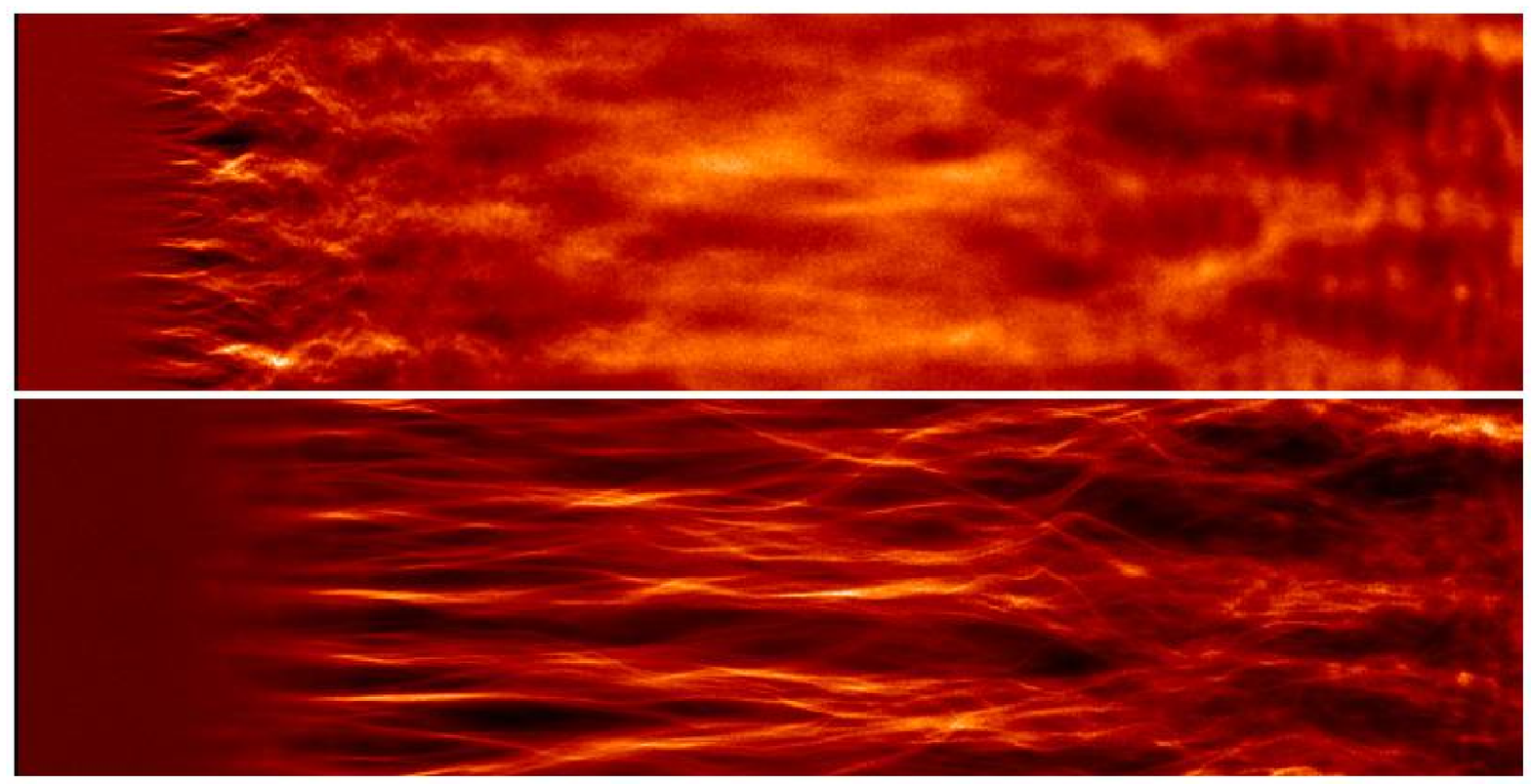,width=0.49\textwidth}
\caption{
Left: The initial conditions for our experiment.
Right: Electron (top) and ion (bottom) currents, averaged 
over the $x$-direction. The plasma is streaming from left
to right.}
\label{fig:cc}
\end{center}
\end{figure}

Chapter \ref{chap:field} was published by Frederiksen, Hededal,
Haugb{\o}lle and Nordlund \cite{bib:frederiksen2004}.
Using three-dimensional particle
simulations we report on  the evolution of an ion-electron
dominated counter-streaming collisionless shock. 
Our experiment consists initially of two populations. An in-streaming population,
with a Lorentz boost of $\Gamma=3$ upstream of the shock interface, and 
a population at rest downstream of the shock interface (see 
Fig.~\ref{fig:cc} to the left).
It is predicted theoretically, that colliding collisionless plasmas are
susceptible to the Weibel-- or two-stream instability.
Microscopic fluctuations in the magnetic field deflect the particles
that in turn enhance the fluctuation and an exponential growth of
the fluctuations results in the generation of strong current channels.
In our simulations this is confirmed and we observe the instability
develop in the shock interface.
In Fig.~\ref{fig:cc} to the right is shown the current densities at late
times. Associated with the current channels is a strong
transversal magnetic field. The magnetic field energy density reaches a
few percent of the kinetic energy in the in-coming beam.
For an ion-electron plasma this is in fact a two stage process.
Initially when the electrons encounter the shock interface, being
the lighter particles they are rapidly deflected into
first caustic surfaces and then current channels.
The magnetic field
keeps growing in scale and strength, until the ions undergo
the same process and similarly ion current channels are formed.
Because of charge separation, the electrons will be attracted
to the ions, and the electron instability is quenched.
Instead the electrons start to Debye shield the ions, forming a
fuzzy cloud around the ion channels (see Fig.~\ref{fig:cc}).
The Debye shielding partly neutralises the ion channels, and helps
stabilise the evolution. The electrons are fully thermalised,
but the ions are only slightly deflected from their initial distribution,
due to the strong shielding of the electrons, and thermalisation might be
significantly slower than predicted simply by extrapolating with
the mass ratio.
The ion current channels mutually attract each other and a
self similar merging process commences, where neighbouring
channels merge to form larger channels.
With the capacity of current computers, the ions cannot be followed
all the way to thermalisation, and merging of current channels is
ongoing when they reach the end of the box and stream out at the
open boundary.

To generate the radiation seen
in observations of GRB afterglows, a magnetic field
 containing $10^{-5}-10^{-1}$ of the kinetic energy is required.
The two-stream instability seen to occur in our experiments is
a strong candidate for explaining the generation of this magnetic
field, since it is unavoidable in collisionless shocks with low degrees
of magnetisation.
It then follows that the magnetic field cannot be taken as a free
parameter, but is a consequence
of the parameters of the shock, such as the inflow velocity and the
density contrast. These findings
do not only pertain to GRB afterglows, but also
imply that magnetic field generation may be an important ingredient
in all weakly magnetised collisionless shocks, and therefore occurs
in a range of objects from supernovae remnants to internal shocks
in outflows from AGN, all harbouring collisionless shocks.

\section{Non-Fermi Powerlaw Acceleration in Astrophysical Plasma Shocks}
In Chapter \ref{chap:acc} I present the results published by
Hededal, Haugb{\o}lle, Frederiksen and Nordlund \cite{bib:hededal2004}.
We study highly relativistic charged ion-electron particle dynamics
in collisionless shocks. 
The numerical experiment reported on here is different from the one in
Chapter \ref{chap:field} in that the in-streaming
plasma has a higher Lorentz factor ($\Gamma=15$) and the computational
box employed is about 3 times longer in the streaming direction, enabling us
to follow the process further downstream of the shock interface and for a
longer period of time, until the shock structure has been more fully
developed.

We find a powerlaw distribution
of accelerated electrons, which turns out to originate from an
acceleration process that is a direct consequence of the two-stream
instability observed in Chapter \ref{chap:field} and is local in nature.
The electrons are accelerated and decelerated when passing through
the cores of the ion current channels generated by the two-stream
instability, and the process is fundamentally different from
recursive acceleration processes, such as Fermi acceleration.
We find a powerlaw slope of $2.7$, in concordance with that
inferred from observations of the afterglow in gamma ray bursts,
and the process may explain
more generally the origin of part of the non-thermal radiation from relativistic
jets, supernovae remnants and shocked inter-- and circum-stellar
regions.

When two collisionless plasmas interpenetrate, current channels are
formed through the two-stream instability. The
ion current channels dominate the dynamics, due to the heavier mass
of the ions, and downstream of the shock the channels merge
in a hierarchical manner forming increasingly stronger patterns.
The electrons act to Debye shield the channels yielding charge neutrality
at large distances. At distances less than the Debye length
the ion channels are surrounded by an intense transverse electric field
that accelerate the electrons toward the channels and then decelerate
them, when they move away from the channel.
\begin{figure*}[!th]
\begin{center}
\epsfig{figure=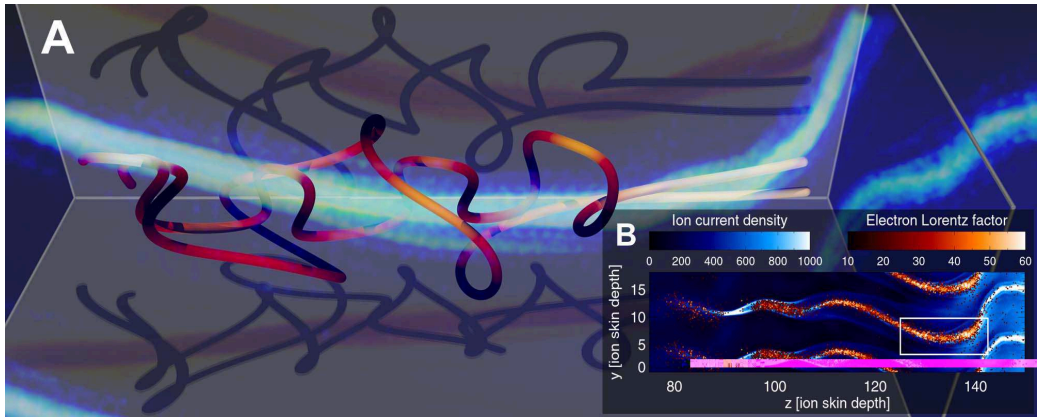,width=\textwidth}
\caption{(A) Ray traced electron paths (red) and current density (blue).
The colours of the electron paths reflect their four velocity according 
to the colour table in the inset (B). The shadows are
equivalent to the $x$ and $y$ projections of their paths. The ion
current density is
shown with blue colours according to the colour table in the inset.
The inset also
shows the ion current density (blue) integrated along the
$x$ axis with the spatial distribution of fast 
moving electrons (red) over plotted.}
\label{fig:accfig}
\end{center}
\end{figure*}
This can be seen in Fig.~\ref{fig:accfig}, where we in part (A) have ray traced
two selected electron paths and colour coded them according to the velocity and
in part (B) have shown the spatial distribution on the fastest electrons in
the box overplotted on top of the ion current distribution. Notice
the strong correlation between fast moving electrons and high ion current
density.

\begin{figure}[!ht]
\begin{center}
\epsfig{figure=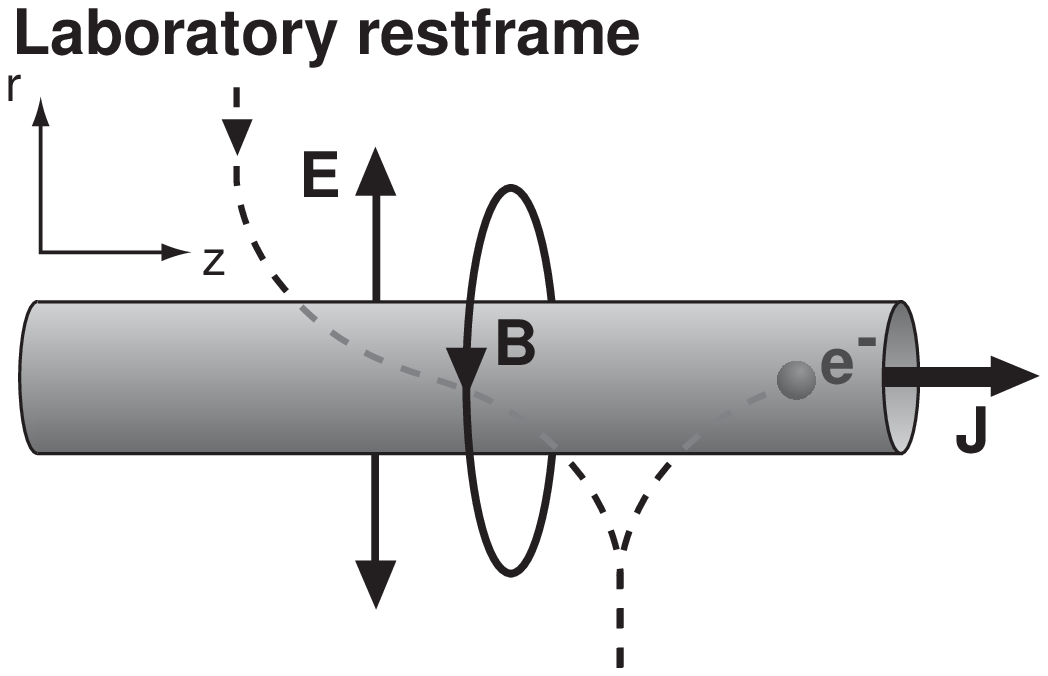,width=0.49\textwidth}
\epsfig{figure=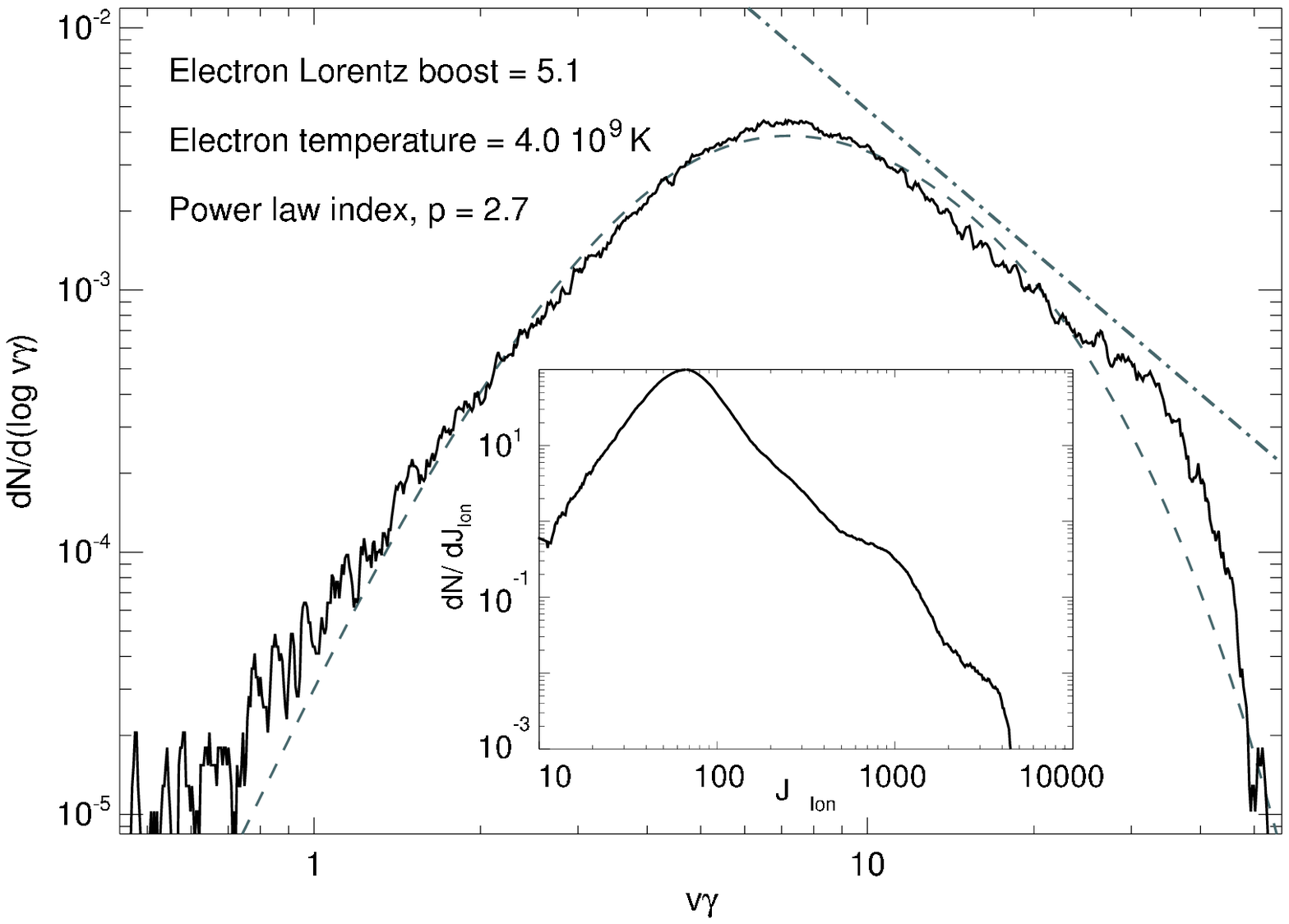,width=0.49\textwidth}
\caption{Left: A toy model of the acceleration process.
Electrons in the vicinity of the current channels are subject
to an electromagnetic force, 
working to accelerate them along the ion flow.
Crossing the centre of the channel the process
reverses leading to an oscillating movement along the channel.
Right: The normalised electron particle distribution function downstream of
the shock. The dot--dashed line is a powerlaw fit to the non--thermal high
energy tail. The inset shows a similar
histogram for ion current density
sampled in each grid point in the same slice as the electrons.
}
\label{fig:toymodel}
\end{center}
\end{figure}

To analyse the process quantitatively we have constructed a
toy model, idealising the ion channel as a solid cylinder of moving ions.
Given an electron we can calculate the maximal
energy gained in the acceleration towards the cylinder (see 
Fig.~\ref{fig:toymodel} to the left). We have compared the acceleration
predicted by this model with the acceleration observed in the experiment
and find good agreement.

To the right in Fig.~\ref{fig:toymodel} we have plotted the particle distribution
function for the electrons in a small slice in the box. We observe a powerlaw
distribution. This should be understood as consequence of 1) the acceleration
mechanism described above that directly relates the maximum kinetic energy of the
electrons to the local ion current density and 2) the powerlaw distribution
of the ion currents, as a consequence of the two-stream instability, seen
as an inset in the figure. The maximum acceleration observed is around
$v\gamma\approx 80$. Using the toy model and rescaling the ion to electron
ratio of 16, used in the experiment, to the real value of $1836$, we find
the maximum energy gained by the electrons to be around $5GeV$.

The presented acceleration mechanism is essentially due to the electrons
oscillating in a potential, though as seen in Fig.~\ref{fig:accfig} the
true paths of the electrons are more complicated, and the radiative
efficiency can be very high, because there are no free high energy electrons
carrying away the kinetic energy such as in recursive acceleration scenarios.
Moreover the properties of the process depend primarily on the local conditions
of the plasma.

In the chapter we estimate the thermalisation length for the ions,
and find by extrapolating the fractional thermalisation observed at the boundary
of the box, that the ions should thermalise in approximately 1500 ion skin depths.
Using typical values for density in a gamma ray burst afterglow this is
equivalent to $10^8$ m. We emphasise that the thermalisation length depends
on the inflow velocity and mass ratio of ions to electrons among others,
and a parameter study is necessary to uncover the true interdependence
of parameters.

Even though the two-streaming shock interface is estimated to be relatively thin,
the high radiative efficiency implies that the non-thermal radiation
observed in gamma ray burst afterglows and relativistic jets in general
could be emitted from such a thin shell.

\section{The Global Structure of Collisionless Shocks}
Collisions in ``collisionless shocks'' are
mediated by the collective electromagnetic field, and the scattering of the
particles on the field slowly heats the particles. At some
point the two-stream instability cannot be sustained, and the current
channels become unfocused and decay, due to the thermal
motion of the individual particles, 
which creates a warm turbulent medium with no significant large scale magnetic
field. In Chapters \ref{chap:field} \& \ref{chap:acc} it is shown how
magnetic field
generation and particle acceleration are integral parts of relativistic
collisionless shocks
in the case of weak or absent large scale magnetic fields.

To understand the impact on observations it is essential to investigate how far
down stream of the initial shock that the two-stream unstable region
extends. With this in mind, in Chapter \ref{chap:global} I
discuss the global structure of collisionless shocks.
A range of experiments are presented, both three-dimensional models of pair
plasmas and two-dimensional models of ion-electron plasmas.
There is a fundamental difference between ion-electron shocks, where the
mass difference leads to the ions dominating the dynamics and the electrons
stabilising the ion channels, and a pair plasma, where the electrons and
positrons form channels on the same timescale, and no shielding occurs.
In the latter case the two-stream unstable
region is significantly smaller than in the case of ion-electron shocks.

In the three-dimensional computer experiments we observe that the electrons and
positrons thermalise fully, and the medium contains five different regions:
The unperturbed upstream medium coming in from the left of the box; the
first discontinuity in the velocity, with a two-stream unstable region; a warm
thermalised region that is separated into a high and a low density state;
another two-stream unstable discontinuity, where the warm shocked medium
collides with the unperturbed downstream medium; and finally the
unperturbed downstream medium. To verify that I have in fact resolved the full
shock structure in a satisfactory manner, and the jump conditions have been
established, I compare the experiment with a fluid simulation and
find good agreement. 
From this experiment we can estimate that the two-stream unstable regions
for electron-positron plasmas decay after 50-100 electron skin depths.

\begin{figure*}[!t]
\begin{center}
\epsfig{figure=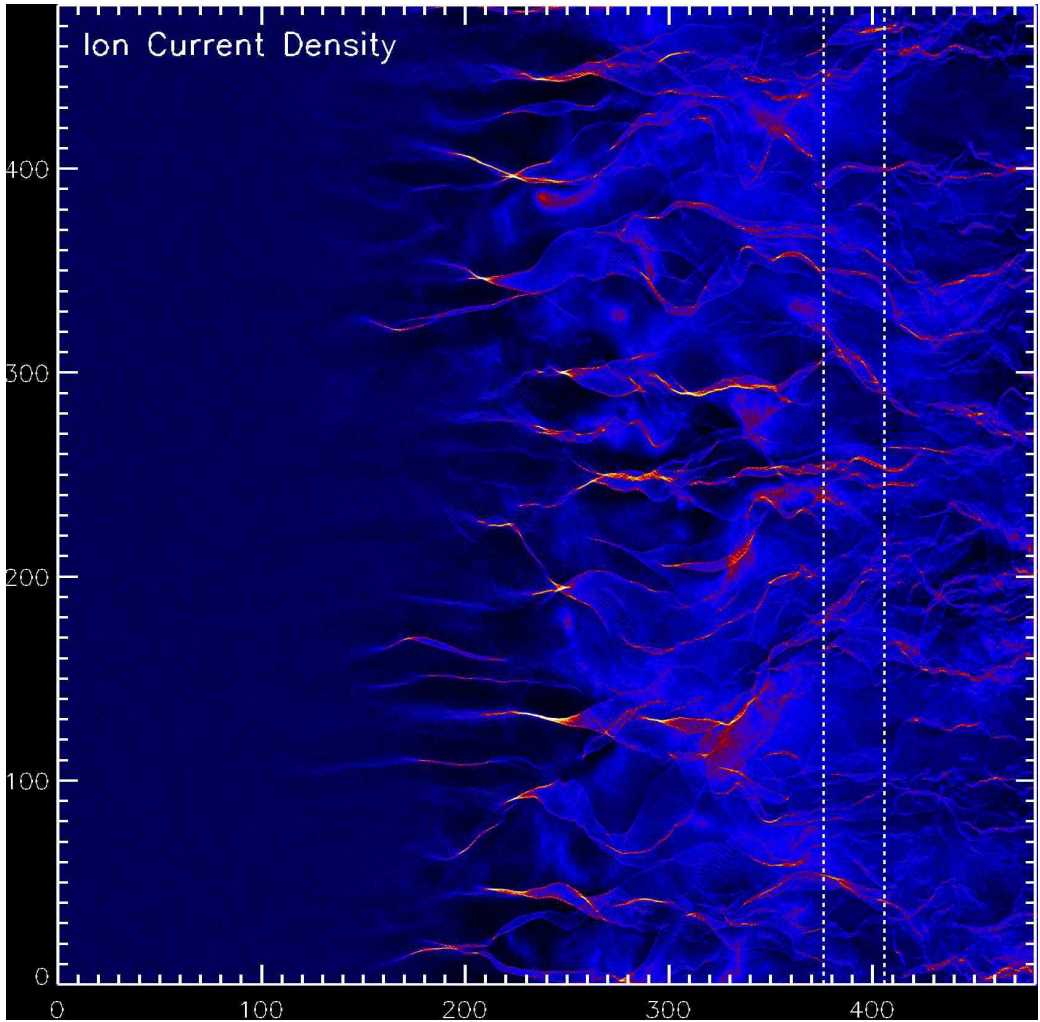,width=\textwidth}
\caption{The current density of the ions in a high resolution two-dimensional
experiment. The dashed lines indicate the region used for constructing particle
distribution functions. Length units
are given in electron skin depths.}
\label{fig:final2d-intro}
\end{center}
\end{figure*}

In the second part of Chapter \ref{chap:global} I consider the global structure
of ion-electron dominated collisionless shocks. With current computer capacities
it is impossible to correctly model the global structure of an ion-electron
shock in three dimensions. 
Two-dimensional collisionless shocks, being less costly computationally,
remain a promising alternative, and I have investigated the applicability
to understanding real three-dimensional models by performing large scale
two-dimensional experiments (see Fig.~\ref{fig:final2d-intro}), comparing
them to the three-dimensional
experiment discussed in Chapter \ref{chap:acc}.

\begin{figure*}[!t]
\begin{center}
\epsfig{figure=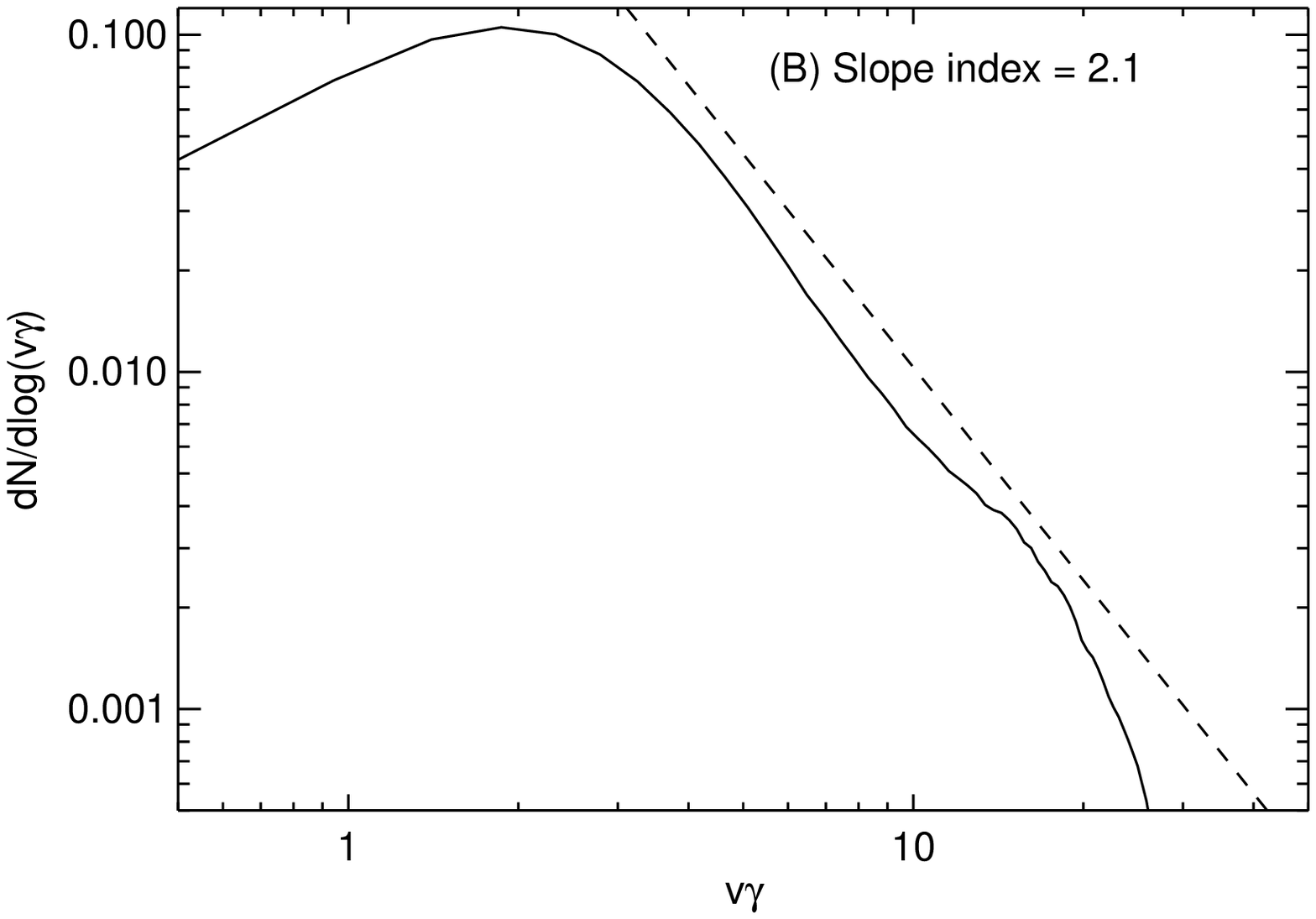,width=0.49\textwidth}
\epsfig{figure=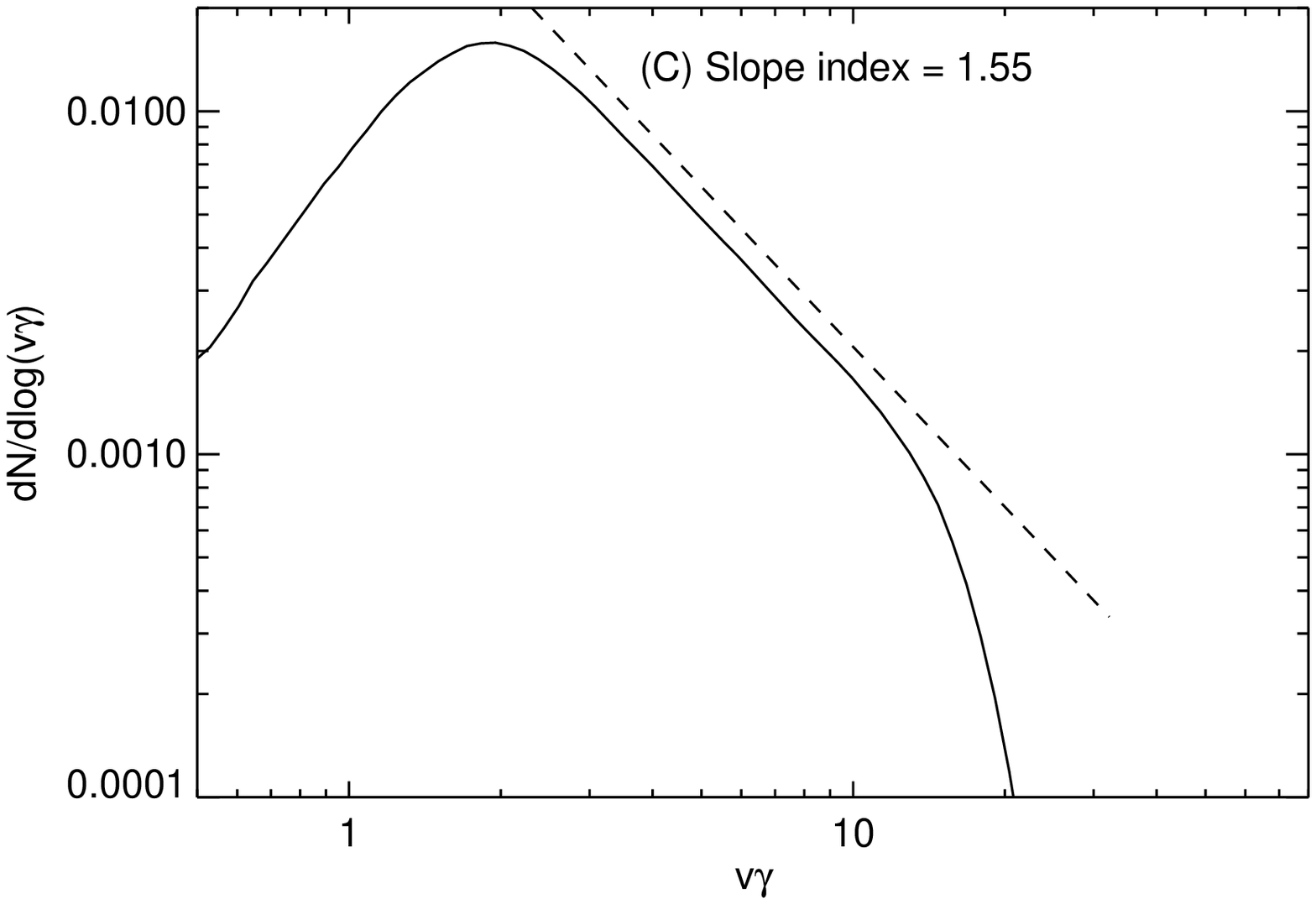,width=0.49\textwidth}
\caption{Particle distribution function for the electrons in a slice
indicated on Fig.~\ref{fig:final2d-intro}.
To the left is shown the PDF for the largest
two-dimensional experiment, while to right the PDF for the three-dimensional
experiment is shown.} 
\label{fig:pdfelec}
\end{center}
\end{figure*}
The particle distribution functions (PDFs) of the electrons
for the two-dimensional
and three-dimensional experiments are compared in Fig.~\ref{fig:pdfelec}. 
The slope indicated in Fig.~\ref{fig:pdfelec} depends on the
amount of heating in the upstream population, impacting the high energy
part of the spectrum, and the down stream population, impacting the low energy
part of the spectrum.
A warmer upstream population will be broader in phase space, and consequently
the maximum is lower, giving rise to a steeper slope. The two-dimensional
experiments have a slope index of $2.1$, while the three-dimensional
experiment has a slope index of $1.55$. The difference in heating rates
is understood in terms of the toy model, introduced
above in section 1.4 and discussed in Chapter \ref{chap:acc},
as a consequence of the different geometries. 

The physical significance of the the two-stream instability remains directly
related to the extent of the two-stream unstable region, and caution should be
voiced about uncritically generalising results from two-dimensional experiments to
three dimensions. My experiments seem to indicate that one will observe a
faster thermalisation rate in two-dimensional experiment than what may be
expected from three-dimensional experiments.

\section{A Next Generation PIC Code}
In Chapter \ref{chap:photonplasma} I present, together with C. Hededal,
the first results from a new particle-in-cell code in development.
The particle code that has been used to obtain the
results described in Chapters \ref{chap:field}--\ref{chap:global}
is limited to modelling the dynamics of charged particles under the
influence of electromagnetic fields.
In the new code, the concept of particles is generalised; most
notably we have introduced photons, and we consider microphysical
interactions such as scatterings, decay, annihilation
and pair production.

Even though work still has to be done before we may start to investigate
non trivial astrophysical scenarios, solid progress has already been made,
and to test the infrastructure of the new code we have implemented
Compton scattering as a simple scattering mechanism.
The results are very promising; there is excellent
agreement between theory and the numerical experiment.

The new code will enable us to target problems that reside in the grey zone
between the MHD and collisionless plasma domains.
This grey zone covers many astrophysical scenarios of great interest,
among others internal shocks in gamma-ray bursts, solar flares
and magnetic substorms, compact relativistic objects, and aspects of supernova
remnants.

\chapter{General Relativistic Magneto-Hydrodynamics}\label{chap:GrMHD}
Electromagnetic fields are ubiquitous ingredients in most astrophysical
objects. In the
case of very compact objects or at cosmological scales, not only do
electromagnetic fields
interact with matter directly, but they also become a source of
energy-momentum and
impact on the metric curvature. Several general relativistic
magnetohydrodynamics (GrMHD) computer codes have been developed and implemented
recently for the study of
compact relativistic objects and their surroundings 
\citep[e.g.][]{2003ApJ...589..458D,2003PhRvD..67j4010K,1999ApJ...522..727K,
               bib:anton05,bib:delzanna03,bib:fragile05},
using both conserved and non-conserved formulations of the basic equations of
motion. They are well-suited for their different purposes, but most of
the implementations above are designed for static space time backgrounds
with diagonal spatial terms.

In this chapter I present the analytic
basis for and numerical implementation of a code to solve the GrMHD equations.
My approach is inspired by the
pioneering work of Koide et al.~\cite{1999ApJ...522..727K} and related in
spirit to the methods of Ant\'on et al.~\cite{bib:anton05} and
Pons et al.~\cite{bib:pons98}.
From the beginning it has been designed
to be general enough to solve the GrMHD matter evolution equations on
any general time--dependent metric. This is an essential requirement if
the code ultimately is to  be coupled with numerical codes solving the
Einstein equations, which evolve
the metric. As far as the implementation is concerned I have currently
implemented a fully parallelised 3D version
of special relativistic MHD and a general relativistic
extension of the hydrodynamics.

In the following section I describe some of my motivations for
developing the code. In
section \ref{sec:2.3} I present the fundamental equations for GrMHD
and adapt them to our specific approach. The equations are well known
(e.g.~\cite{1982MNRAS.198..339T}), but I make an effort to rewrite
them in a form that is suited for my numerical purpose. For clarity
I first consider hydrodynamics and discuss the question of artificial
viscosity and imperfect fluids, to then extend the system to include
electromagnetic fields. In section \ref{sec:2.4}, I present the numerical algorithm that
I have chosen to implement the equations with. Section \ref{sec:2.5} contains a large
test bed of demanding problems. Section \ref{sec:2.6} contains some astrophysics
related tests of the code and finally, in section \ref{sec:2.7} I consider
the crucial aspects of performance and scalability among others.

\section{Motivation}
An important motivation for developing this kind of code is to make it
possible to study the
evolution of cosmological magnetic fields in the primordial universe,
taking into account the metric back reaction and coupling of gravitational
waves with the electro magnetic field. The WMAP satellite has already
detected the first polarisation signal in the cosmic microwave background
radiation (CMBR) \cite{bib:WMAP}. The Planck
satellite and ground/balloon based experiments will
improve the quality of the signal further in the coming years. Even though
primordial magnetic fields make a
very small contribution to the CMBR, in contrast to other imprints,
they source vector perturbations and hence it may be possible to disentangle the
weak signal from other sources through its unique character
\cite{bib:grasso00,bib:pogosian03,bib:naselsky04}.
Turbulent primordial magnetic fields can arise naturally during a
phase transition, such as the transitions from an electroweak plasma
and from the quark gluon phase to normal matter \cite{bib:vachaspati:01}.
Alternatively, they may be produced during inflation \cite{bib:ashoorioon04}.
If a signal from primordial magnetic fields is indeed detected, we
would have yet another probe to understand early universe physics.
Galaxies and clusters of galaxies at high redshift have been
observed to contain magnetic fields comparable to present day
galaxies. They have only rotated a few times during their short life,
and this is difficult to explain without invoking primordial magnetic
fields at some level. Dynamo theory alone does not
seem to be enough \cite{bib:grasso00,bib:jedamzik03}.
MHD simulations of turbulent helical fields have shown that an inverse
cascade process operates which transfers small scale power to
larger scales, changing the simple energy decay due to the
expansion of the universe \cite{bib:christensson01}.
Until now, except from purely analytical analyses, the question of
evolving magnetic fields in the early universe
has primarily been tackled in two different ways. 1) Simple 3D
turbulence experiments have been made, using existing non-relativistic
MHD codes to
address the possibility of inverse cascades which could alter significantly
the longevity of large scale primordial fields; 2) Semi analytical arguments
have been used to explore the couplings between primordial
magnetic fields and the metric, neutrinos, effects from Silk-dampening,
etc \cite{bib:lewis04}
If imprints in the cosmological microwave background from primordial magnetic
fields are detected, it will be crucial to understand the evolution of the fields
in a realistic manner, in order to constrain possible generation scenarios.
I have verified the results by Christensson et al
\cite{bib:christensson01} using a purely special relativistic version of the code.
With the code developed here, these questions may be addressed 
in a unified way, by performing
large scale 3D experiments including general relativistic effects and couplings
between the magnetic field and the metric perturbations.

Another strong motivation for developing a GrMHD code
is the fact that it provides the perfect complement to the particle-
and photon plasma codes, presented in the subsequent chapters,
for the study of extreme astrophysics around compact objects and
in jets. To understand the complex physics, we need to consider
processes happening at many different time and length scales.
A GrMHD code can be used to model the large scale dynamical flow and,
as detailed in Chapter \ref{chap:photonplasma}, provide realistic
boundary conditions for microphysical studies of plasma instabilities
and radiative processes.

We note that the first
results of coupling the full Einstein equations to the MHD equations has been
published \cite{bib:shapiro05} only very recently, and that the field is still
in its infancy.

\section{The GrMHD equations}\label{sec:2.3}
\subsection{3+1 Formulation of general relativity}
In numerical relativity it has proven very fruitful to exploit the so called
3+1 split of the metric. Instead of working with a four dimensional
manifold and the Einstein equations in the form of an elliptic
nonlinear set of partial differential equations, an explicit split
between temporal and spatial dimensions is imposed
(though see \cite{bib:meier03b} for
an alternative four dimensional approach). Assuming that we can construct a 
foliation of space time --- usually a very reasonable condition except maybe 
for near (naked) singularities --- it is then possible to rewrite the
Einstein equations
as a hyperbolic set of evolution equations, some elliptic constraint equations
and an associated Cauchy data set describing the initial conditions.
This formulation lends itself easily to a numerical implementation and
has been named the 3+1 approach.

The standard way of writing the metric in 3+1 form\footnote{Up to a plus or
minus sign and a
factor of $\alpha^{-1}$ for $\beta$} is:
\begin{equation}\label{metric1}
    ds^2 = - \alpha^2 dt^2 + \gamma_{ij}
          \left(dx^i + \beta^i dt\right)\left(dx^j + \beta^j dt\right),
\end{equation}
where $\alpha$ is called the lapse function, $\beta$  is the shift or shear and
$\gamma$ is the spatial 3-metric. The contravariant version of the metric
$g^{\mu \nu}$ is written
\begin{equation}
g^{\mu \nu} = \left(
\begin{array}{cc}
  -\frac{1}{\alpha^2} & \frac{\beta^i}{\alpha^2} \\
  \frac{\beta^j}{\alpha^2} & \gamma^{ij}\\
\end{array}
\right).
\end{equation}
This form of the metric has the same number of degrees of freedom,
namely ten, as in the obvious form $g_{\mu\nu}$. Here they are spread out as
one for the lapse, three for the shear and finally six in the spatial curvature.
Therefore, any metric which is not null may be written in this form.

In this thesis I only consider the evolution of matter and fields in 
a background
space time although through the Einstein equation they are
sources for the metric fields. Thus, it is important to leave $\alpha$, $\beta$
and $\gamma_{ij}$ unspecified, making the design ready for
integration with evolving metric fields.

\subsection{Different coordinate systems}
The global coordinate system \Eq{metric1} is often called the \emph{star fixed
coordinate system} ({\SFCS}), because in most applications
it is asymptotically flat and, therefore, connected to inertial
observers at infinity.
If we consider instead local observers who do not observe any shear
and measure time in terms of local clocks, their line element must be given as
\begin{equation}\label{metric2}
    ds^2 = - d\th^2 + \gamma_{ij} d\xh^i d\xh^j.
\end{equation}
This coordinate system is denoted the \emph{local laboratory frame} ({\LOLA} frame), and
I write any quantity in this coordinate system with a hat.
In the {\LOLA} frame in many interesting cases $\gamma_{ij}$ is almost diagonal
and one could then easily rescale the problem as done by Koide et
al.~\cite{1999ApJ...522..727K} to evolve matter and fields as seen by local
observers, or FIDOs\footnote{FIDOs are fiducial observers whose metric
is defined as that seen by observers in local inertial frames.} instead.

I have done so but to keep my approach general, I have exploited the idea
to always rescale the diagonal in the metric, even though it may well be
non diagonal.
Because the off diagonal terms in the spatial part of the metric often
are comparable in size to the diagonal ones, I have effectively normalised
the metric. Since the metric is almost a FIDO metric I have 
named it the \emph{pseudo FIDO frame} ({\PFIDO}) frame. In this frame the
metric tensor is given as 
\begin{align}\label{eq:metric3}
ds^2 &= -d\t{t}^2 + \tilde{\gamma}_{ij} d\t{x}^id\t{x}^j\,, \\
\tilde{\gamma}_{ij} &= \frac{\gamma_{ij}}{\sqrt{\gamma_{i\,i}\gamma_{jj}}}\,,
\end{align}
and there are only three non-trivial terms in the {\PFIDO} metric, because
all but the non diagonal terms in \Eq{eq:metric3} have been normalised.

The central idea of my numerical scheme is to use the {\PFIDO} frame
to measure all physical quantities. The {\PFIDO} frame is only
defined locally and we still need to use the global coordinates connected
to the {\SFCS} to measure distances. The general way to construct
an equation is first to derive it in the {\SFCS} and then to transform
the tensors and vectors from the {\SFCS} to the {\PFIDO} frame, while
keeping the derivatives and differentials with respect to the {\SFCS}.
It is central that the transformation from the {\SFCS} to the {\PFIDO}
frame is completely linear and simple, even for generally evolving
coordinates. Had we, instead, chosen to go all the
way to a FIDO frame in 
the general case, we would have had to invert a matrix, the metric, at every
point for every time step. The {\PFIDO} frame is
a healthy in-between, which gives us almost all of the advantages of
the FIDO frame but at a much lower cost.

Intuitively it is clear that when going to a local frame of reference 
the curvature
of space only manifests itself as extra external Coriolis--like forces, giving
some extra terms in the evolution equations below. From a numerical view point
there is an added benefit when we consider space times with a strong shear or
frame dragging,
ie.~points where $\beta^i$ is large. The standard example of this is the Kerr
metric in Boyer-Lindquist coordinates.
Inside the ergosphere, from the point of view of the {\SFCS}, everything is
rotating 
around the black hole in the same direction as the spin of the hole. The closer
we are to
the event horizon, the faster the rotation induced by the shear. From a local
observers point of view in the {\PFIDO} frame though, there is no shear and the
locally defined velocity is much smaller. The locally defined velocity is the
truly interesting velocity, since it arises due to physical processes, while
the apparent high velocity seen by an observer attached to the {\SFCS} is partly
due to the geometrical structure of the background space time and partly due
to physical processes, thus, a result of the
chosen reference frame. Near the horizon the shear-induced frame dragging
velocity can be much greater than the local velocity, and we can run into problems
with numerical cancellations smearing out variations in the local
velocity. Yet this is avoided by choosing to work in the {\PFIDO} frame.

From the line elements \Eq{metric1} and \Eq{eq:metric3} we may derive the
transformation laws. In particular we have
\begin{align}
                      \alpha dt & = d\t{t}\,, \\
\sqrt{\gamma_{ii}} (dx^i + \beta^i dt) & = d\t{x}^i\,,
\end{align}
and coordinate differentials are contravariant vectors. Then, any
contravariant vector $U^\mu$ transforms like
\begin{equation}\label{eq:trans1}
\t{U}^t = \alpha U^t\,, \quad \t{U}^i = \gaii \left(U^i + \beta^i U^t\right).
\end{equation}
It is a matter of linear algebra to show that covariant
vectors transform like
\begin{equation}\label{eq:trans2}
\t{U}_t = \frac{1}{\alpha}\left(U_t - \beta^i U_i \right)\,, \quad 
\t{U}_i = \frac{1}{\gaii} U_i\,.
\end{equation}
Tensors transform as the product of vectors by their very definition.
We refer the reader to App.~\ref{chap:appa} for a complete list of
transformation relations that have proven useful when deriving the
equations in this chapter.

\subsection{Basic equations}
The basic fluid equations follow from conservation laws.
The conservation of the baryon current gives
\begin{equation}\label{eq:baryoncons}
\nabla_\mu \left( \rho U^\mu \right)=0\,,
\end{equation}
where $\nabla_\mu$ is the covariant derivative, $\rho$ is the rest mass density
and
$U^\mu$ is the four velocity in the {\SFCS} coordinate system.
The conservation of the energy--momentum tensor $T^\mu_\nu$ leads to a similar 
expression
\begin{equation}
\nabla_\mu T^\mu_\nu = 0.
\end{equation}
The version that we have chosen
to use of the energy--momentum tensor for a fluid is given as
\begin{equation}\label{eq:tmunu}
T^\mu_{(HD)\nu}=\rho h U^\mu U_\nu + \delta^\mu_\nu P-2\eta\sigma^\mu_\nu\,,
\end{equation}
where $P$ is the pressure, $h = 1 + e_{int}
+ P/\rho$ is the relativistic enthalpy,
$e_{int}$ is the internal energy, 
 and $\eta\sigma^\mu_\nu$ is the shear
viscosity. It has the definition
\begin{equation}\label{eq:shear}
\sigma^{\mu \nu} = \frac{1}{2}\left( h^{\mu\alpha}\nabla_{\alpha} U^\nu
                                   + h^{\nu\alpha}\nabla_{\alpha} U^\mu \right)\,,
\end{equation}
where $h^{\mu\nu}$ projects into the fluid rest frame
\begin{equation}
h^{\mu\nu} = U^\mu U^\nu + g^{\mu\nu}.
\end{equation}
We consider the energy--momentum tensor in mixed form as the basic
hydrodynamical object to evolve, because even for general
metrics the pressure term disappears in \Eq{eq:tmunu} for off-diagonal
components \citep{2003ApJ...589..444G}.
This is not the case for the purely co-- or
contravariant versions.

The energy momentum tensor of the electromagnetic field is
\begin{equation}
T^\mu_{(EM)\nu} = F^{\mu\sigma} F_{\sigma\nu}
                 - \frac{1}{4}\delta^\mu_\nu F^{\kappa\sigma} F_{\kappa\sigma}\,,
\end{equation}
where $F^{\mu\nu}$ is the electromagnetic field strength tensor.

We can simplify the covariant derivatives significantly by using
the following identities
\begin{subequations}\label{coderiv}
\begin{align}
\nabla_\mu f U^\mu &= \frac{1}{\sqrt{-||g||}} 
                            \partial_\mu \left(\sqrt{-||g||}\, f U^\mu \right)\,, \\
\nabla_\mu T^\mu_\nu &= \frac{1}{\sqrt{-||g||}} 
                            \partial_\mu \left(\sqrt{-||g||}\, T^\mu_\nu \right)
             - \frac{1}{2}T^{\kappa \sigma} \partial_\nu g_{\kappa \sigma}\,, \\
\nabla_\mu F_\nu^{\phantom{\nu}\mu} &= 
       \frac{1}{\sqrt{-||g||}} 
             \partial_\mu \left(\sqrt{-||g||}\, F_\nu^{\phantom{\nu}\mu} \right)\,,
\end{align}
\end{subequations}
where $f$ is a scalar function, $U^\mu$ a vector, $T^\mu_\nu$ is any
symmetric tensor, $F_\nu^{\phantom{\nu}\mu}$ any antisymmetric tensor
and $||g||$ is the determinant of the metric. 

\subsection{Selecting evolution variables}
We have chosen our field variables with respect to the
{\PFIDO} frame and the basic evolution variables
take the form
\begin{align}\label{eq:D}
D  &= \ga \rho {\t U}^t = \ga \rho W\,, \\ \label{eq:E}
\E &= - \ga {\t T}^t_t - D = \ga \left( \rho h W^2 - P - \rho W \right)\,, \\
\label{eq:Pi}
\P_i &= \gaii \ga {\t T}^t_i = \gaii \ga \rho h W {\t U}_i\, ,
\end{align}
where $W={\t U}^t$ is the Lorentz factor of the fluid with respect to the
{\PFIDO} frame and $\ga=\sqrt{||\gamma||}$ is the square root of the determinant
of the spatial metric. 
Looking at \Eq{coderiv} and \Eq{eq:trans1} we see that the reason for
choosing the factor $\ga$ in front of the variables in
Eqs.~(\ref{eq:D})--(\ref{eq:Pi})
is to cancel out $\sqrt{-||g||}$ in \Eq{coderiv}.
The subtraction of the relativistic mass density in the
definition of the total fluid energy density is done in order to cancel
the rest mass energy
density, which could otherwise jeopardise a numerical implementation when the
flow is non--relativistic. 

\subsection{Hydrodynamic equations}
In order to highlight the physical content I first write 
down the equations of motion in the
case where there are no electromagnetic fields:
$T^{\mu\nu}=T^{\mu\nu}_{(HD)}$. To find the equations of motion, we use 
\Eqs{eq:trans1}{eq:trans2} and their extension to mixed typed
two-tensors (see App.~\ref{chap:appa}) together with the rules for covariant derivatives \Eq{coderiv} and
the fundamental equations of motion in the {\SFCS} \Eq{eq:baryoncons} and
\Eq{eq:tmunu}
\begin{align}\label{eq:Deqofm}
\partial_t D &= -\partial_j D \vb^j\,, \\ \nonumber
\partial_t \left[\E + \Sigma^t_t\right]   
 &= -\partial_j\left[\left(\E + \ga P\right)\vb^j + \Svb_t^j\right] \\ \nonumber
&\phantom{=}+\frac{1}{\alpha}\left[\P_i\left(\partial_t+\vb^j\partial_j\right)
                    +\Sigma^t_i\partial_t + \Svb_i^j\partial_j\right]\beta^i
\\ \nonumber &\phantom{=}
        -\left[DhW\left(\partial_t+\vb^j\partial_j\right)
                   +\ga P \left(\partial_t-\beta^j\partial_j\right)\right.
\\ \nonumber &\qquad\qquad\qquad\qquad\left.
                   +\Sigma^t_t\partial_t+\Svb_t^j\partial_j\right]\ln\alpha
\\ \label{eq:Eeqofm} &\phantom{=}
       -\partial_j\left(\ga P \beta^j\right) + \left( \beta^i\M_i-\M_t \right)\,,
       \\ \label{eq:Peqofm}
\partial_t \left[\P_i + \Sigma^t_i \right] &= 
       -\partial_j \left[ \P_i \vb^j + \Svb^j_i \right] 
                   -\partial_i \left[ \alpha\ga P \right] + \alpha\M_i\, ,
\end{align}
where the normal three-velocity has the usual definition
\begin{equation}
{\t v}^\mu = \frac{{\t U}^\mu}{{\t U}^t} = \frac{{\t U}^\mu}{W}\,,
\end{equation}
the transport velocity is the three velocity seen from the {\SFCS}
\begin{equation}
\vb^i = \frac{\alpha}{\gaii} {\t v}^i - \beta^i = \frac{U^i}{U^t}\,,
\end{equation}
the geometrical terms $\M_\mu$ are
\begin{equation}
\M_\mu = \frac{1}{2} \ga T^{\alpha\nu}\partial_\mu g_{\alpha\nu}\,,
\end{equation}
and the viscosity terms are
\begin{align}
\Sigma^t_t &= - \ga {\t \sigma}^t_t\,, \\
\Sigma^t_i &= \gaii \ga {\t \sigma}^t_i\,, \\
\Svb^j_t   &= - \ga \left[ \frac{\alpha}{\gamma_{jj}} {\t \sigma}^j_t
                           -\beta^j {\t \sigma}^t_t \right]\,, \\
\Svb^j_i   &= \gaii \ga \left[ \frac{\alpha}{\gamma_{jj}} {\t \sigma}^j_i
                           -\beta^j {\t \sigma}^t_i \right]\,.
\end{align}
Even though the evolution equation for the energy has become a bit more
complicated than in the special relativistic case ($\alpha=\ga=\gaii=1$,
$\beta=0$), it represents a substantial simplification in that relations between
the different variables reduce almost to the special relativistic form.
Hence for example the Lorentz factor $W$ may be computed as 
\mbox{$W=[1+{\t \gamma}_{ij}\t U^i\t U^j]^{1/2}$} bearing in mind
that the diagonal is already normalised. Let us consider a
space time without any off-diagonal spatial components but
with an arbitrary shear. For example Boyer Lindquist coordinates in an
extreme astrophysics context or the uniform curvature gauge 
in a cosmological context. In these examples, the shear viscosity is identical
to the special relativistic form. This is because the {\PFIDO} frame
reduces to a FIDO frame of reference. To handle
coordinate systems that penetrate the event horizon of a black hole, for
example the Kerr--Schild coordinates, we need at least one off-diagonal
spatial component \cite{bib:cook00}. In this case extra terms 
in the shear tensor arise, but changes are minimal. 

\subsection{Artificial viscosity}\label{sec:av}
It was argued by Anninos \& Fragile \cite{2003ApJS..144..243A} that 
in order to make a consistent
relativistic finite difference code with artificial viscosity (AV)
it is crucial to use a viscosity that has been defined in a physically
sensible manner,
otherwise it will break down for flows with high Lorentz factors.
An efficient AV should be covariant in its definition, such
that, the code can easily be adapted to general relativity, be physically
meaningful, respect energy conservation, and reduce to some normal Newtonian
AV formulation in the non relativistic limit.
We know of no implementation so far that has respected all of the
above points.
Indeed it seems that the prevalent thing is constructing a mock-up 
``viscous pressure''
using the prescription $P \rightarrow P + Q_{visc}$ and then include a
directional dependence such that the effective energy-momentum tensor takes the
form
\begin{equation}\label{eq:viscp}
T^{\mu\nu}_{(HD)} = (\rho h + Q_{visc})U^\mu U^\nu + g^{\mu\nu} P + Q^{\mu\nu}\,.
\end{equation}
Such a viscosity may be able to deal
with mildly relativistic shocks but it does not even reduce properly
in the non relativistic limit.

A general imperfect fluid
energy--momentum tensor may be written 
\begin{align}
T^{\mu\nu}_{(HD)} &= \rho h U^\mu U^\nu + g^{\mu\nu} P + Q^{\mu\nu}\,, \\
Q^{\mu\nu}        &= - 2\eta\sigma^{\mu\nu} -\xi\theta h^{\mu\nu} \,,
\end{align}
where $\eta$ and $\xi$ is the shear and bulk viscosity coefficients, 
\mbox{$\theta = \nabla_\mu U^\mu$} 
is the expansion of fluid world lines, and $\sigma^{\mu\nu}$
is the spatial shear tensor (see \Eq{eq:shear}).
In the non relativistic limit we find that
\begin{align}
T^{tt} - D &\rightarrow \frac{1}{2}\rho v^2 + \rho e_{int}\,, \\
T^{ti}     &\rightarrow \rho v^i\,,
\end{align}
which shows that any consistent shear viscosity should reduce as
\begin{align}
\left(Q^{ti},Q^{ij}\right) &\rightarrow \left(v^j \tau_{ij},\tau_{ij}\right) \\
\tau_{ij} &= \nu_{ij} \left( \partial_i v^j + \partial_j v^i\right)
\end{align}
in the non relativistic limit. Here $\nu_{ij}$ is some viscous constant,
which could depend on the numerical
grid spacing $dx^i$, the local sound speed
and other factors. Neither the viscous pressure
formulation (\Eq{eq:viscp})
nor the bulk viscosity $\xi\theta h^{\mu\nu}$ reduce properly in the limit.
Only the shear viscosity $\eta \sigma^{\mu\nu}$ does so.
The shear viscosity is included directly in the energy-momentum tensor
and it is by construction covariant and preserves energy and momentum.

\subsection{Electromagnetic fields}
The $3+1$ formulation of Maxwell's equations was originally calculated by
Thorne \& MacDonald \cite{1982MNRAS.198..339T} and may be written (see also
Baumgarte \& Shapiro \cite{2003ApJ...585..921B})
\begin{align}
\partial_i \ga E^i &= 4\pi \ga \rho_e\,,\\ \label{eq:solenoid}
\partial_i \ga B^i &= 0\,,\\
\partial_t \ga E^i &= \epsilon^{ijk}\partial_j(\alpha \ga B_k)-4\pi\alpha\ga J^i
		    + \partial_j\left[\beta^j\ga E^i-\beta^i\ga E^j\right]\,,\\
\partial_t \ga B^i &= -\epsilon^{ijk}\partial_j(\alpha \ga E_k)   
		    + \partial_j\left[\beta^j\ga B^i-\beta^i\ga B^j\right]\,,
\end{align}
where $E^i$, $B^i$, $\rho_e$ and $J^i$ are the electric field, magnetic field,
charge density and current density as seen by observers in the {\SFCS}
frame. With the goal of simplifying the equations, we absorb 
the determinant of the 3-metric
in the definition of the different fields. Furthermore we use the fields as seen
by observers in the {\PFIDO} frame. The Maxwell equations then become
\begin{align}
\partial_i \E^i &= 4\pi \brho\,,\\
\partial_i \B^i &= 0\,,\\
\label{eq:Ampere}
\partial_t \E^i &= \epsilon^{ijk}\partial_j(\alpha \B_k)-4\pi\alpha\bJ^i
		 + \partial_j\left[\beta^j\E^i-\beta^i\E^j\right]\,,\\
\label{eq:Faraday}
\partial_t \B^i &= -\epsilon^{ijk}\partial_j(\alpha \E_k)   
		 + \partial_j\left[\beta^j\B^i-\beta^i\B^j\right]\,,
\end{align}
where $\B^i=\frac{\ga}{\gaii}\t B^i=\ga B^i$, $\B_i=\gaii\ga\t B_i=\ga B_i$,
$\E^i=\frac{\ga}{\gaii}\t E^i=\ga E^i$, $\E_i=\gaii\ga\t E_i=\ga E_i$, 
$\brho=\ga\t\rho_e$ and $\bJ^i=\frac{\ga}{\gaii}\t J^i$. Except for
the shift terms and some lapse factors, this equation set is identical to the
special relativistic Maxwell equations.

The energy and momentum equations are modified in the presence of 
electromagnetic fields, reflecting the transfer between fields and fluids.
\begin{equation}
\nabla_\mu T^\mu_{(HD)\nu} = - \nabla_\mu T^\mu_{(EM)\nu} = F_{\nu\mu}\J^\mu\,,
\end{equation}
where $\J^\nu$ is the four current vector. After some algebra we find
\begin{align}\nonumber
\partial_t\E &=\ldots+\ga \left[\beta^i F_{i\mu}\J^\mu-F_{t\mu}\J^\mu\right]
              =\ldots+\frac{\alpha}{\ga} \bJ^i \E_i \\ \label{eq:energy}
	     &= \ldots+\frac{\alpha}{\ga} \bJ \cdot \vec\E\,, \\ \nonumber
\partial_t\P_i &=\ldots+\alpha \ga F_{i\mu}\J^\mu
                =\ldots+\frac{\alpha}{\ga}\left[\epsilon_{ijk}\bJ^j\B^k
		                                + \brho\E_i\right] \\ \label{eq:momentum}
               &=\ldots+\frac{\alpha}{\ga}\left[\bJ\times\vec\B
	                                     +\brho\cdot\vec\E\right]_i\,.
\end{align}
It is worth noticing that the result practically reduces to special
relativity except
for the prefactor $\alpha \ga^{-1}$.

\subsection{Ohm's Law}
If we consider relativistic MHD, we have to supply an Ohm's law to link
the electric and magnetic fields with the current density. A relativistic
version of the standard non-relativistic Ohm's law may be written
\cite{bib:lichnerowicz67,2003ApJ...585..921B,bib:meier04}
\begin{align}\nonumber
\eta_c J_i &= U^\nu F_{i\nu} \\
             &= \alpha E_i U^t + \epsilon_{ijk}\left(U^j + \beta^j U^t\right)
                B^k\,, 
\end{align}
where $\eta_c$ is the resistivity. Using \Eq{eq:trans1} it reduces to
\begin{align}\nonumber
\E_i &= \frac{\eta_c}{W}\bJ_i - \frac{1}{\gaii}\epsilon_{ijk}{\t v}^j\B^k\\
\label{eq:ohmslaw}
     &= \frac{\eta_c}{W}\bJ_i - \frac{1}{\gaii}{\t v}^j\times\vec{\B}^k.
\end{align}
Except for the Lorentz factor $W$ and the single geometric factor, this is
identical to the standard non relativistic result.

In this thesis the ideal MHD condition will not be used directly,
since resistivity is applied in the code. However, taking
$\eta_c=0$ and assuming the ideal MHD condition
Faradays law \Eq{eq:Faraday} in the {\SFCS} may be reduced to
\citep{2003ApJ...585..921B}
\begin{equation}
\partial_t\ga B^i = \partial_j\left((U^t)^{-1}U^i\ga B^j
                                   -(U^t)^{-1}U^j\ga B^i\right)\,,
\end{equation}
which in our notation is
\begin{equation}\label{eq:idealmhd}
\partial_t\B^i = \partial_j\left(\vb^i\B^j
                                   -\vb^j\B^i\right)\,.
\end{equation}

\section{The Numerical Algorithm}\label{sec:2.4}
I have used the equations of motion Eqs.~(\ref{eq:Deqofm}),
(\ref{eq:energy}), (\ref{eq:momentum}), (\ref{eq:Faraday}) together
with \Eq{eq:Ampere} for the current density and an Ohms law \Eq{eq:ohmslaw}
as a basis for the general relativistic code, but even though many mathematically
equivalent forms of the equations of motion exist, they may lead to numerical
implementations with radically different success rates. In this section,
I detail some of the concepts I have used to deal with the problems
that inevitably arise when solving a set of equations numerically.

The most important choice is to determine if we want to exploit the
characteristic structure of the equations or just directly use finite
differencing to solve the equations. In keeping with the tradition in Copenhagen
I have chosen the latter. This has helped to develop the code
in a relatively short time span and I am indebted in my reuse of
techniques and tricks from the non relativistic codes developed in Copenhagen.

The next fundamental
choice is the form of the equations. Either we can use a flux conservative
or a non conservative formulation. There are benefits to both: In the flux
conservative formulation, the Rankine-Hugoniot jump conditions are automatically
satisfied across shock fronts even if the model does not resolve the shocks
entirely. This is not the case for a non conservative formulation.
On the other hand: In a flux conservative formulation, one of the conserved
variables is the total energy. It contains contributions both from the
fluid and from the electromagnetic fields. If the plasma is strongly dominated
by the electromagnetic fields, the internal energy, the difference
between the total and electromagnetic energies, can be swamped by numerical
noise and round off. Another problem --- albeit technical --- is that the
conservative variables in the MHD case are related algebraically to the
so called primitive variables through a sixth order polynomial. There is no
analytical solution to the problem, and an expensive numerical root finder
method has to be used.

I have chosen a cross breed solution: I use conservative variables for
the hydrodynamics, while in the case of MHD, I do not include the
magnetic energy and momentum in the total energy $\E$ and covariant
momentum $\P_i$. The basic reason for not
using conservative variables is due to the problems with magnetically
dominated plasmas. As an added benefit, I circumvent the problems of
finding primitive variables through non analytical methods. Nonetheless,
still at every time step it is necessary
to find the four velocity $\t U^\mu$ and enthalpy $h$ from the
total hydrodynamic energy $\E$ and covariant momentum $\P_i$.

\subsection{Primitive variables}
Given the dynamical variables $D$, $\E$ and $\P_i$ in 
Eqs.~(\ref{eq:D})-(\ref{eq:Pi}) together with the equation of state
for an ideal gas
\begin{equation}
P = (\Gamma - 1)\rho e_{int} = \frac{\Gamma - 1}{\Gamma} \rho (h-1)\,,
\end{equation}
where $\Gamma$ is the adiabatic index, I define two derived quantities
\begin{align}\label{eq:X}
X &\equiv \frac{\E}{D} 
  = (h-1)W + W - 1 - \frac{\Gamma-1}{\Gamma}\frac{h-1}{W}\,, \\
Y &\equiv \frac{\P_i \P^i}{D^2} \label{eq:Y}
  = h^2 (W^2 - 1).
\end{align}
Using \Eq{eq:Y} to solve for $W$ and inserting the solution into \Eq{eq:X}
a fourth order polynomial in $h_m=h-1$ may be constructed, which only contains
$X$, $Y$ and $\Gamma$ in the coefficients, viz.
\begin{align}\nonumber
h_m^4 + 2[\Gamma+1]\,h_m^3 + 
[1 + \Gamma(4 - \Gamma X(2 + X) + 2Y) ]\,h_m^2 + \quad\quad \\ \nonumber
[1 - \Gamma X(X + 2) + (1 + \Gamma) Y ]\,h_m  + \quad \\ \label{eq:root}
\Gamma^2 (1 + Y)(Y - X^2 - 2 X) & = 0.
\end{align}
When the desired root has been found, it is trivial from \Eq{eq:Y} to
obtain $\t U^i\t U_i = W^2 - 1$ and then any other desired quantity.
Fourth order polynomials
may be solved iteratively using a range of different root finder methods,
such as the Newton--Raphson method. I tried this, and even though it most
often worked flawlessly and was fast, for certain corner cases, it is
both unstable
and slow. Slowness in a few cases may be acceptable, but if the method crashes,
the simulation crashes. Stability is the key. An alternative is to use
an analytic formula for the roots, but
great care has to be taken. In any na{\"\i}ve implementation, for example
taking directly the output from Mathematica, the coefficients will cancel
numerically at the slightest difference in scale of the four velocity and
the Lorentz boost
and the result will be imprecise. In the end I settled on a method detailed
in \cite{bib:stegun} to reformulate the problem in terms of roots in one
third order and four second order polynomials. I find the roots using
stable formulae, which guard for cancellations, from \cite{bib:numrecip}.
With this approach the code runs most tests using single precision variables and
only for the most extreme cases (high Lorentz boost and very low pressure),
we have to fall back to double precision. The solver is not only rock solid
but also very fast. Properly implemented with no if-lines and all calculations
vectorised, it takes approximately 20\% of a time step, and therefore
does not, in any way, dominate the problem. Note that a related approach
has been reported in \cite{bib:delzanna03}.

\subsection{Artificial viscosity}
I do not try to solve, neither exactly nor approximately, the Riemann problem at
cell boundaries. Instead, I use finite difference derivatives. To stabilise 
the algorithm it is critical to add AV. During the development of the code I have
tried many different formulations both inspired by the non relativistic codes
developed in Copenhagen, classical formulations of AV and the self consistent AV
detailed in \cite{2003ApJS..144..243A}. In the end I settled for an AV based on a
physical model of shear viscosity derived from the energy momentum tensor
of an imperfect fluid (see section \ref{sec:av}). To determine the viscosity
coefficient $\eta$ in front of the shear viscosity in \Eq{eq:tmunu} I
have extended the prescription already used in the non relativistic
codes in Copenhagen \cite{bib:stagger}, and use a Richtmeyer--Morton type
hyper viscosity that depends on the local conditions in the fluid:
\begin{align}\label{eq:visceta}
\eta_{ij}     &= \Delta x_{ij} \left[ \nu_1 c_s + \nu_3 |\vb| + \nu_2 \Delta l
                                 |\partial_\mu \t U^\mu|_{< 0}\right]\,, \\
\Delta x_{ij} &= \frac{1}{2} Dh \left[\Delta x^i + \Delta x^j \right]\,,
\end{align}
where $c_s$ is the relativistic sound speed, $\Delta l = \max(\Delta x^i)$ and
$|\cdot|_{<0}$ means that the strong shock viscosity only is operative
where there is a compression of the fluid. Except for the sound speed,
the only other changes in the coefficient $\nu_{ij}$ compared
to \cite{bib:stagger} are the use of
$Dh$, as seen by an observer in the local {\PFIDO}, frame instead of the
mass density $\rho$, and the use of
the divergence of the four velocity in the relativistic case compared
to the normal divergence of the spatial three velocity in the non
relativistic case. It is non trivial to find the time derivative of the
Lorentz boost $W$. We found by experimenting with different, mathematically
equivalent prescriptions, that by far the most stable formulation is
\begin{equation}
\partial_t W = \frac{1}{2W}\partial_t \t U^i \t U_i\,.
\end{equation}
The shear viscosity, given in \Eq{eq:shear}, contains time
derivatives of the four velocity too. In the code I use a third order
Runge--Kutta integrator for the normal dynamical variables.
I evaluate the four velocity derivatives by explicit
derivatives, storing old velocities three sub time steps back in time.
This way I get third order correct time derivatives. Unfortunately they are
not correctly time centred and I speculate that some of the problems
I see in the test problems below for high Lorentz boosts may be due to
the time derivatives lagging approximately half a full time step compared to the
rest of the terms. In the energy and the momentum equations (\ref{eq:Eeqofm})
and (\ref{eq:Peqofm}) AV terms arise both on the right hand side and
in the time derivative. I have currently not included the time derivative
of the shear viscosity in the code. 

\subsection{The magnetic field}
The equations are evolved on a staggered mesh (see below) and
the divergence free condition \Eq{eq:solenoid} of the magnetic field
is naturally conserved.
 To raise the entropy in magnetically driven shocks I use the exact 
same formulation
as in \cite{bib:stagger} for the resistivity $\eta_c$ since the Maxwell
equations by construction comply with special relativity, and the only change
has been to substitute a relativistic correct expression for the fast mode
speed.

Ohms law \Eq{eq:ohmslaw} and Amperes law \Eq{eq:Ampere} are used to derive
the electric field and the current density respectively. We use an explicit
time derivative to evaluate the displacement current. Even though it is lagging
behind with half a time step, like the time derivatives of the four velocity,
it has proven very effective in limiting the magnetically driven wave speeds
except when the Alfv\'en velocity becomes close to the speed of light. 
The magnetic part of the code is calculated following the scheme
\begin{itemize}
  \item{Calculate the resistivity $\eta_c$. It is proportional to $\nu_B \nu_3$.}
  \item{Estimate the electric field: 
            $\E_i^\star = - \frac{1}{\gaii}{\t v}^j\times\vec{\B}^k$.}
  \item{Calculate ${\E^\star}^i$ and find the displacement current
        using an explicit time derivative.}
  \item{Calculate an estimate for the current
            $\alpha {\bJ^\star}^i = \epsilon^{ijk}\partial_j(\alpha \B_k)
             + \partial_j\left[\beta^j{\E^\star}^i-\beta^i{\E^\star}^j\right]$.}
  \item{Lower the current and find the final electric field
            $\E_i = \frac{\eta_c}{W}\bJ_i^\star + \E_i^\star$.}
  \item{Use the displacement current to update the current
             $\bJ^i = {\bJ^\star}^i - \frac{1}{\alpha}\partial_t {\E^\star}^i$.}
  \item{Proceed calculating Faradays law \Eq{eq:Faraday} and the energy and
        momentum sources Eqs.~(\ref{eq:energy}) and (\ref{eq:momentum}).}
\end{itemize}
I have tested different variations of the scheme above using the full 
version of the current density $\bJ^i$, including the displacement current,
to find the final electric field. Even though formally better, it turned out to be
less stable, giving short wave oscillations and essentially the same results.

\section{Testing the Code}\label{sec:2.5}
I have implemented an MHD version of the above equations, currently
restricted to special relativity. A pure HD version has been made
for general relativity with diagonal metrics.
To test the code, I have applied a battery of tests that are presented
below. In all tests I have used a 3 dimensional version of the code.
The boundary conditions are implemented
in the $y$ direction by design, and therefore our 1D box has the
size $(1,N_y,1)$. If not stated otherwise,
in all runs, the weak shock viscosity coefficients are
$\nu_1 = \nu_3 = 0.029$, the strong shock viscosity coefficient is
$\nu_2 = 0.55$,
the magnetic resistivity coefficient (see \cite{bib:stagger}) is $\nu_B=1$  and
the Courant limit is $C_{dt}=0.3$. The code can handle more extreme problems
by tuning the different numbers, but I feel that it is important that the code
``just works''; in real physical applications the results should not rely too
much on
the tuning of these technical parameters, since that would question
the validity of the
results. As an example, by just decreasing the Courant limit 
and the weak viscosity $\nu_1$ I am able to run the wall shock test with
a $W_{inflow}=5$ and obtain
satisfactory results. Only in two of the magnetic tests, I have tuned the
coefficients to facilitate the comparison with other codes.

\subsection{Hydrodynamical tests}
The code has been developed without extending any preexisting relativistic fluid
dynamics code, and it is important to demonstrate that it can solve correctly
a variety of purely hydrodynamical problems. Fortunately, the analytic solution to
hydrodynamic shock tubes is known \cite{bib:pons00,bib:marti94,bib:thompson86}. I
have used the \verb|RIEMANN| program published by Mart\'i and M\"uller
\cite{bib:marti03} to generate the analytic solutions.

\subsubsection{Blast waves}
The blast wave is a problem with two domains initially at rest with a
discontinuous jump in the density and pressure. 
A blast wave is launched at the interface
with a very thin shell of high density. 
The fluid separates in five different states.
Two initial states at the left and right boundary, a rarefaction wave, the contact
discontinuity and a shock wave. This setup is ideal for testing how diffusive
the scheme is, since the shock wave, for suitable parameters, is very thin.
The initial states for the three problems we consider are shown in Table I. 
\begin{center}
\begin{tabular}[htb]{llllllll}
\hline\hline
Table I     & \multicolumn{2}{c}{Problem I}  & 
              \multicolumn{2}{c}{Problem II} & 
              \multicolumn{3}{c}{Problem III} \\ 
Blast waves & Left & Right & Left & Right & Left & Center & Right \\
\hline
Pressure    & 13.33& 0.001 & 100 & 0.1  & 100 & 0.01 & 100 \\
Density     & 10  & 1     & 1    & 1  & 1 & 1 & 1 \\ 
Gas Gamma   &\multicolumn{2}{c}{5/3} &
             \multicolumn{2}{c}{1.4} &
             \multicolumn{3}{c}{5/3} \\
\hline
\end{tabular}
\end{center}
\begin{figure}[th]
\begin{center}
\epsfig{figure=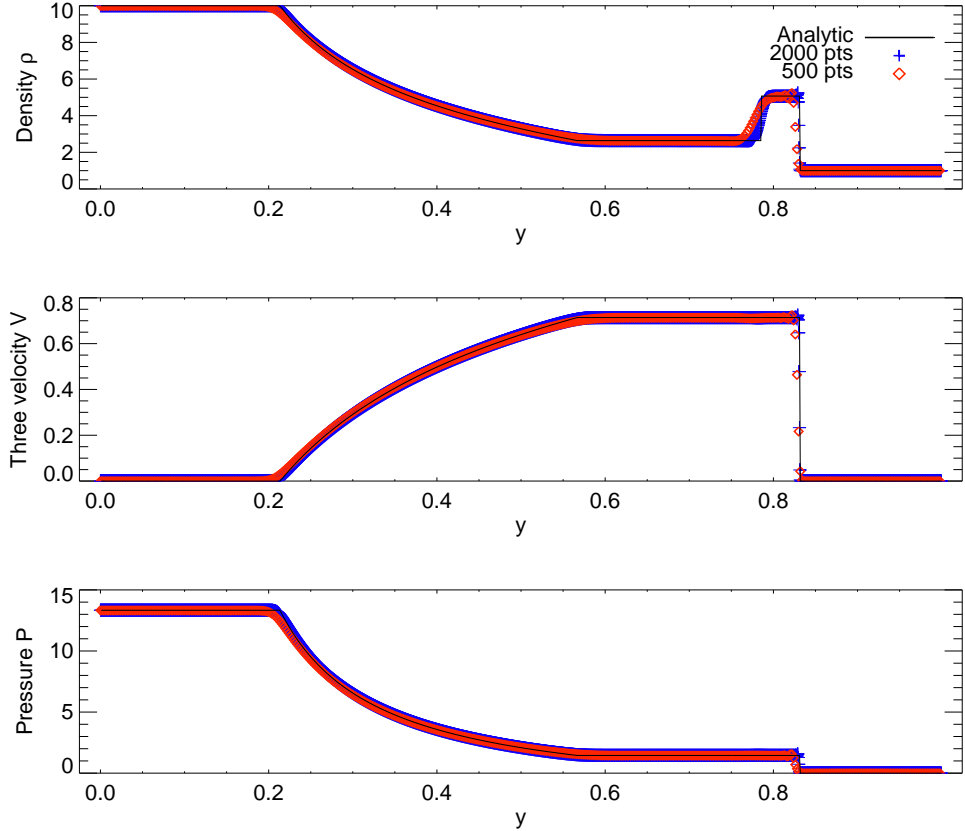,width=\textwidth}
\caption{Problem I: A mildly relativistic blast wave problem. Notice the slight
oscillation at the edge of the shock front. This is due to the large jump in
pressure at that point.}
\label{fig:probI}
\end{center}
\end{figure}
\begin{figure}[th]
\begin{center}
\epsfig{figure=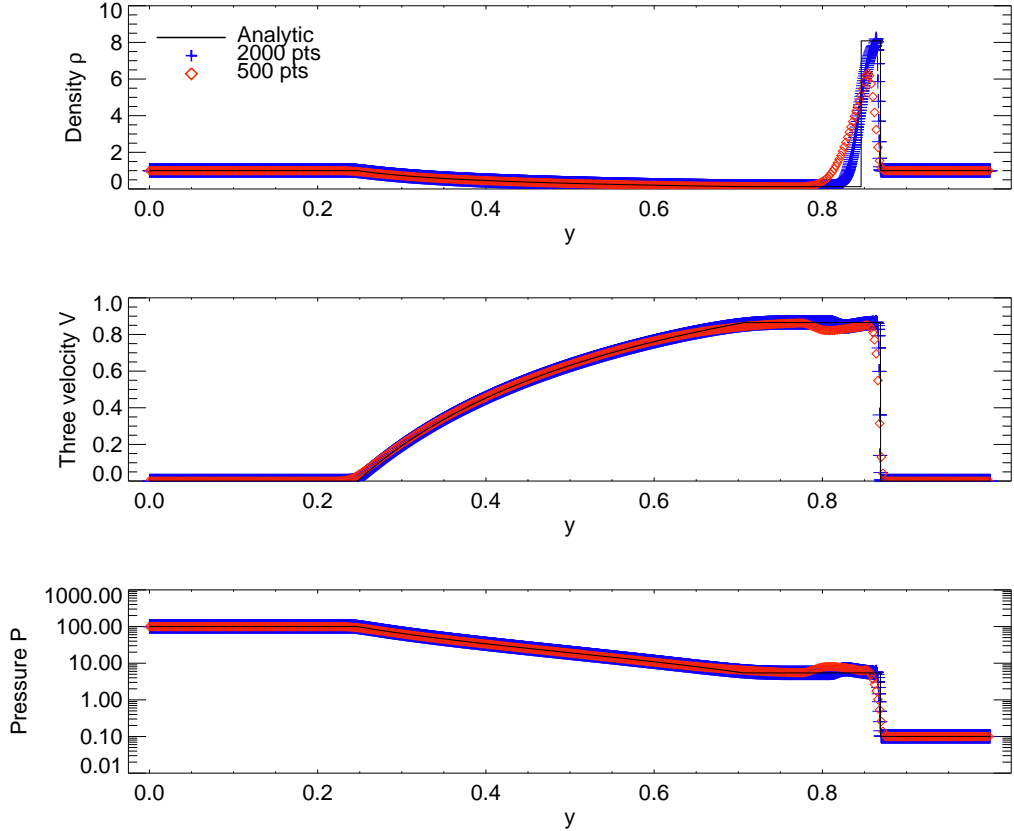,width=\textwidth}
\caption{Problem II: A relativistic blast wave problem. Our code has
some problems with maintaining sharp contact discontinuities at points,
where a high density blob is moving away from a low density area, such as
just behind the high density shell.}
\label{fig:probII}
\end{center}
\end{figure}
\begin{figure}[tphb]
\begin{center}
\epsfig{figure=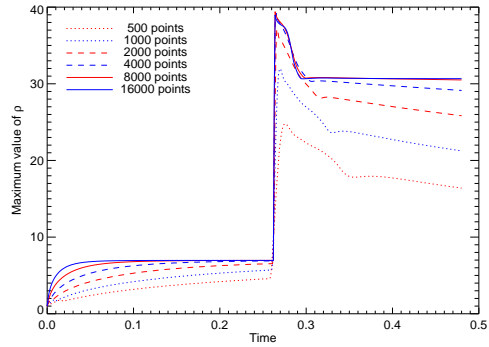,width=0.5\textwidth}
\caption{Problem III: Colliding blast waves. The evolution of the maximum
in density as a function of time is shown.
A resolution of 8000 points is needed to resolve the very thin shell
of high density that is created when the two blast waves collide, and to
accurately calculate the post shock profile, while with 2000 points
we marginally resolve the preshock solution at $t=0.26$.}
\label{fig:probIII}
\end{center}
\end{figure}
\begin{figure}[tphb]
\begin{center}
\epsfig{figure=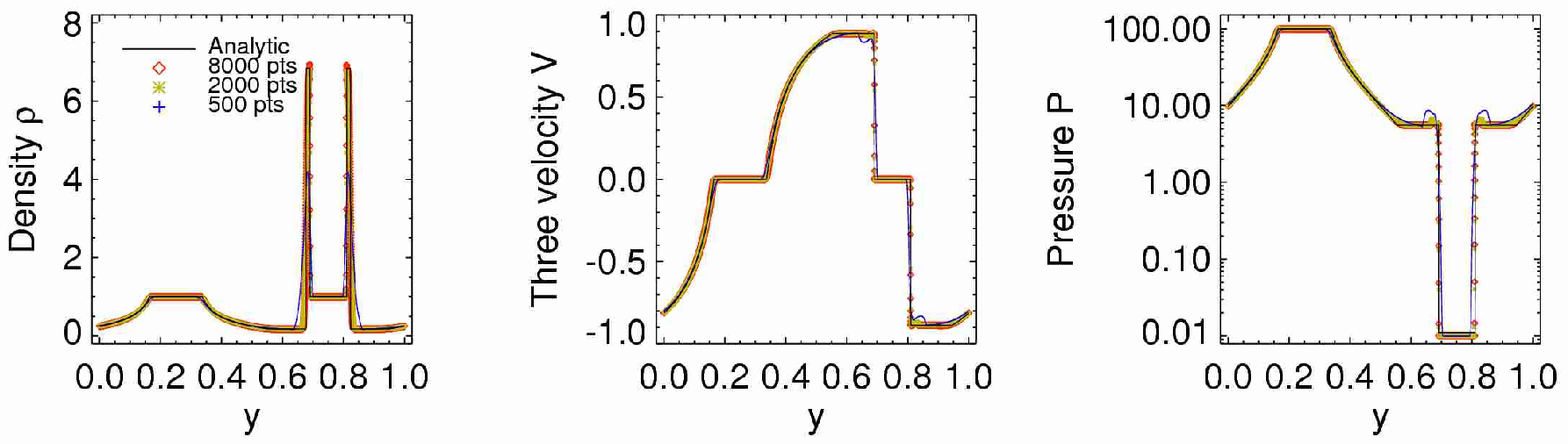,width=\textwidth}
\caption{Problem III: The solution at $t=0.2$, just before the two
shock waves collide.}
\label{fig:probIIIa}
\end{center}
\end{figure}
\begin{figure}[tphb]
\begin{center}
\epsfig{figure=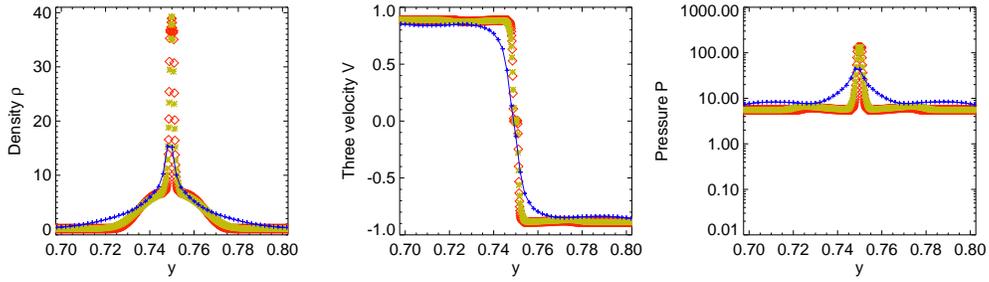,width=\textwidth}
\caption{Problem III: The system, at the collision at $t=0.265$.
Notice we have changed the scale of both the $x$-- and $y$--axis
to reflect the the large change in density, and visualise the thin structures.
 See Fig.~\ref{fig:probIIIa} for legend.}
\label{fig:probIIIb}
\end{center}
\end{figure}
\begin{figure}[tphb]
\begin{center}
\epsfig{figure=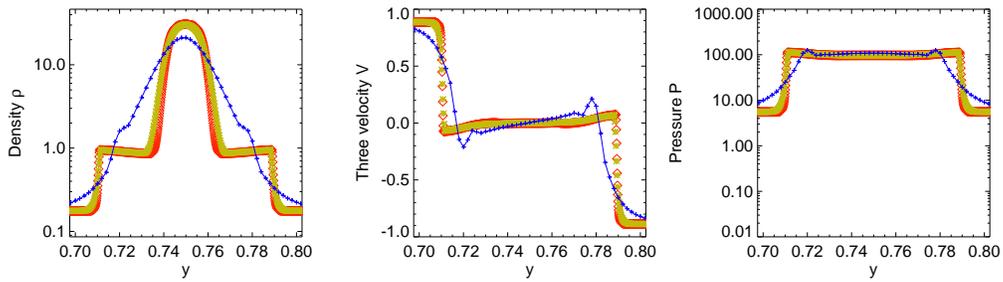,width=\textwidth}
\caption{Problem III: The system, after the collision at $t=0.3$.
 See Fig.~\ref{fig:probIIIa} for legend.}
\label{fig:probIIIc}
\end{center}
\end{figure}
\begin{figure}[tphb]
\begin{center}
\epsfig{figure=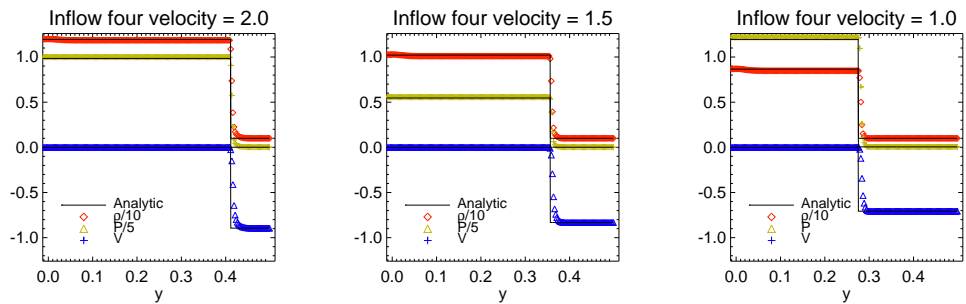,width=\textwidth}
\caption{Problem IV: The wall shock problem. The solution is shown at $t=2$ and
the resolution is 200 points.}
\label{fig:probIV}
\end{center}
\end{figure}
\begin{figure}[tphb]
\begin{center}
\epsfig{figure=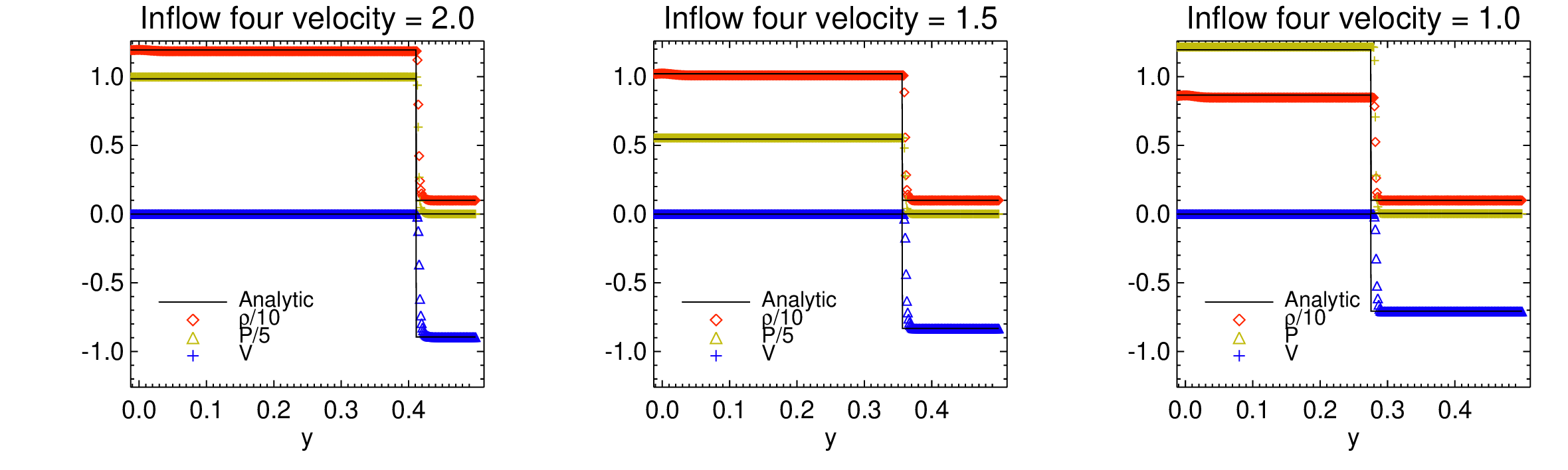,width=\textwidth}
\caption{Problem IV: The same as in Fig.~\ref{fig:probIV}, but the
resolution is 400 points. Notice that the number of points in the shock
interface stays the same for different resolutions; about 3, 2 and 1 1/2
points for the different velocities.}
\label{fig:probIVb}
\end{center}
\end{figure}

Problem I, shown in Fig.~\ref{fig:probI}, is a classic shock tube,
that most relativistic codes
have been tested against (see \cite{bib:marti03} for a compilation). Ideally
the right state should have zero pressure but due to numerical reasons
we have set it to $0.001$. A small weakness of the code is already
visible in this test. When a high mass density region separates from
a low density region, such as at the contact discontinuity in
Fig.~\ref{fig:probI}, there
is a certain amount of stickiness. It is in fact a feature to avoid
low density regions to develop into true vacuums, making the code crash, but
it also makes advecting high density blobs become more diffusive at the
trailing edge. The shock velocity is maintained to a very high precision
and the rarefaction wave is near perfect too, even at low resolutions.

Problem II is a more relativistic variation of problem I. The shock wave
is propagating
with $0.92 c$. At $t=0.4$ the shell has a thickness of $\Delta y=0.023$
or 11 grid zones at a resolution of 500 points. The AV spreads out the
discontinuity over 6 points and this explains why the shock wave is under resolved
at this resolution. 2000 points are needed to get a reasonable solution at
$t=0.4$. Notice also, that the diffusion in the density impacts on the flat
profiles of pressure and velocity.

Problem III is the most extreme shock tube. To make a different
setup I have removed the rigid boundaries and instead imposed periodic
boundaries (see Fig.~\ref{fig:probIIIa}). A similar problem was considered
by Mart\'i and M\"uller \cite{bib:marti94}.
Compared to problem II, the pressure in the right zone is also lowered, and the
equation of state is more sensitive to the pressure.

When the two shock waves collide at $t=0.26$ a very dense shell is created.
To track the evolution in an easy way, I have plotted the maximum density as a
function of time in Fig.~\ref{fig:probIII} for different resolutions.
To resolve the preshock state reasonably well, at least 2000 points are needed,
while 8000 points are necessary to resolve the high density region and
the post shocks. In \cite{bib:marti94} 4000 points were
needed using a shock capturing PPM method to accurately model their problem.

\subsubsection{The wall shock}
The last hydrodynamical problem I have tested against is the wall shock. A cold
fluid comes in from the right and hits a wall at the left edge where it is
reflected.
The inflow density is $\rho=1$ and the adiabatic index is $\Gamma=4/3$.
When reflected a warm dense medium builds up. Figs.~\ref{fig:probIV} and 
\ref{fig:probIVb} shows the solution at different resolutions and
time $t=2$ for mildly relativistic
velocities of $v_s=0.9$ and downwards. The analytic
solution to the wall shock problem may be found in \cite{2003ApJS..144..243A}
and \cite{bib:marti03}.\\[2ex]

It is clear from the above tests that the code is working very well up to
a Lorentz factor of about $W=2.5$. For higher Lorentz factors the current
artificial viscosity implementation
becomes problematic. I believe there are two problems with the current
implementation: We use explicit time derivatives for the four
velocities, but exactly because
they are explicit, for a given time step $t$ they are found at $t-\frac{1}{2}dt$,
and if the fluid is highly relativistic this will make a difference. In the
wall shock, I observe that only decreasing the Courant limiter from
the stock $0.3$ to $0.01$, I can
reach an inflow velocity with a Lorentz boost of $3.5$. Anninos \& Fragile
\cite{2003ApJS..144..243A} have developed, to our best knowledge,
the only explicit AV based code that can handle high Lorentz factors. This is
possible, because they include the time derivatives of the viscosity.

\subsection{Magnetohydrodynamical tests}
To validate the magnetic aspects of the code, I have performed
a range of tests. Unfortunately,
in relativistic MHD, no analytic solution is known to the Riemann problem,
and I have to rely on comparison with tests considered by other groups
using different codes and methods. Komissarov published in 1999 a
testbed \cite{bib:komissarov99} (hereafter K99) with different shock tubes.
Unfortunately there were some errors in the tables, which are corrected in
\cite{bib:komissarov02}. Some of the tests were used by De Villiers and
Hawley \cite{2003ApJ...589..458D} and Gammie et al \cite{2003ApJ...589..444G}
to validate their respective GrMHD codes. I have continued this trend by
performing the same tests as in \cite{2003ApJ...589..458D}. They augmented
the testbed of K99 with an Alfv\'en pulse test that tests for correct wave
speed of Alfv\'en waves at different background fluid speeds and degrees of
magnetisation, and a
more complete set of magnetosonic shocks. Presented below are tests of
magnetosonic shocks, magnetised shock tubes and similar Alfv\'en pulses.
\begin{figure}[tphb]
\begin{center}
\includegraphics[width=0.4\textwidth]{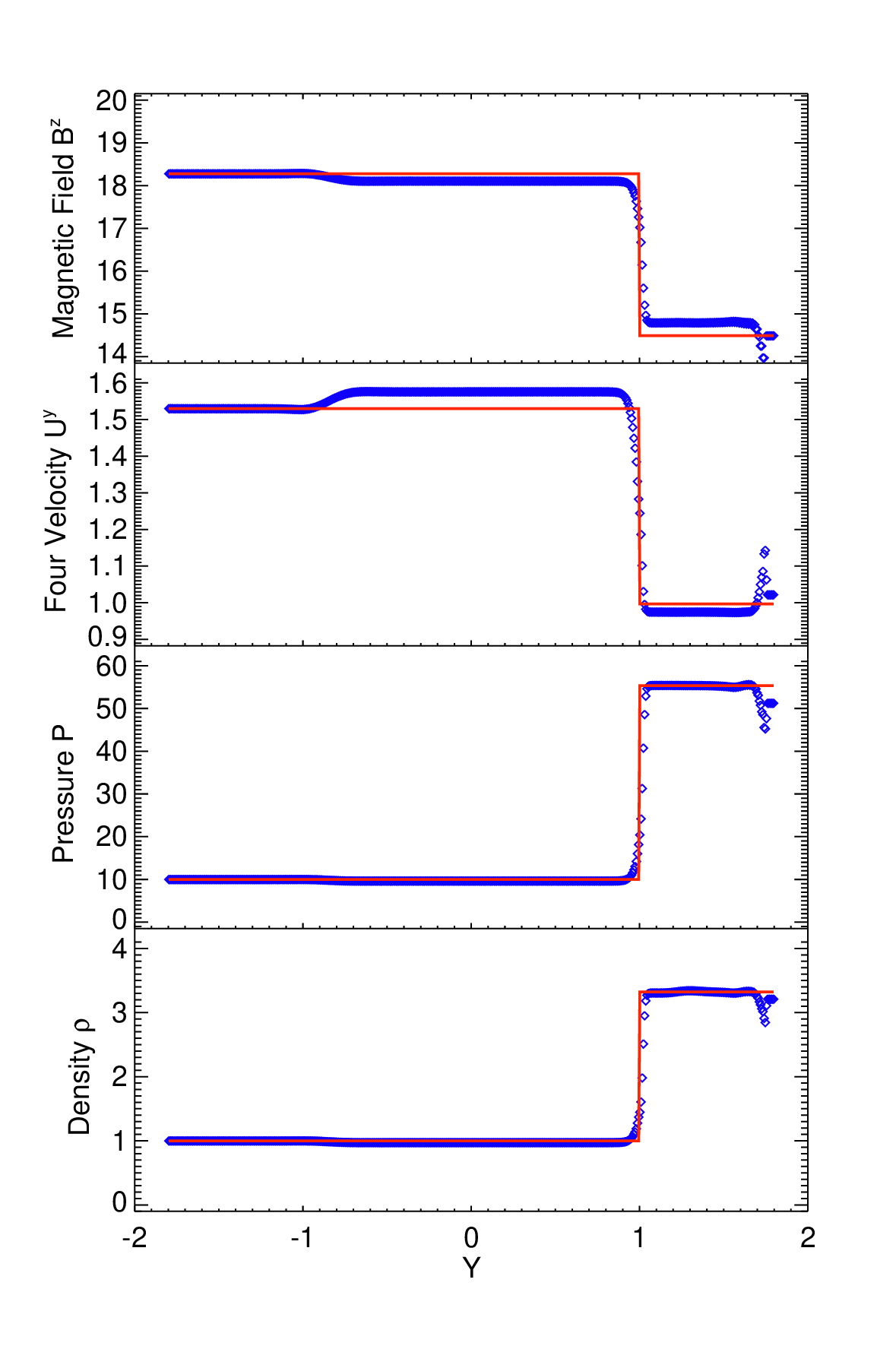}
\includegraphics[width=0.4\textwidth]{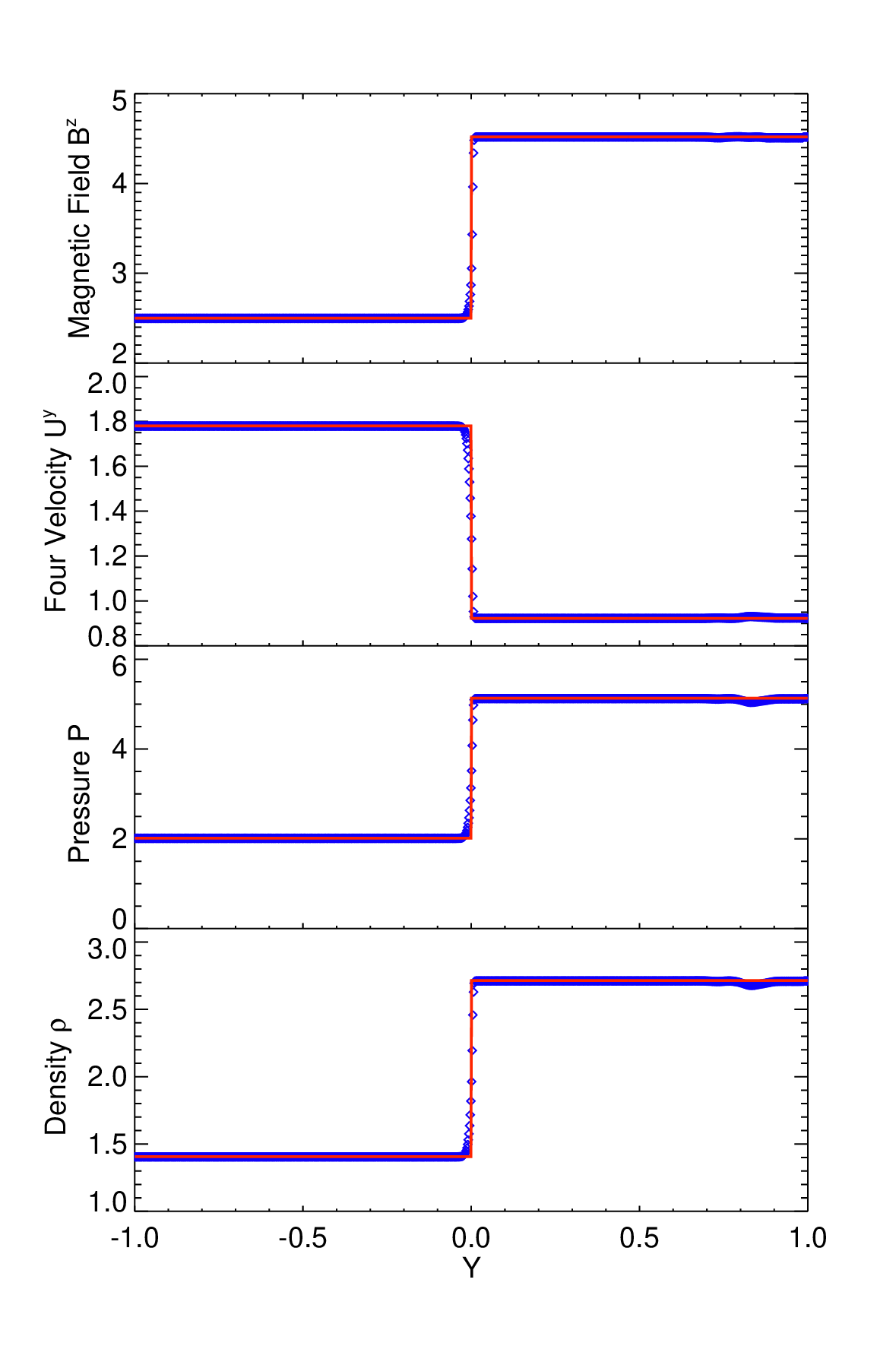}
\includegraphics[width=0.4\textwidth]{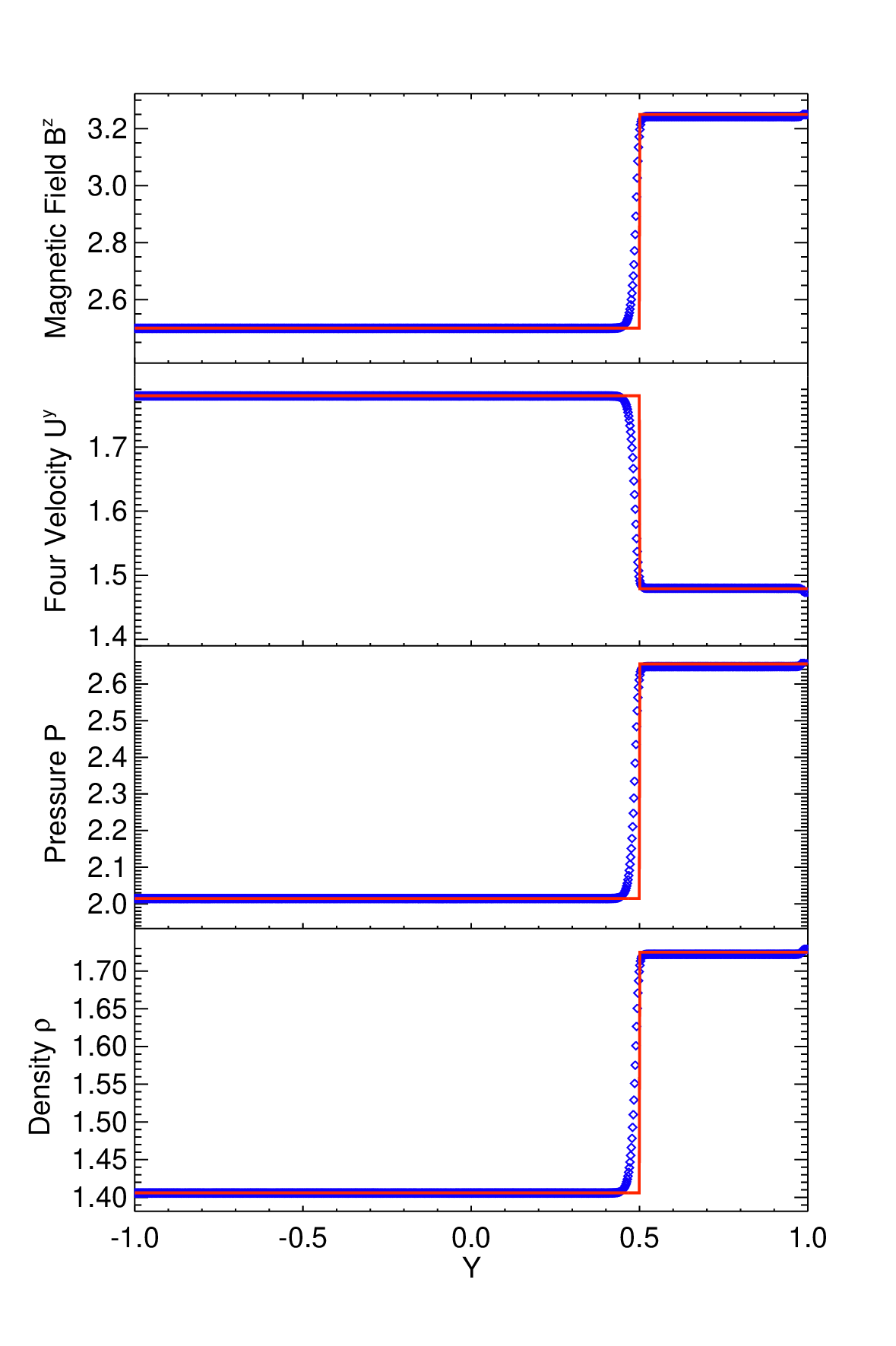}
\includegraphics[width=0.4\textwidth]{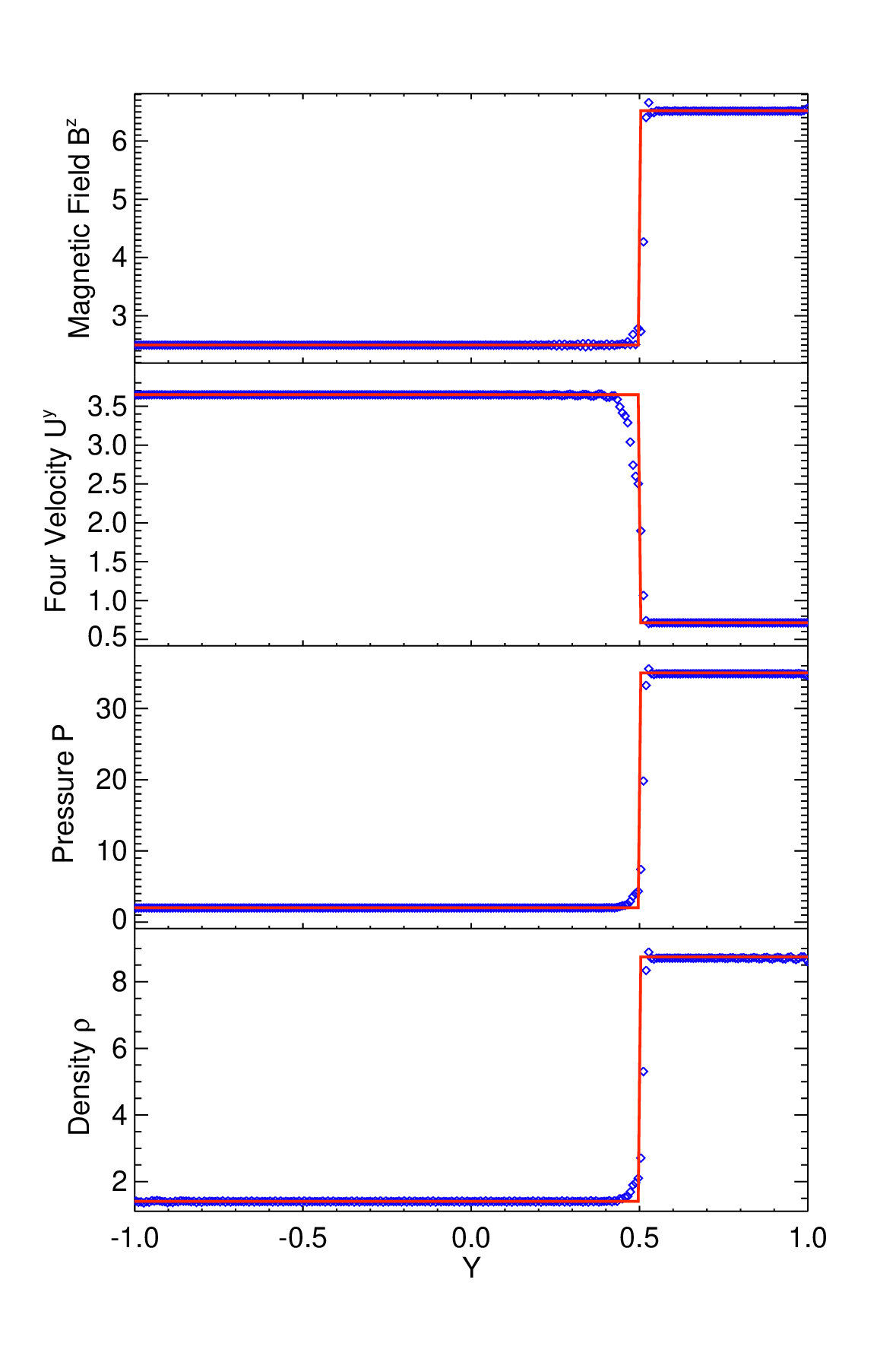}
\caption{Problem V: Magnetosonic shocks. The slow shock is top left,
fast shock I is top right, fast shock II bottom left and fast shock III is
bottom right. The buildup in the right side of the slow shock is due to
interaction with the boundary. In the other shocks, the solution is close
to perfect and buildup does not occur.}
\label{fig:probV}
\end{center}
\end{figure}
\begin{figure}[tphb]
\begin{center}
\includegraphics[width=0.4\textwidth]{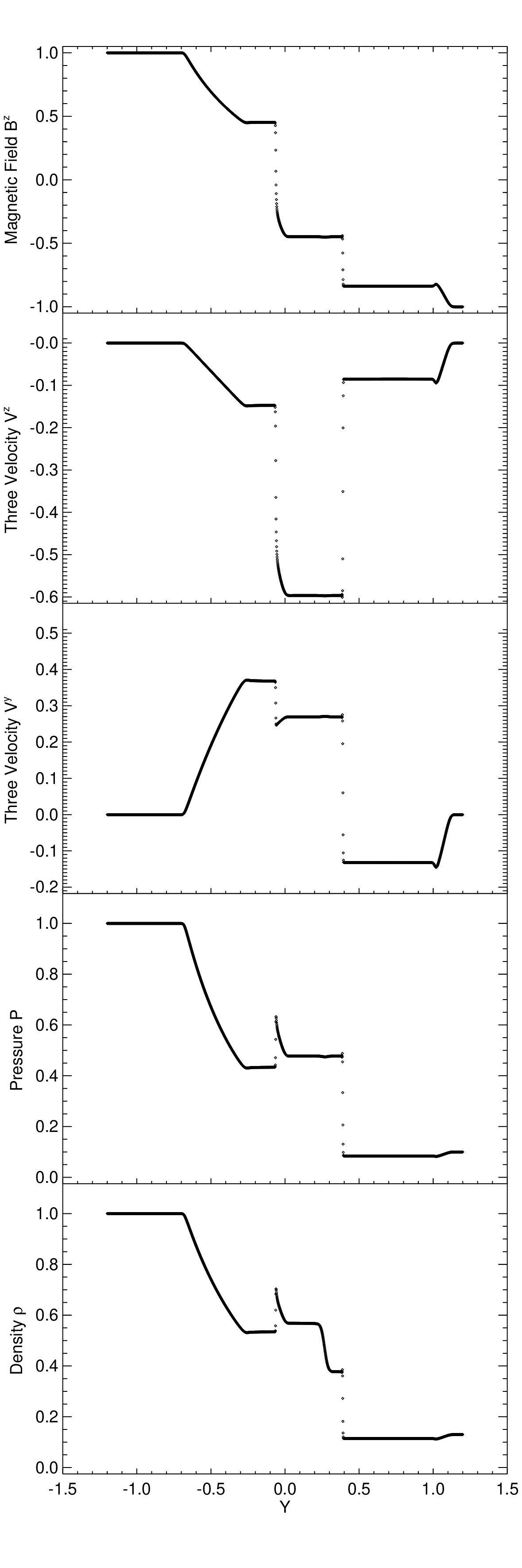}
\includegraphics[width=0.4\textwidth]{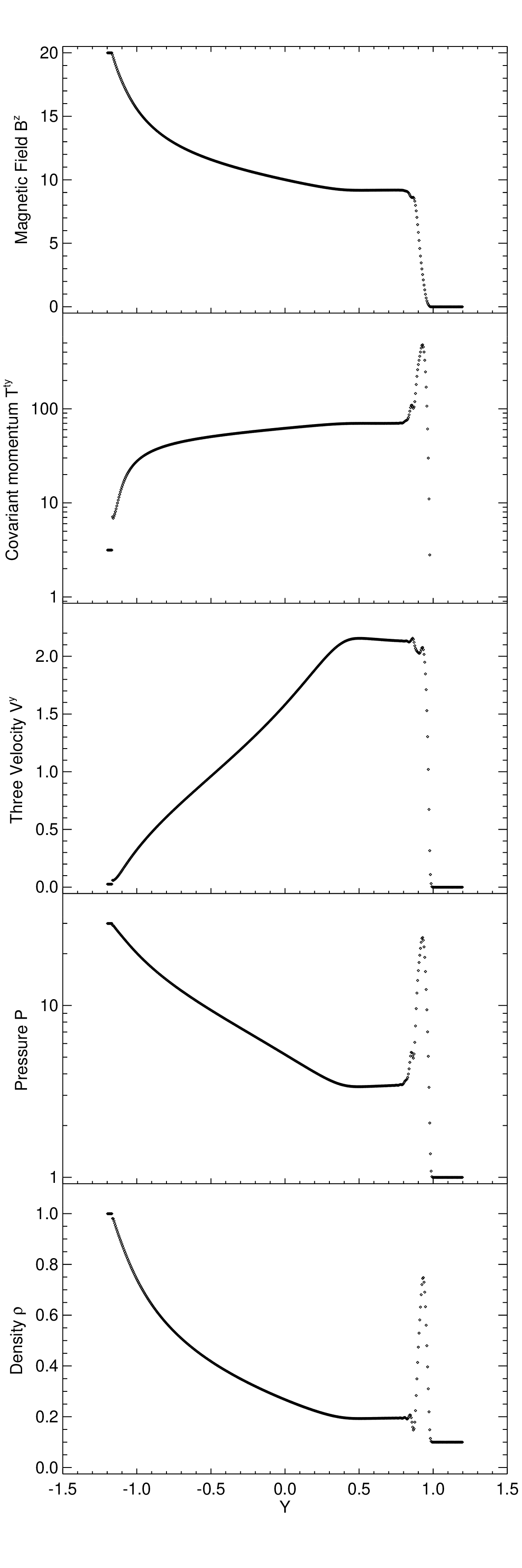}
\caption{Problem VI: Magnetised shock tubes. To the left is the relativistic
version of the Brio \& Wu shock tube, to the right the K99 shock tube 2. Compared
to Figs.~6 and 7 in \cite{2003ApJ...589..458D} and Fig.~6 in
\cite{bib:komissarov99} it is clear that most of the different waves have
the correct amplitude, but there are problems with too high wave speed
and therefore errors in the rarefaction wave. This is most pronounced
for the K99 shock tube to the right.}
\label{fig:probVI}
\end{center}
\end{figure}
\begin{figure}[tphb]
\begin{center}
\includegraphics[width=0.8\textwidth]{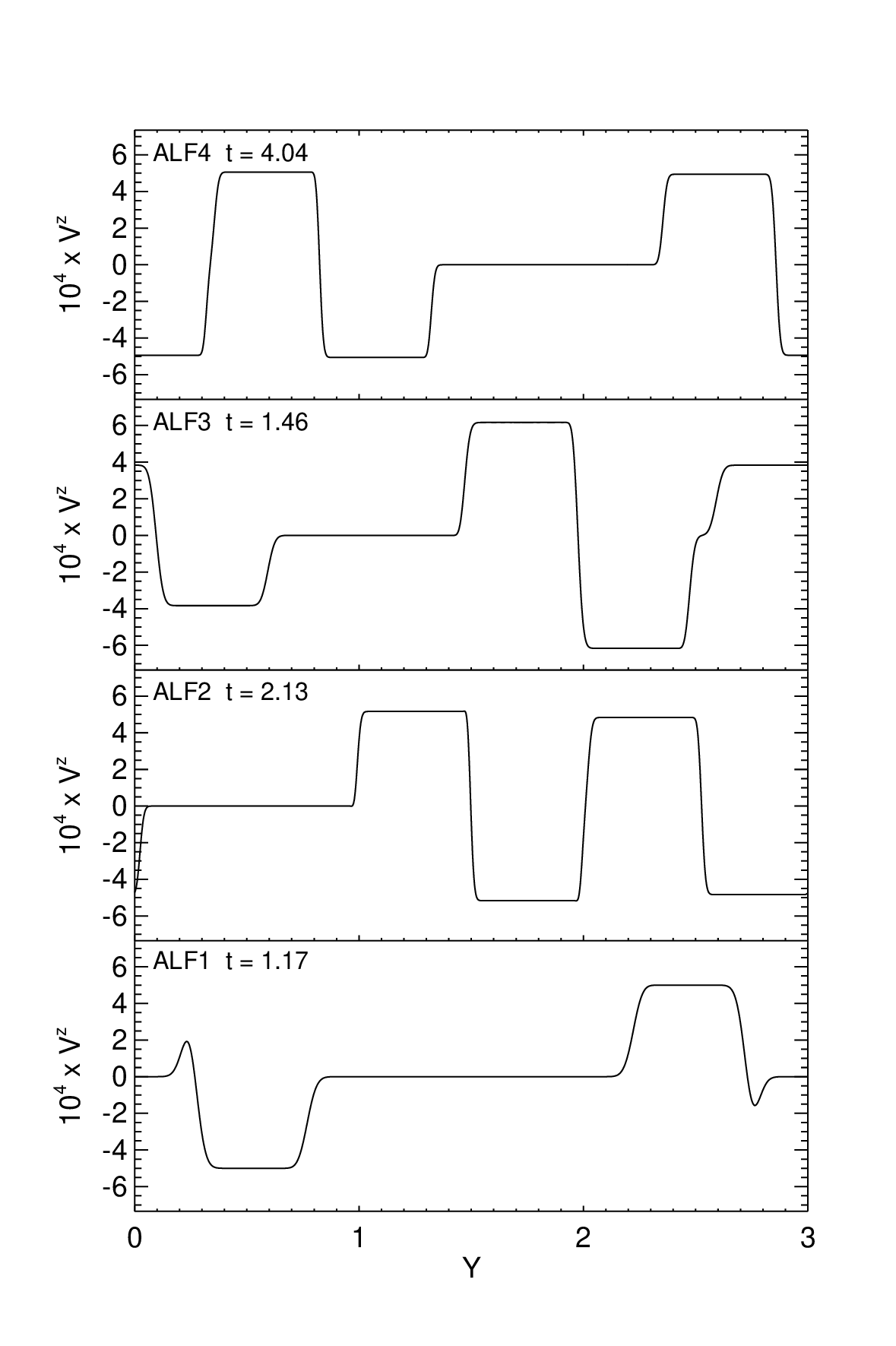}
\caption{Problem VII: Alfv\'en pulse test. We start two Alfv\'en pulses
at $y=1.5$. The wave speeds depend on the background fluid velocity
and the degree of magnetisation. We begin to get significant errors when
$v_A \gtrsim 0.7c$. In all figures, the time is selected to have the two
waves line up. This is not the case for ALF1 and ALF3.}
\label{fig:probVII}
\end{center}
\end{figure}
\subsubsection{Magnetosonic shocks}
In Fig.~\ref{fig:probV} I present a collection of four different standing
magnetosonic shock waves. The parameters of the different waves may be found in
table II and have been taken from \cite{2003ApJ...589..458D}.
In the most extreme shock, the Fast Shock III, we had to decrease the Courant
limit to $C_{dt}=0.1$ and the shock viscosities to $(\nu_1,\nu_2)=(0.001,0.03)$.

For all cases the solution is in excellent agreement with the analytical
solution. Only in the case of the slow shock a slight over density has built
up and is propagating away from the shock wave. This might be due
to a relaxation of slightly imperfect initial conditions, and the solution has
instead settled to a new static solution with a small difference in the
parameters.
In the cases of the fast shocks initially there is a perturbation too,
but only as a small temporary ripple. In the cases of the Fast Shock II and III
(the lower plots in Fig.~\ref{fig:probV}) the ripple has already been advected
out of the box, while in the case of the Fast Shock I it can still be seen at
the right edge of the figure.

\begin{figure}
\begin{center}
\begin{small}
\begin{tabular}[htb]{lllll}
\hline\hline
Table II    & \multicolumn{2}{c}{Slow Shock ($V_s = 0.5$)}    & 
              \multicolumn{2}{c}{Fast Shock I ($V_s = 0$)}  \\
            & Left & Right & Left & Right \\
\hline
Pressure    & 10 & 55.33 & 2.015 & 5.135 \\
Density     & 1  & 3.322 & 1.406 & 2.714 \\
Four Vel &(0,1.53,0)  &(0,0.957,-0.682)&(0,1.78,0.114) &(0,0.922,0.403)\\
Mag Field&(0,10,18.28)&(0,10,14.49)&(0,3.33,2.5)&(0,3.33,4.52)\\
Gamma   &\multicolumn{4}{c}{4/3}\\
$t_{\textrm{final}}$ &\multicolumn{2}{c}{2.0} 
            &\multicolumn{2}{c}{2.5}\\
Grid size   &\multicolumn{2}{c}{512} 
            &\multicolumn{2}{c}{1024}\\
\hline
\hline
            & \multicolumn{2}{c}{Fast Shock II ($V_s = 0.2$)} &
              \multicolumn{2}{c}{Fast Shock III ($V_s = 0.2$)} \\
            & Left & Right & Left & Right \\
\hline
Pressure    & 2.015 & 2.655 & 2.015 & 34.99 \\
Density     & 1.406 & 1.725 & 1.406 & 8.742 \\
Four Vel &(0,1.78,0.114)&(0,1.479,0.28)&(0,3.649,0.114)&(0,0.715,0.231)\\
Mag Field&(0,3.33,2.5)&(0,3.33,3.25)&(0,3.33,2.5)&(0,3.33,6.52)\\
Gamma   &\multicolumn{4}{c}{4/3}\\
$t_{\textrm{final}}$ &\multicolumn{2}{c}{2.5} 
            &\multicolumn{2}{c}{2.5}\\
Grid size   &\multicolumn{2}{c}{1024} 
            &\multicolumn{2}{c}{512}\\
\hline
\hline
            & \multicolumn{2}{c}{Relativistic Brio \& Wu} &
              \multicolumn{2}{c}{Shock tube 2 from K99} \\
            & Left & Right & Left & Right \\
\hline
Pressure    & 1.0 & 0.1 & 30 & 1.0 \\
Density     & 1.0 & 0.13 &1.0 & 0.1 \\
Mag Field&(0,0.75,1.0)&(0,0.75,-1.0)&(0,0,20)&(0,0,0)\\
Gamma   &\multicolumn{4}{c}{4/3}\\
$t_{\textrm{final}}$ &\multicolumn{4}{c}{1.0} \\
Grid size   &\multicolumn{2}{c}{2048} 
            &\multicolumn{2}{c}{512}\\
\hline
\end{tabular}
\end{small}
\end{center}
\end{figure}
\begin{figure}
\begin{center}
\begin{small}
\begin{tabular}[htb]{lllllllll}
\hline
\hline
\multicolumn{8}{c}{Table III: Alfv\'en pulse tests}\\
Test  &$\beta$&$v^y$&$v_a^+$&$v_a^-$&$10^4\times A^+$&$10^4\times A^-$&time&$B^z$\\
\hline
ALF1  &0.05&0.0&1.04(0.85)&-1.04(-0.85)&5.0(5.0)&5.0(5.0)&1.17&7.8\\
ALF2  &0.315&0.249&0.48(0.47)&0.00(0.00)&4.8(4.7)&5.2(5.3)&2.13&5.4\\
ALF3  &0.1&0.8&1.09(0.95)&0.33(0.34)&3.8(2.5)&6.2(7.5)&1.46&10.7\\
ALF4  &0.315&0.088&0.33(0.33)&-0.17(0.17)&0.49(0.49)&0.51(0.51)&4.04&5.0\\
\hline
\end{tabular}
\end{small}
\end{center}
\end{figure}
\subsubsection{Magnetised shock tubes}
I have performed two magnetised shock tube tests and the parameters
may be found in table II. The first is a relativistic
version of the classic shock tube of Brio and Wu \cite{bib:briowu88}.
The shock tube is not very extreme and with a resolution of 2048 points,
just like in \cite{2003ApJ...589..458D} we clearly resolve all shock waves.
The solution is shown at $t=1$. Comparing with Fig.~7 in
\cite{2003ApJ...589..458D}
we see that the wave speeds are wrong. The right rarefaction wave has reached
$y=1.1$ and is superluminal while it should have reached $y=0.9$.
The left rarefaction wave is in good agreement with Fig.~7
in \cite{2003ApJ...589..458D} propagating with $v\approx0.68$.

I was only able to obtain a stable solution of shock tube 2 of K99
by lowering the viscosity to $\nu_1=0.001$ and enhancing the magnetic resistivity
to $\nu_B=3.0$. The shock tube is on the limit of the codes capability; small
oscillations in the density $\rho$ just behind the shock wave in
Fig.~\ref{fig:probVI} are evident and there are large errors in the
rarefaction wave, which propagates superluminally at $v=1.2c$.
The forward going shock wave is only slightly wrong propagating with
$v\approx 1$, where it should be going with $v=0.95$.

\subsubsection{Alfv\'en Pulse test}
The test is conceptually very simple. In a background with constant magnetic
field and velocity in the $y$ direction we set up a small square pulse
in the perpendicular velocity component $v^z$. It splits into two waves
that travel with the Alfv\'en velocity. The test is presented in
\cite{2003ApJ...589..458D} and although simple
in concept it will easily reveal any errors in the wave speed. Since we use
a direct finite difference technique to solve for the magnetic field it
is critical that the displacement current is calculated correctly when
the Alfv\'en speed approaches the speed of light. It is already
evident from the shock tube test above that this is not always
the case, and this test has been invaluable during the implementation,
for assessing different schemes to calculate the displacement current.

Initially there is a constant background magnetic field
$\B^y$ and a constant background fluid velocity $v^y$. On top of
that a small square pulse with transverse velocity $v^z$ is
superimposed. The pulse will split in two waves travelling with
the Alfv\'en velocity, given by \cite{2003ApJ...589..458D}
\begin{equation}\label{eq:pulse}
v_a^{\pm}=\frac{v^y\pm\xi\sqrt{\xi^2 + W^{-2}}}{1+\xi^2}\,,
\end{equation}
where $\xi^2 = |b|^2/(\rho hW^2)$ and $b^\mu$ is the magnetic field measured in
the fluid rest frame. The size of the magnetic field in the fluid rest frame
in a flat space time is related to $\B^i$ as
\begin{equation}
|b|^2 = \frac{1}{W^2}\B^2 + \left[v^i \B_i\right]^2.
\end{equation}
Notice that there is a factor of $4\pi$ in difference
with \cite{2003ApJ...589..458D},
due to different conventions for $\B^i$.

We may parametrise the problem,
by the using the usual definition of $\beta=\sqrt{2P/|b|^2}$ as the ratio of
gas to magnetic pressure in the fluid rest frame. For an ideal equation of
state, in terms of $\beta$ and $P$, $\xi$ is written
\begin{equation}\label{eq:xi}
\xi^2 = \frac{2P}{\rho + \frac{\Gamma}{\Gamma - 1} P}\frac{1}{\beta^2 W^2}.
\end{equation}
To facilitate comparison, I have used the same box size, $0<y<3$, pressure
$P=1/3\,\, 10^{-2}$, background density $\rho=1$ and amplitude of the
perturbation,
$A_0=10^{-3}$, as in \cite{2003ApJ...589..458D}. The adiabatic index is
relativistic with $\Gamma=4/3$. The pulses are set up with a square formed
wave in $v^z$
\begin{equation}
v^z = \left\{ \begin{array}{ll}
A_0  & \textrm{if }1\le y<1.5 \\
-A_0 & \textrm{if }1.5\le y<2 \\
0    & \textrm{elsewhere}
\end{array}\right.
\end{equation}
and for fixed $\rho$ and $P$ the Alfv\'en velocities $v_a^\pm$ only depend on
$\beta$ and $v^y$. The parameters are given in table III and
Fig.~\ref{fig:probVII} shows $v^z$ at the time given in table III.
The times have been selected to those moments in time where the two pulses
line up exactly one after the other, and by visual inspection it is easy to see
how the test fares.

The amplitudes of the two waves are inversely proportional to their
Lorentz factors \cite{2003ApJ...589..458D}
\begin{equation}\label{eq:apm}
\frac{A^+}{A^-} = \frac{W(v_a^-)}{W(v_a^+)}
\end{equation}
and because the starting amplitude $A_0$ is very small, the waves should not
interact with each other.
Then, the sum of the amplitudes is equal to the initial amplitude
$A_0 = A^+ + A^-$. In table III, I have given the measured velocities
and amplitudes together with the expected ones derived from 
Eqs.~(\ref{eq:pulse}) and (\ref{eq:apm}).

The tests are selected
to highlight different regimes of \Eq{eq:pulse}. In ALF1, we have a very low
$\beta$ and consequently the Alfv\'en velocity is close to the speed of
light. The code does not fare well, showing $22\%$ disagreement witht the
expected value. In ALF2 the background fluid velocity is selected such that
one pulse is frozen. It can be verified from the figure that the test is
passed. In ALF3 $v^y=0.8$ and both pulses are travelling to
the right. For the fast moving pulse, again the wave speed is too high with
a 15\% overshoot and the amplitudes are furthermore wrong.
In ALF4 I have adjusted $v^y$ to yield two pulses with $v^+_a = -2 v^-_a$,
and there are no problems with the test.

The tests indicate that the code begins to significantly overestimate
the Alf\'en velocity when $v_a \gtrsim 0.75$, but in all cases the sum of the
amplitudes is conserved. This is in accordance with the results  from the
shocktubes, where correct jumps where observed albeit propagating with
different velocities.\\[2ex]

The many tests presented in this section document both the strengths and
weaknesses of the code. It is essential to know the limits of the code, not
only in terms of stability, but also when to trust the physical models
produced using it.

It is clear that there are some stability
problems with high Lorentz boosts, it is too viscous in the advection of
high density blobs away from low density areas, and that it
overestimates the Alfv\'en speed, when it is relativistic.
The cures to these problems are twofold:
\begin{itemize}
 \item{The time derivatives of four velocities and the electric field have
       to be properly centred.}
 \item{The time derivatives of the shear viscosity in Eqs.~(\ref{eq:Eeqofm}) and
       (\ref{eq:Peqofm}) have to be included.}
\end{itemize}

On the positive side the results all show flux conservation and reproduction of
the proper jump conditions across discontinuities both in HD and MHD tests.
We can successfully model problems with severe pressure and density contrasts
and in most cases faithfully resolve sharp
features with very few points. This is done without showing oscillatory
behaviour. Even though the largest fraction of the CPU time is spent
calculating the shear viscosity, the gains in stability and the sharpness
of discontinuous features increased fundamentally when I shifted from using a
``mockup viscosity'' to a full physically motivated one.

The two points above are not fundamental or unsurmountable
in any way and will be addressed in future work.

\section{Astrophysical Applications}\label{sec:2.6}
We can already apply the code to the understanding of
mildly relativistic phenomena. Here I present first
results from two applications
related to the areas which motivated the development of the code.

\subsection{Decaying magnetic fields in the early universe}
In the introduction, we considered the evolution of magnetic fields
in the early universe. Many analytical studies show that,
at best, it will be very hard to find traces or fingerprints of
primordial magnetic fields in the cosmic microwave background radiation,
but these analyses do not take into account the non linear coupling
between the different wave modes and the possibility of inverse cascades
transferring energy from the small to the large scales.

Christensson et al \cite{bib:christensson01,bib:christensson02} argued
that, in fact, if a turbulent helical magnetic field was created, for
example at the electro weak phase transition, it would undergo
an inverse cascade, while a non-helical field would not.

As a nontrivial 3D test of the code I have initialised a simple
turbulent non-helical magnetic field and a turbulent velocity
field with power spectra given as
\begin{align}
P_\B(k) &= \left< |\B_k|^2\right> = P_0 k^{n_B} 
              \exp\left[-\left(\frac{k}{k_c}\right)^4\right]\,, \\
P_v(k) &= \left< |v_k|^2\right> = P_0 k^{n_v} 
              \exp\left[-\left(\frac{k}{k_c}\right)^4\right]\,,
\end{align}
where the index $k$ indicates the Fourier transform and due to causality,
the exponents are constrained to $n_v\ge0$, $n_B\ge2$. In accordance with
\cite{bib:christensson01}, I have taken them to be at their minimal value.
The cut--off $k_c$ is introduced to limit numerical noise near the Nyquist
frequency. In a $96^3$ run, with a box size of $[0,2\pi]^3$, where $k=1$
corresponds to the largest mode in the box, I found that a value of 
$k_c=10$ was sufficient to quench the numerical noise. To generate
proper divergence free initial conditions, I first calculate the
corresponding vector potential and then take the curl. The initial
magnetic and kinetic energy are both $5\times10^{-3}$ and the average density is
$\rho=1$. The internal energy is initialised such that the sound speed
is relativistic, $c_s^2 = 1/3$.

\begin{figure}[!t]
\begin{center}
\includegraphics[width=0.49\textwidth]{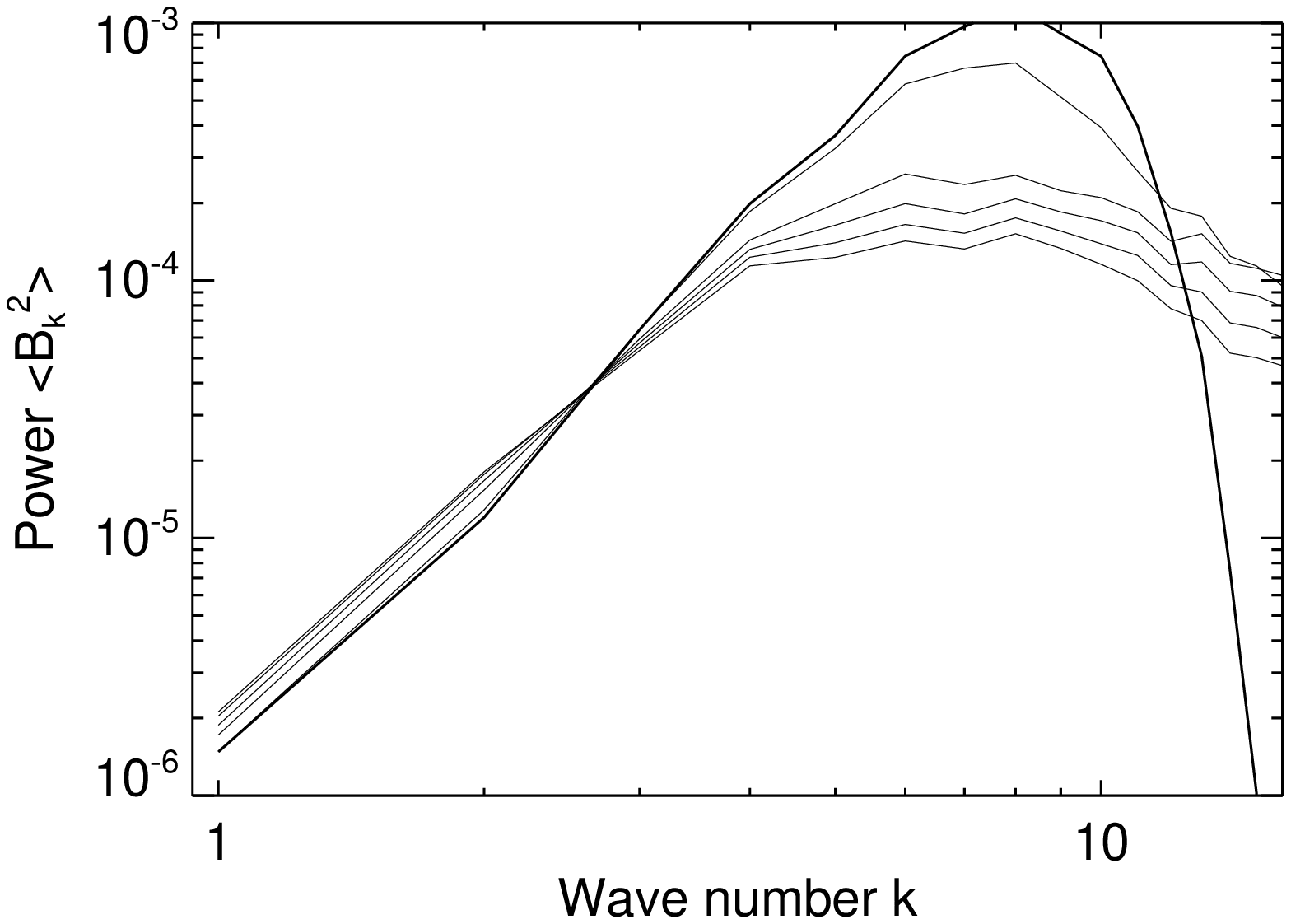}
\includegraphics[width=0.49\textwidth]{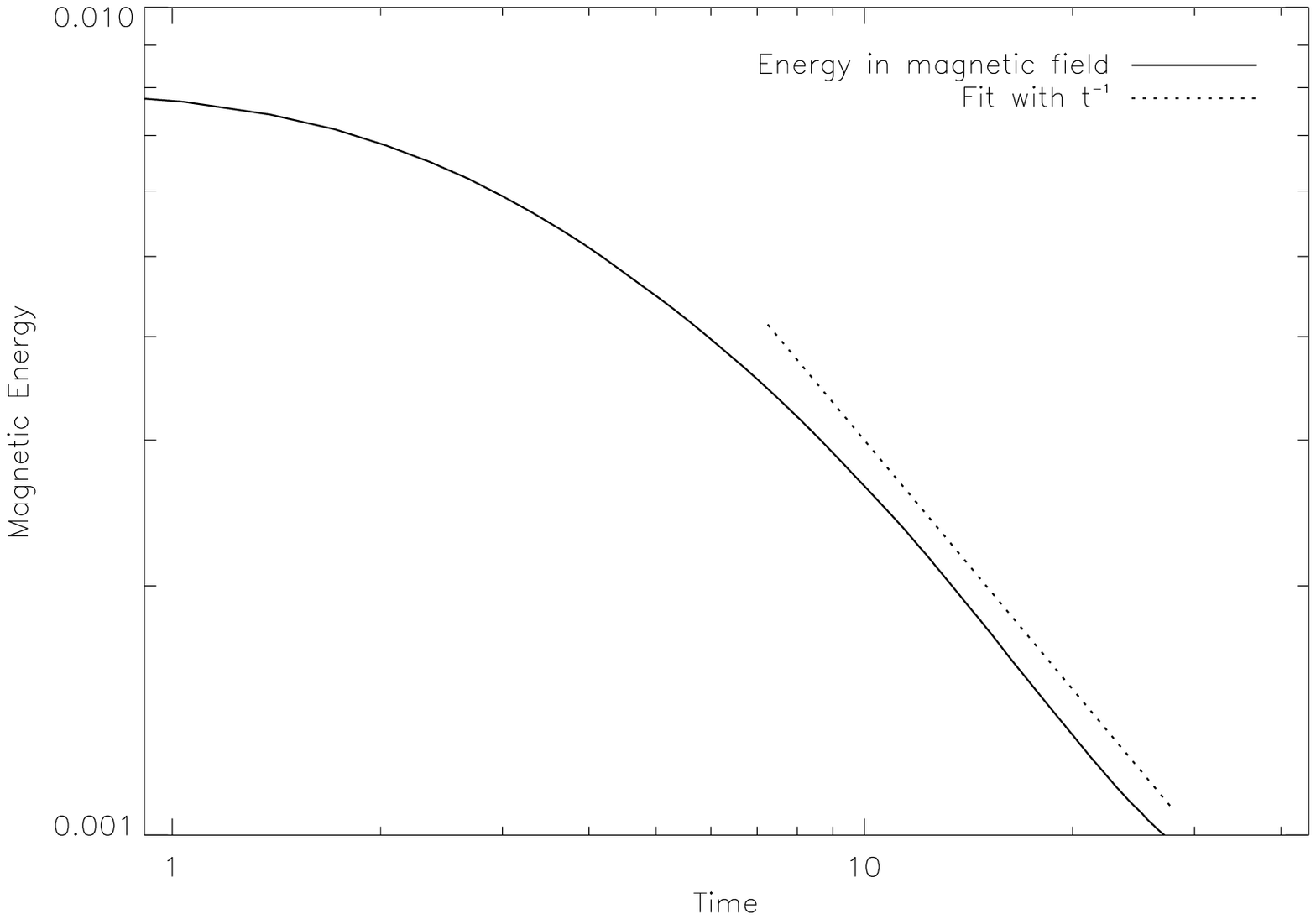}
\caption{Evolution of the power spectrum and the total energy
density for a turbulent magnetic field. The curves to the left are, in
decaying order, for $t=[0,3,9,12,15,18]$.}
\label{fig:pspec}
\end{center}
\end{figure}
Simulations of turbulence are very
sensitive to the type and amount of viscosity used, since it can alter
the long time evolution of high frequency modes significantly. We have
made a series of runs with less and less viscosity and noticed, that
with coefficients $\nu_{1,3}<0.003$ and $\nu_2 < 0.06$ there was no change
in the decay rate of the spectrum. To the left in Fig.~\ref{fig:pspec}
is shown the magnetic power spectrum at different times for a run
with $\nu_{1,3}=0.0003$ and $\nu_2=0.006$. Correspondingly, to the right
is shown the evolution of the magnetic energy in the box that I find
decays as $E_M \propto t^{-1}$.
Comparing my results with \cite{bib:christensson01} I find good
agreement. They found a $E_M \propto t^{-1.1}$ scaling
law. The main difference between my runs and theirs is that they evolve
the non relativistic equations, while I use the relativistic equations.

\subsection{Relativistic jets}
\begin{figure}[tphb]
\begin{center}
\includegraphics[width=\textwidth]{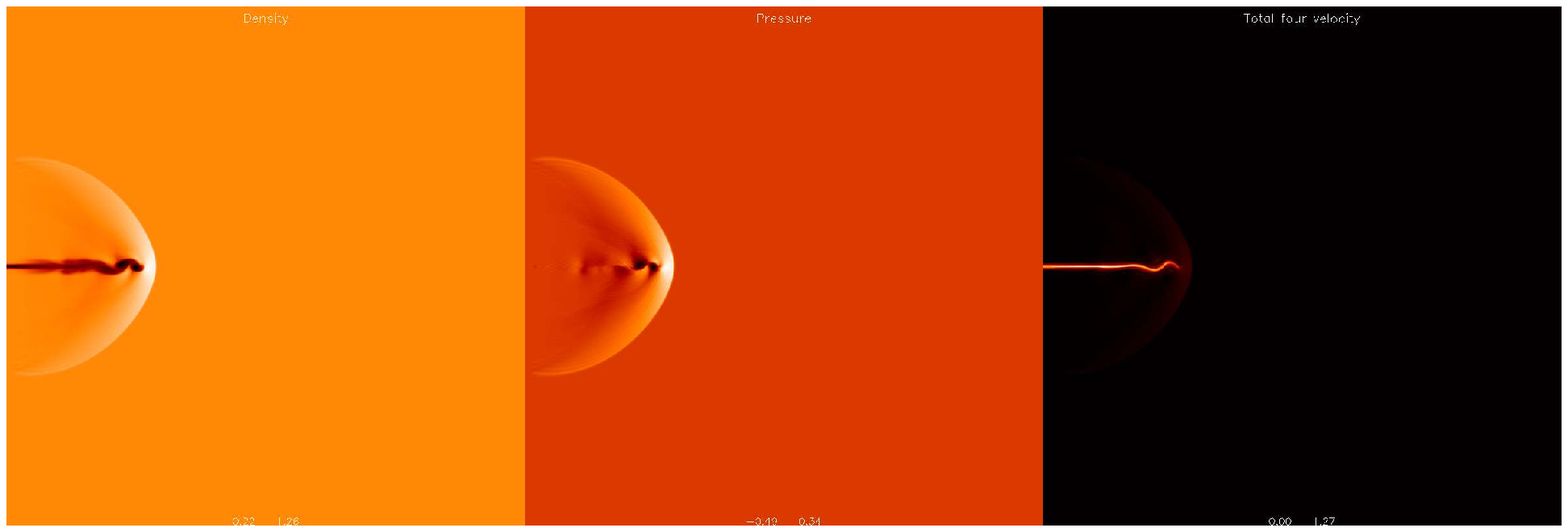}
\includegraphics[width=\textwidth]{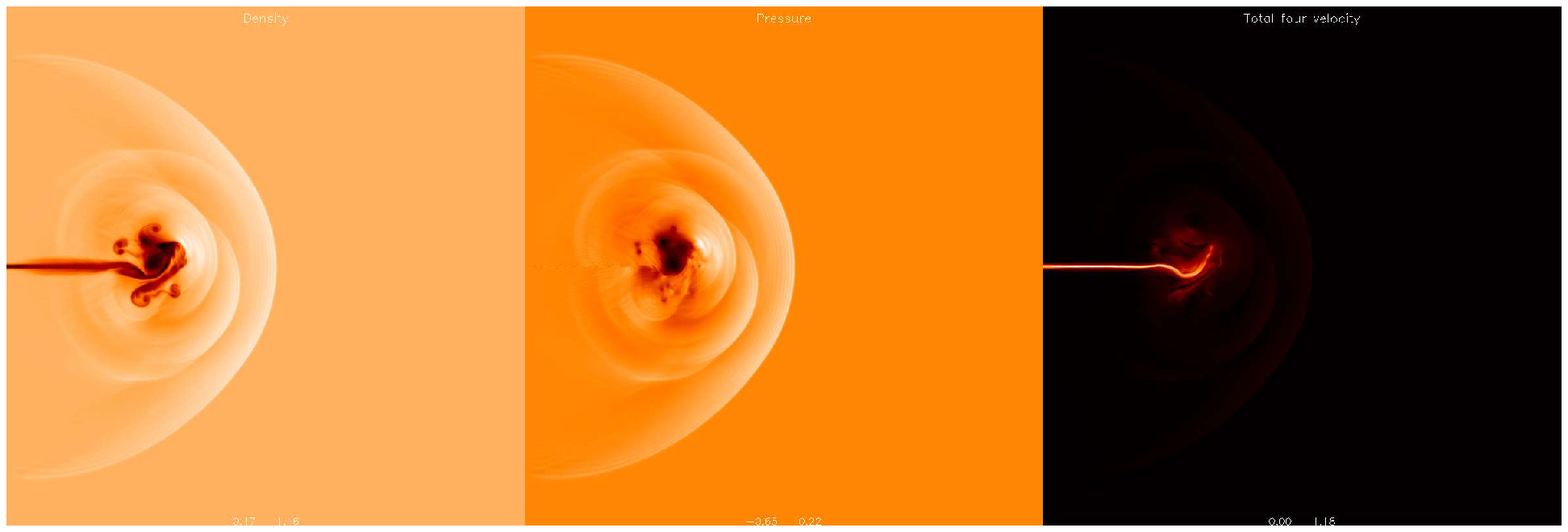}
\includegraphics[width=\textwidth]{jet11_150_final.eps}
\includegraphics[width=\textwidth]{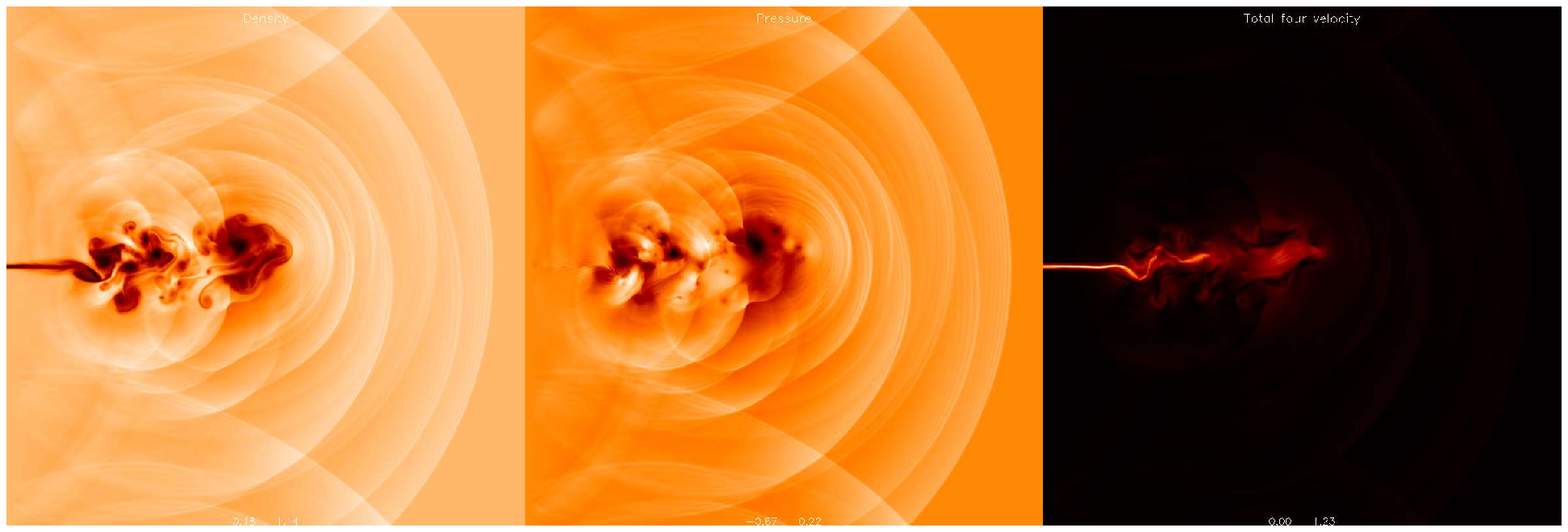}
\caption{From left to right: Density, pressure and four velocity.
From top to bottom: The jet at $t=500,1000,1500,2000$.}
\label{fig:jet}
\end{center}
\end{figure}
Relativistic jets seem ubiquitous in the universe, to be found over a
large range of scales, from sub parsecs to kilo parsecs 
\cite{bib:marti03}. It is one of
the purest displays of special relativity at work. While the other
application tested the impact of the viscosity in the code, a jet
is an excellent test of the codes capability to handle strong shocks.

Taking into account the large computer resources a
3D jet takes, I have chosen to make a 2D jet.
At the moment, only Cartesian geometry is implemented in the code and
I have constructed a slab jet, which is periodic in the $z$-direction.
The injection happens in the $y$-direction,
where rigid boundaries are imposed,
while there are periodic boundaries in the $x$-direction. To avoid
significant collision of the bow shock with itself, the computational
domain is a square with $(N_x,N_y)=(800,800)$.
I have tried both to inject the jet in an purely hydrodynamic medium void of
magnetic fields and in a medium with a parallel magnetic field $\B^y$. As
expected the main difference was further collimation of the jet due to
magnetic confinement. Similar experiments have been reported by other authors
\cite{bib:komissarov99,bib:koide96}. The jet has an injection radius of
$R_j = 5$ cells, the density contrast is $\rho_{ambient}/\rho_{jet}=10$ and
the pressure is $P=1$. There is pressure equilibrium between the jet and
the ambient medium. We inject the jet with a Lorentz factor of $W=1.5$
and the adiabatic index is set to be relativistic with $\Gamma=4/3$.
In Fig.~\ref{fig:jet} a sequence of snapshots are shown. 
The large resolution and thin injection
radius makes it possible to follow the jet until it becomes unstable
and decays. 
At $t=500$ we see a classic jet. At the jet head there is a Mach shock,
and material that has passed through the head is slowly
forming a backflow along the jet, building up a shear layer.
Furthest out is the bow shock. The jet is unstable, and at later
times the jet head disintegrates into a number of vortices and looses
most of the kinetic energy. The perturbation runs backwards, slowly
unwinding the spine of the jet.

\section{Code Implementation: Performance and Scalability}\label{sec:2.7}
To successfully exploit modern massively parallel computers and
clusters of computers, 
which at national centres of supercomputing often consist of hundreds of CPUs,
the numerical
implementation has to be carefully crafted. The program has to run at the
optimal speed for small problems on a single CPU on a variety of architectures,
while at the same time,
it is important to distribute the workload evenly over all CPU's in the
machine (be it a large shared memory computer or a cluster of off-the-shelf workstations).
\subsection{The stagger code}
\begin{figure}[thb]
\begin{center}
\epsfig{figure=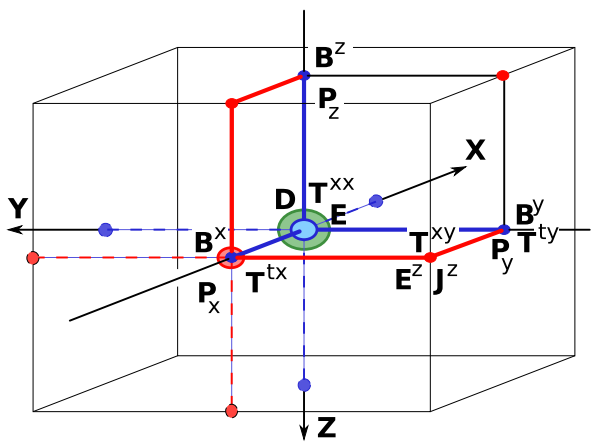,width=0.7\textwidth}
\caption{The basic staggering of different quantities on the grid. The figure
was adapted from \cite{bib:trier}. To make the figure visually easier to
understand I have on purpose drawn a left handed coordinate system.
I use a right handed coordinate system in the code.}
\label{fig:yee}
\end{center}
\end{figure}

All of the fluid dynamics codes currently in use in Copenhagen are based on or derived from
a common base code, the so called \emph{stagger code}. The first version was made by
Nordlund and Galsgaard in 1995
\cite{bib:stagger}. The GrMHD code makes use of the same basic principles. The
equations of motion Eqs.~(\ref{eq:Ampere}), (\ref{eq:Faraday}), (\ref{eq:energy}),
(\ref{eq:momentum}) are solved with finite differences through direct differentiation, and the variables
are staggered on the grid. Scalar variables $D$ and $\E$, and derived scalar
quantities are centred in each cell. The
primary vector quantities $\P_i$ and $\B^i$ are calculated on the faces of the cell while
the derived vector quantities $\E^i$ and $J^i$ are calculated on the edges (see
Fig.~\ref{fig:yee}). The boundary conditions are implemented as in \cite{bib:stagger}.

The differentiation operators are sixth order in space. Derivatives are
calculated at half grid points and use a stencil of 6 points. In many cases this
gives a natural placement of the variables, since the different quantities already
are staggered in space. For example, the electric current $J^i$ is found located at the
edges and according to Eq.~(\ref{eq:Ampere}) is the curl of $\B^i$ (for
the sake of simplicity in this example we disregard the displacement current, $\alpha$ and shear $\beta^i$). $\B^i$ is located at the face of each
cell, and in the code the calculation of the current can be packed into three simple lines:
\begin{align}\nonumber
J^x &= \textrm{ddydn}(\B^z) - \textrm{ddzdn}(\B^y)\\ \label{eq:curlB}
J^y &= \textrm{ddzdn}(\B^x) - \textrm{ddxdn}(\B^z)\\ \nonumber
J^z &= \textrm{ddxdn}(\B^y) - \textrm{ddydn}(\B^x)
\end{align}
In some cases, most notably the complicated viscosity operator, the differentiation does not
place the variables at the desired position on the grid, and interpolation has to be done.
The corresponding interpolation operator is of fifth order.
It also uses a stencil of 6 points \cite{bib:stagger}. A crucial addition to
the original method described in \cite{bib:stagger}, which later has been
employed in most of the stagger based codes is the use of
exponentials and logarithms to produce geometric interpolation. As an analogy, the
geometric mean of two numbers may be rewritten in terms of the arithmetic mean of the 
logarithms
\begin{equation}
G(a,b)=\sqrt{a\cdot b} = \exp\left[\frac{1}{2}\left(\ln a +\ln b \right)\right]
      = \exp\left[H(\ln a, \ln b)\right].
\end{equation}
Geometric interpolation has two very appealing qualities, when dealing with
discontinuities across shocks. First of all, positive definite quantities,
such as the density and the energy, stay positive. Secondly, geometric
interpolation is a much better measure
when the density or pressure is changing with orders of
magnitude over a few points. This happens at shock fronts and
surface transitions.

To make the code easily readable and hide all the loops,
where the different interpolations
and differentiations are done, all operators are hidden in
a set of subroutines. In fact,
Eq.~(\ref{eq:curlB}) corresponds exactly to the simple version of the code. In the
production version, we make an effort to optimise memory references and reuse the cache
memory at each CPU, but even with full parallelisation
the $\nabla\times\B^i$ term
only expands to

\begin{small}
\begin{verbatim}
!-----------------------------------------------------------------
!  Electric current I = curl(B)
!-----------------------------------------------------------------
  do kk=ks,ke
    call ddydn_set(Bz,Jx) ! Jx = ddydn(Bz) - ddzdn(By) 
    call ddzdn_sub(By,Jx)
    call ddxdn_set(By,Jz) ! Jz = ddxdn(By) - ddydn(Bx)
    call ddydn_sub(Bx,Jz)
    call ddzdn_set(Bx,Jy) ! Jy = ddzdn(Bx) - ddxdn(Bz)
    call ddxdn_sub(Bz,Jy)
  end do
\end{verbatim}
\end{small}

The reason why sixth order differentiation and fifth order interpolation
operators are used in the stagger code is a question of balance between
precision and computational load.
The highest wavenumber a given method can resolve depends not only on the
resolution of the mesh, but
also on the order of the interpolation and differentiation operators. It was found
empirically by Nordlund and Galsgaard \cite{bib:stagger} that sixth order 
gives effectively a better resolution than fourth order operators, even
after the somewhat larger computational cost of the 6th order
operations are considered.
Maron \cite{bib:maron04} made a formal investigation of the effective resolution
of different methods and orders and found that
a fourth order scheme can resolve up to 0.24 of the maximal wave number
$k_{max}$, while a sixth order scheme resolves waves with up to 
0.34 $k_{max}$. Going to eighth
order the maximal wave number is 0.4 $k_{max}$. At higher wave numbers the gain is
negligible if one takes into account the added communication 
and amount of ghost zones that
have to be allocated.

\subsection{The paper code: Optimal parallelisation and cache reuse}
Together with J.~Hunsballe \cite{bib:jaab} I performed what essentially
amounts to a complete rewrite of the stagger code.
We still retain the qualities that have been described above. The basic
physical equations are the same, and artificial viscosity is implemented
in the same manner.
We use the same high order interpolation and differentiation operators.
The difference is
in the technical details: Our goal has been to
produce a very high level object oriented code, that is easily readable,
runs at the highest possible speed and scale to hundreds of CPUs.

\begin{figure}[thb]
\begin{center}
\epsfig{figure=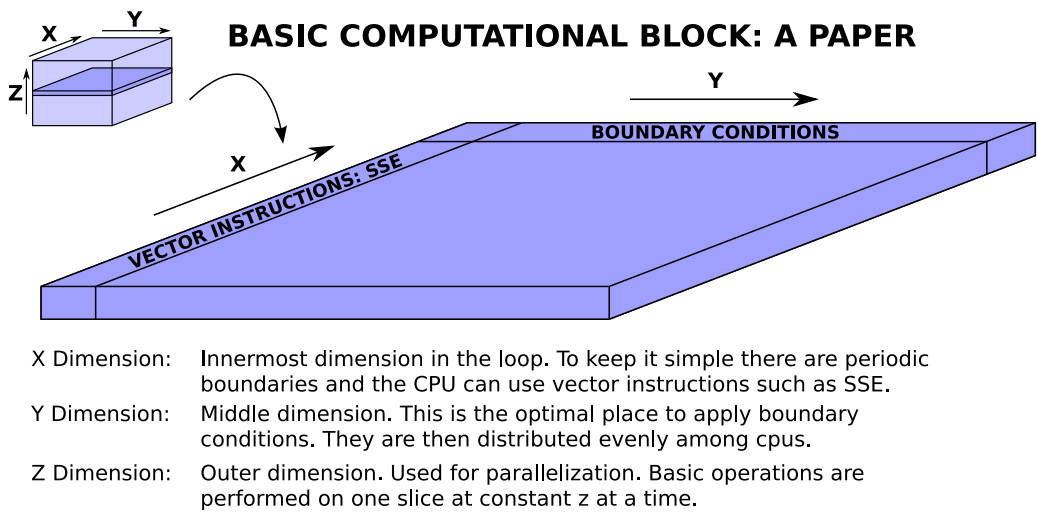,width=\textwidth}
\caption{The basic scope for any calculation in the new Paper Code is the $x-y$ plane giving
optimal reuse of cache, vectorisation and simple implementation of boundary conditions.}
\label{fig:cache}
\end{center}
\end{figure}

In the stagger code the basic scope for any operator (such as a differentiation operator)
has been the full three dimensional array. For example, to interpolate the density to face
values one would write:
\begin{verbatim}
  call xdn_set(rho,xdnr)
  call ydn_set(rho,ydnr)
  call zdn_set(rho,zdnr)
\end{verbatim}
The problem with this approach is that, on modern computers, the bandwidth between the main
memory and the CPU is much lower than the computational power. There is also a big latency
involved. It can easily take 200 clock cycles from the moment the CPU asks for a specific
block of memory until it actually is delivered. To alleviate this problem, there is a small
amount of very fast memory, the cache, often integrated directly on the CPU. On current
high end architectures, such as the Itanium, Power, Alpha, Sparc and Mips based machines the
cache size is of the order of 3-8 MB per CPU, while a normal Opteron or Pentium based CPU
has 1 MB cache. On a small problem, of todays standards, such as a $128^3$ mesh
per CPU, the amount of cache taken by just one array is already 8 MB. That means
the \emph{stagger code} is already beginning to be memory
bandwidth limited, and not limited by the speed of
the CPU at this problem size. In the new Paper Code, the basic scope is instead a slice in
the $x-y$ plane (see Fig.~\ref{fig:cache}); hence the name. The above lines of code would
then be written:
\begin{verbatim}
  do kk=ks,ke
    call xdn_set(rho,xdnr)
    call ydn_set(rho,ydnr)
    call zdn_set(rho,zdnr)
  end
\end{verbatim}
where the \verb|kk|-loop runs over the papers.
The code is almost identical, since we hide the loop index in a global 
variable, but the
characteristics are radically different. Because we use a sixth order scheme 
we now only required to store 5 ``papers'' for the z-operator that needs
values from different
papers of $\rho$, in the cache to keep the CPU running at maximal speed. Even for a $512^3$
problem 5 papers only take up 5 MB of memory, and by testing on an SGI Altix machine with 3 MB of cache, we have found that performance starts to decrease 
around $400^3$, while at
$1024^2\times20$ performance has fallen to $2/3$ of maximum.
All modern CPUs are able to vectorise and pipeline simple instructions. 
By default, we have therefore chosen the
innermost dimension, the $x$-direction, to be as simple as possible, with 
periodic boundary conditions. Then, the compiler will be able to schedule
essentially all operations as SIMD instructions. The middle dimension, 
the $y$-direction, does not have any significance and is
the best place to calculate boundary conditions, for problems that only 
contain boundaries in one direction.
This way any computational load from the boundary is spread evenly over
all the papers.

\begin{figure}[thb]
\begin{center}
\epsfig{figure=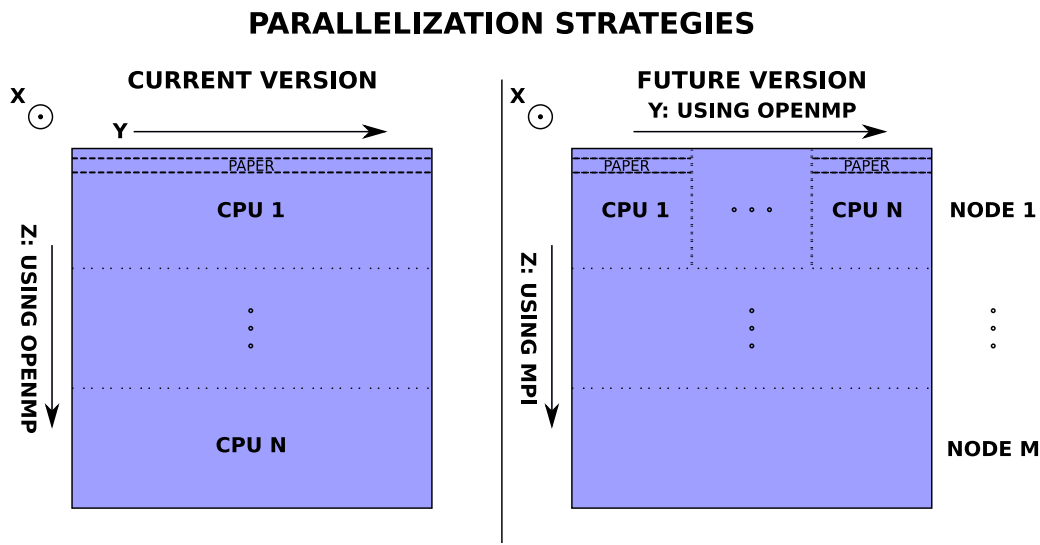,width=\textwidth}
\caption{Parallelisation strategies. We have demonstrated perfect scalability
up to 128 CPUs with our current OpenMP implementation. Future implementations
will be based on a hybrid OpenMP/MPI model. An added benefit of a hybrid model is 
the improved cache reuse for very large box sizes (i.e. $1024^3$ and beyond).}
\label{fig:parallel}
\end{center}
\end{figure}

So far in Copenhagen we have had good access to shared memory machines. 
By far the easiest way to parallelise the code is then to use OpenMP.
However, shared memory machines are relatively expensive and limited in size.
A few versions exist of the stagger code that use MPI to run on clusters.
One of the major current technology trends is the integration of
two (Intel, AMD, Sun) or more (IBM) CPU cores on a single piece of
silicon. We can only expect that all CPUs in the future will be massively
multi threaded. Then the optimal approach to parallelisation will be 
a hybrid one with OpenMP inside a single CPU node and MPI between nodes.
Current and future parallelisation strategies have been sketched
in Fig.~\ref{fig:parallel}.
\begin{figure}[thb]
\begin{center}
\epsfig{figure=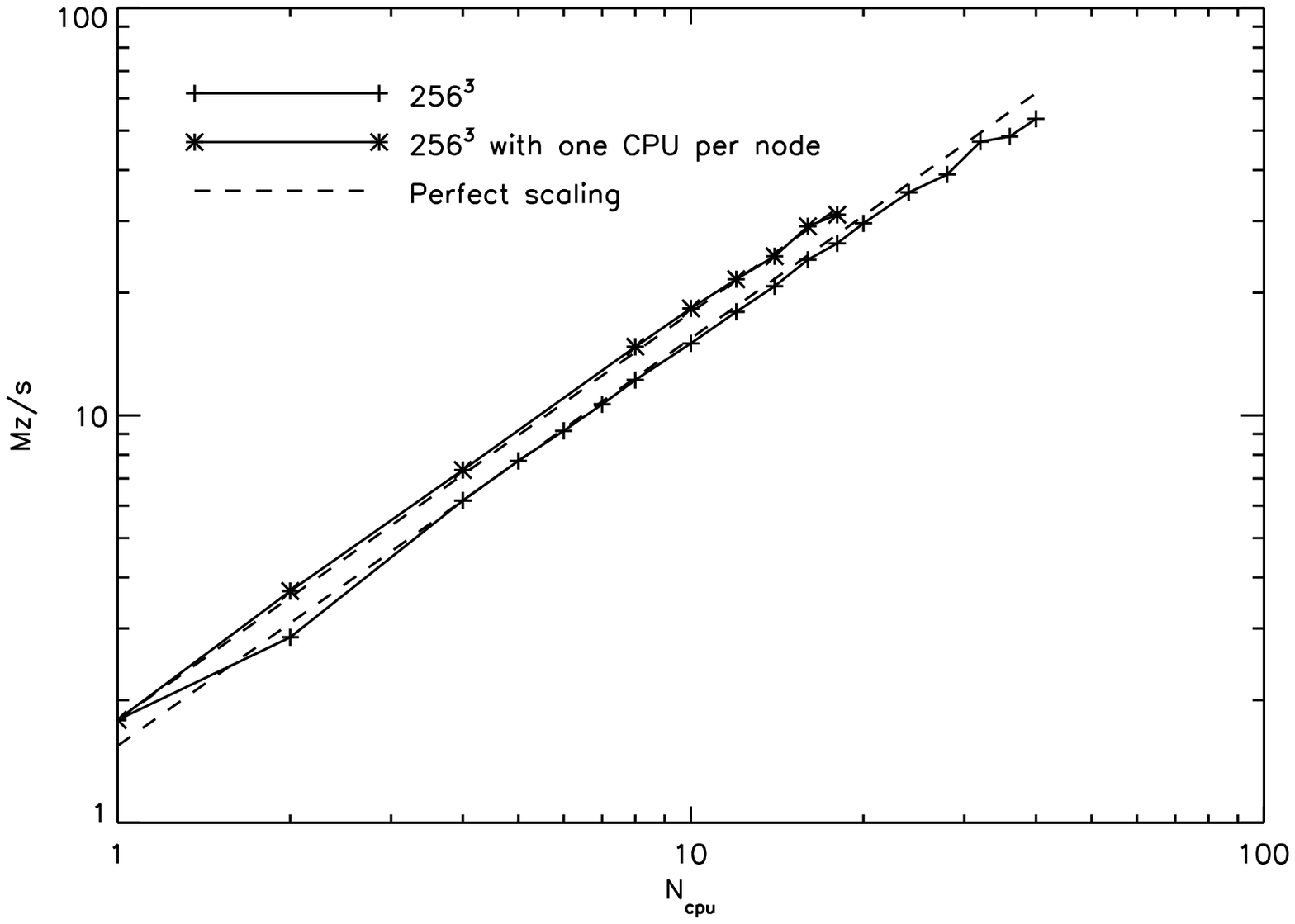,width=0.49\textwidth}
\epsfig{figure=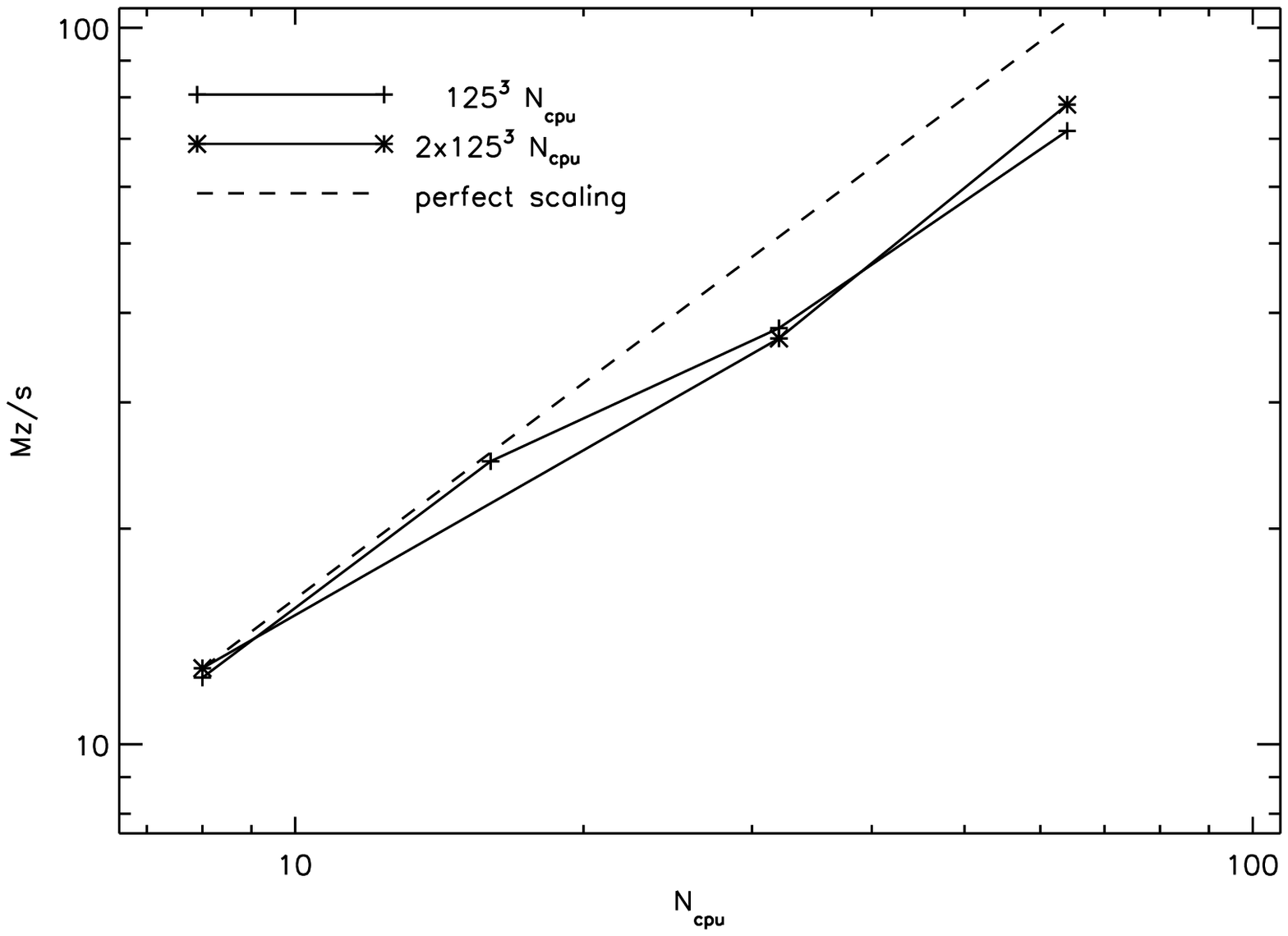,width=0.49\textwidth}
\caption{Scalability of the paper code: Results of scaling a simple HD
experiment on an SGI Altix machine. To the left is the strong scalability
for a $256^3$ experiment on a dedicated machine with a 1.3GHz CPU.
To the right is shown the weak scalability running on a loaded machine with
a 1.6GHz CPU with the experiment size varying according to the number of CPUs.}
\label{fig:scaling}
\end{center}
\end{figure}

In the Paper Code we have effectively hidden the parallel nature of the code.
Each CPU is automatically assigned a number of papers from \verb|ks|
to \verb|ke|, and in the main part of the code, where the physics
are calculated, one only has to consider dependencies in the $z$-direction
and insert synchronisation points as appropriate.
As an example consider the use of geometric means to interpolate the
density to face values
\begin{equation}
\rho_x = \exp(\textrm{xdn}(\ln(\rho))) \,,\,
\rho_y = \exp(\textrm{ydn}(\ln(\rho))) \,,\,
\rho_z = \exp(\textrm{zdn}(\ln(\rho))) \,,
\end{equation}
where $i\textrm{dn}$ denotes interpolation half a point down in the
$i$ direction. This may be coded in two blocks:
\begin{verbatim}
  do kk=ks,ke
    lnr(:,:,kk) = alog(rho(:,:,kk))
  enddo
  !$omp barrier  !<-- Sync: zdnr=zdn_set(lnr)
                 !          depends on non-local papers
  do kk=ks,ke
    call xdn_exp(lnr,xdnr)
    call ydn_exp(lnr,ydnr)
    call zdn_exp(lnr,zdnr)
  end
\end{verbatim}
Notice that geometric interpolation is a common operation and to streamline things
we have made special interpolation operators, that automatically applies the exponential.

A barrier works as a synchronisation point.
The CPUs have to stop at a barrier and wait until
all of them have arrived. When only using a small number of
CPUs the number of barriers
are not very important, but when considering hundreds of CPUs, it is
essential that
the barrier count is minimised. Any small disturbance for one
CPU will make all the
others wait at each barrier. To take an example: If there in each
time step are 100 barriers,
and a CPU is randomly disturbed once during a time step, giving a 
slowdown of 1\%,  this extra
noise will for two CPUS give rise to a 2\% slowdown. For hundred of
CPUs the same disturbance,
since it occurs in random sections, gives on average at least a 50\% slowdown.
By carefully analysing the numeric implementation, we have found that
in a full update of the cells the
minimum number of barriers needed to calculate any part of the code is 6.
The old \emph{stagger code} based MHD was logically structured in different 
sections, according
to the different physics. First, the calculation of velocities from momentum,
then, the pressure, the viscosity, the stress tensor, the MHD equations 
and at last the equation of motion
for the internal energy. Since all parts need between 4 and 6 barriers, 
one ends up having at least 20 barriers. 
With the new code, we have applied a ``principle of origami'' folding the
logical structure of the code. After each barrier, we consider all the different
equations and calculate the maximum amount of physics.
When threading the 5 small MHD parts, 6 viscosity parts etc together
we end up having only 6 barriers. Recently, we had the
opportunity to have the paper code tested on the NASA Columbia supercomputer
and the code scaled efficiently on up to at least 128 CPUs. 

We have implemented a full HD/MHD code, including self gravity \&
turbulent driving,
and a special relativistic HD/MHD version of the code described in this thesis.
Both codes show spectacular performance. The MHD code can update
1.2 million cells per CPU per second and it runs at
30\% of theoretical peak performance on the SGI Altix machine. A normal grid
based MHD code,
even when optimised, performs at
between 5\% and 10\% of peak performance (see \cite{bib:SC04}
for a detailed analysis of five state of the art codes). 
In fact, we believe that the paper code is one of the highest
performing codes of its kind. This is both due to the low absolute cost
of evaluating the MHD equations for a single cell and the effectiveness
with which we have implemented the
algorithm. The special relativistic MHD version runs at 250.000 zone
updates per second, which is also well above quoted numbers in the literature.

\section{Discussion}\label{sec:2.8}
In this chapter I have discussed the theoretical foundation and numerical
implementation of a new general relativistic magnetohydrodynamics code.
When designing a new code without building upon existing work,
it is tempting to use an already existing theoretical basis.
However, instead of this, I have derived a new form of
the equations of motion with global coordinates evolving the dynamical
variables from the point of view of a local observer.
This approach makes it possible to employ a highly sophisticated 
artificial viscosity.
This is just but an example of the
possibilities the new formulation opens up for.
The implication of my approach is that any new physics
that is implemented and working in special relativity in the future,
be it a new equation of state, radiative transfer, or a perturbative
implementation of gravity, may easily be reused in the general
relativistic version of the code.
This may be done because the special and the general relativistic
versions are related through the simple formulas given in App.~\ref{chap:appa}.
When deriving the equations of motion, I have not made any assumptions
about the background metric, so that the design is ready to be coupled with
methods solving the full Einstein equations, such as the CactusCode.

This new GrMHD code has been tested on a variety of demanding problems,
and it has been
demonstrated that it is able to deal with huge pressure and density gradients.
It shows some problems in the case of flows with high Lorentz factors, but
they can be addressed and will be solved in the near future.
The tests carried out include both synthetic benchmarks that tests a certain
aspect of the code, and real astrophysical applications.

The computer code is based on a refinement of the current
infrastructure for fluid dynamics used in Copenhagen, which has
been developed together with J.~Hunsballe. It shows a spectacular
performance on modern computer architectures, exploiting up to
30\% of the theoretical peak performance.
The special relativistic versions of the hydrodynamics and
magnetohydrodynamics codes are three dimensional and have been fully
parallelised.
They have been tested and scale to hundreds of CPUs, making it possible
to exploit massive supercomputers at national centres to the full extent.

I plan to employ the code in combination with the other numerical tools
presented in this thesis in order to understand extreme astrophysics near compact
objects. A first joint application of the particle code and
this code is presented in Chapter \ref{chap:global}. Furthermore, observational
cosmology is reaching a level of quality, where soon not
everything can be addressed in terms of simple one dimensional linear
perturbation theory, and I plan to employ the code in the understanding of
the non-trivial evolution of magnetic fields in the early universe.

\newpage
$\phantom{.}$
\newpage
\chapter{Magnetic Field Generation in Collisionless Shocks; 
Pattern Growth and Transport}\label{chap:field}
In this chapter I present results from three-dimensional particle 
simulations of collisionless 
shock formation, with relativistic counter-streaming ion-electron plasmas
first published in Fredriksen et al.~\cite{bib:frederiksen2004}.  
Particles are followed over many skin depths downstream of the shock.  Open 
boundaries allow the experiments to be continued for several particle crossing 
times.  The experiments confirm the generation of strong magnetic and electric 
fields by a Weibel-like kinetic streaming instability, and demonstrate that the
electromagnetic fields propagate far downstream of the shock.  The magnetic 
fields are predominantly transversal, and are associated with merging ion current 
channels.  The total magnetic energy grows as the ion channels merge, and as the
magnetic field patterns propagate down stream.  The electron populations are 
quickly thermalised, while the ion populations retain distinct bulk speeds in
shielded ion channels and thermalise much more slowly.  The results help us to reveal 
processes of importance in collisionless shocks, and may help to explain the origin 
of the magnetic fields responsible for afterglow synchrotron/jitter radiation from 
Gamma-Ray Bursts.

\section{Introduction}

The existence of a strong magnetic field in the shocked external
medium is required in order to explain the observed radiation in
Gamma-Ray Burst afterglows as synchrotron radiation 
\citep[e.g.][]{bib:Panaitescu+Kumar}.  Nearly collisionless shocks,
with synchrotron-type radiation present, are also common in many other
astrophysical contexts, such as in supernova shocks, and in jets
from active galactic nuclei.  At least in the context of Gamma-Ray
Burst afterglows the observed synchrotron radiation requires the
presence of a stronger magnetic field than can easily be explained by just
compression of a magnetic field already present in the external medium.

Medvedev \& Loeb \citep{1999ApJ...526..697M} showed through a linear
kinetic treatment how a
two-stream magnetic instability -- a generalisation of the Weibel
instability \citep{1959PhRvL...2...83W,bib:YoonDavidson} -- can generate a
strong magnetic field  ($\epsilon_B$, defined as the ratio of magnetic energy to 
total kinetic energy, is $10^{-5}$-$10^{-1}$ of equipartition value) 
in collisionless shock fronts
\citep[see also discussion in][]{2003MNRAS.339..881R}. We
note in passing that this instability is well-known in other plasma
physics disciplines, e.g. laser-plasma interactions
\cite{bib:YangGallantAronsLangdon,1998PhRvE..57.7048C},
and has been applied in the context of pulsar winds
by Kazimura et al.~\cite{bib:Kazimura}.

Using three-dimensional particle-in-cell simulations to study
relativistic collisionless shocks (where an external plasma impacts the
shock region with a bulk Lorentz factor $\Gamma = 5-10$),
Frederiksen et al.~\cite{bib:frederiksen2002},
Nishikawa et al.~\cite{2003ApJ...595..555N}, and
Silva et al.~\cite{2003ApJ...596L.121S}
investigated the generation  of magnetic fields by the two-stream
instability.
In these first studies the
growth of the transverse scales of the magnetic field was limited by the
dimensions of the computational domains.  The durations of the 
Nishikawa et al.~\cite{2003ApJ...595..555N} experiments were less than
particle travel  times through the experiments, while
Silva et al.~\cite{2003ApJ...596L.121S} used periodic 
boundary conditions in the direction of streaming.
Further, Frederiksen et al.~\cite{bib:frederiksen2002}
and Nishikawa et al.~\cite{2003ApJ...595..555N} used electron-ion ($e^-p$) 
plasmas, while experiments reported upon by
Silva et al.~\cite{2003ApJ...596L.121S} were done with $e^-e^+$
pair plasmas.

Here, we report on 3D particle-in-cell simulations 
of relativistically counter-streaming $e^-p$
plasmas. Open boundaries are used in the streaming direction, and experiment 
durations are several particle crossing times. 
Our results can help to reveal the most important
processes in collisionless shocks, and help to explain the observed afterglow
synchrotron radiation from Gamma-Ray Bursts. 
We focus on the earliest development in
shock formation and field generation. Late stages in shock formation will be
addressed in Chapter \ref{chap:global}. 

\section{Simulations}
Experiments were performed using a self-consistent 3D3V electromagnetic
particle-in-cell code originally developed for simulating reconnection 
topologies \citep{bib:HesseKuzenova}, 
redeveloped by Frederiksen \cite{bib:trier}  to obey special relativity
and to be second order accurate in both space and time.

The code solves Maxwell's equations for the electromagnetic
field with continuous sources, with fields and field source
terms defined on a staggered
3D Yee-lattice \citep{bib:Yee}. The sources in Maxwell's equations
are formed by weighted averaging of particle data to the field grid,
using quadratic spline interpolation. Particle velocities and positions are
defined in continuous (${\bf{r}},\gamma{\bf{v}}$)-space, 
and particles obey the relativistic equations of motion.

The grid size used in the main experiment was $(x,y,z)=200\times200\times800$,
with 25 particles per cell, for a total of $8\times10^8$ particles,
with ion to electron mass ratio $m_{i}/m_{e} = 16$.
To adequately resolve a significant number of electron and ion
skin-depths ($\delta_e$ and $\delta_i$), the box size was chosen such that
$L_{x,y} = 10\delta_i \sim 40\delta_e$ and $L_z \sim 40 \delta_i
\sim 160\delta_e$. Varying aspect and mass ratios were used in complementary experiments.

\begin{figure*}[!t]
\begin{center}
\epsfig{figure=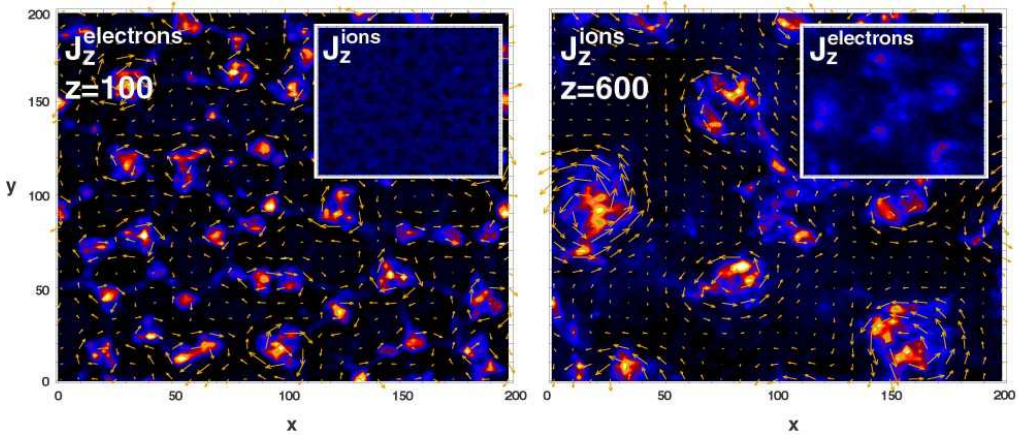,width=\textwidth}
\caption{
The left hand side panel shows the longitudinal electron current
density through a transverse cut at $z=100$, with a small inset showing
the ion current in the same plane.  The right hand side panel shows
the ion current at $z=600=30\delta_i$, with the small inset now instead showing
the electron current. The arrows represent the transverse magnetic field. Both panels are from time $t = 1200$.}
\label{fig:Slice}
\end{center}
\end{figure*}

Two counter-streaming -- initially
quasi-neutral and cold -- plasma populations are simulated. At the two-stream interface (smoothed around $z=80$)
a plasma ($z<80$) streaming in the positive z-direction, with a
bulk Lorentz factor $\Gamma=3$, hits another plasma ($z\ge80$) at rest in
our reference frame. The latter plasma is denser than the former by a factor of 3. 
Experiments have been run with both initially sharp and initially smooth 
transitions, with essentially the same results.  
The long simulation time gradually allows the shock to converge towards 
self-consistent jump conditions.
Periodic boundaries are imposed in
the $x$-- and $y$--directions, while the boundaries at $z=0$ and $z=800$ are open,
with layers absorbing transverse electromagnetic waves.  Inflow
conditions at $z=0$ are fixed, with incoming particles supplied at a
constant rate and with uniform speed.  At $z=800$ there is free outflow of particles.
The maximum experiment duration is 480 $\omega_{pe}^{-1}$ (where $\omega_{pe}$ is the electron plasma frequency), 
sufficient for propagating $\Gamma \approx 3$ particles 2.8 times through the box.

\section{Results and Discussions}

The extended size and duration of these experiments make it possible
to follow the two-stream instability through several stages of development;
first exponential growth, then non-linear saturation, followed by pattern growth 
and downstream advection.  We identify the mechanisms 
responsible for these stages below.

\subsection{Magnetic field generation, pattern growth \\ and field transport} 
\label{field_generation}
Encountering the shock front the
incoming electrons are rapidly (being lighter than the ions) deflected by
field fluctuations growing due to the two-stream instability \citep{1999ApJ...526..697M}. 
The initial perturbations grow
non-linear as the deflected electrons collect into first caustic
surfaces and then current channels (Fig.~\ref{fig:Slice}). Both streaming and 
rest frame electrons are deflected, by arguments of symmetry.
 
In accordance with Ampere's law the current channels
are surrounded by approximately cylindrical magnetic fields
(illustrated by arrows in Fig.~\ref{fig:Slice}), causing
mutual attraction between the current channels.  The current
channels thus merge in a race where larger electron
channels consume smaller, neighbouring channels.
In this manner, the transverse magnetic field
grows in strength and scale downstream. This continues until
the fields grow strong enough to deflect the
much heavier ions into the magnetic voids between
the electron channels. The ion channels are then subjected to
the same growth mechanism as the electrons. When ion channels
grow sufficiently powerful, they begin to experience Debye shielding by the
electrons, which by then have been significantly heated by scattering 
on the increasing electromagnetic field structures. The two electron 
populations, initially separated in $\gamma{\bf{v}}$-space, merge 
to a single population in approximately $20\delta_e$ ($z=80$--$200$) 
as seen in Fig.~\ref{fig:acc}. The same trend is seen for the ions -- albeit 
the merging rate might be significantly slower than predicted by
extrapolating with $m_i/m_e$, since Debye shielding stabilises the
ion channels.

\begin{figure}[!t]
\begin{center}
\epsfig{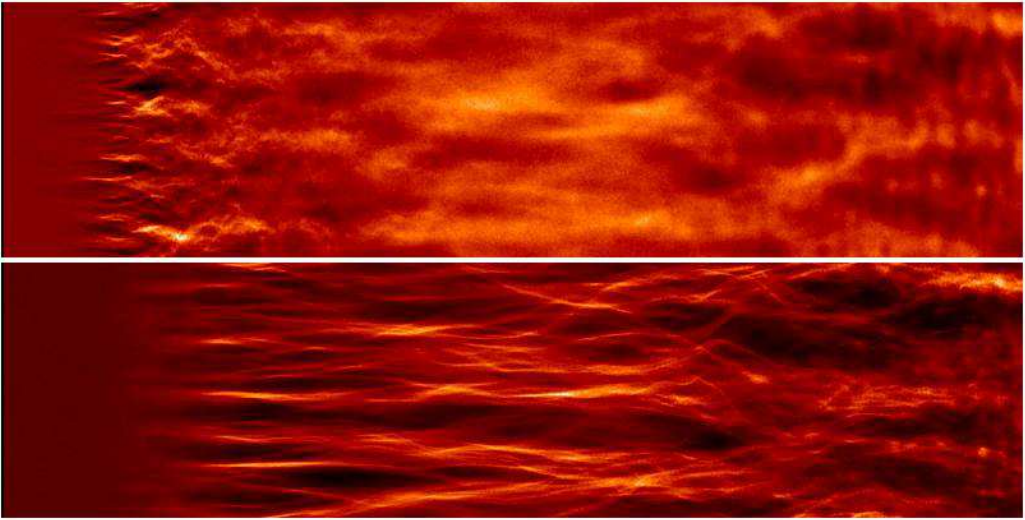}
\caption{
Electron (top) and ion (bottom) currents, averaged over the $x$-direction, at time
$t=1200$.
}
\label{fig:jiz}
\end{center}
\end{figure}
\begin{figure}[!ht]
\begin{center}
\epsfig{figure=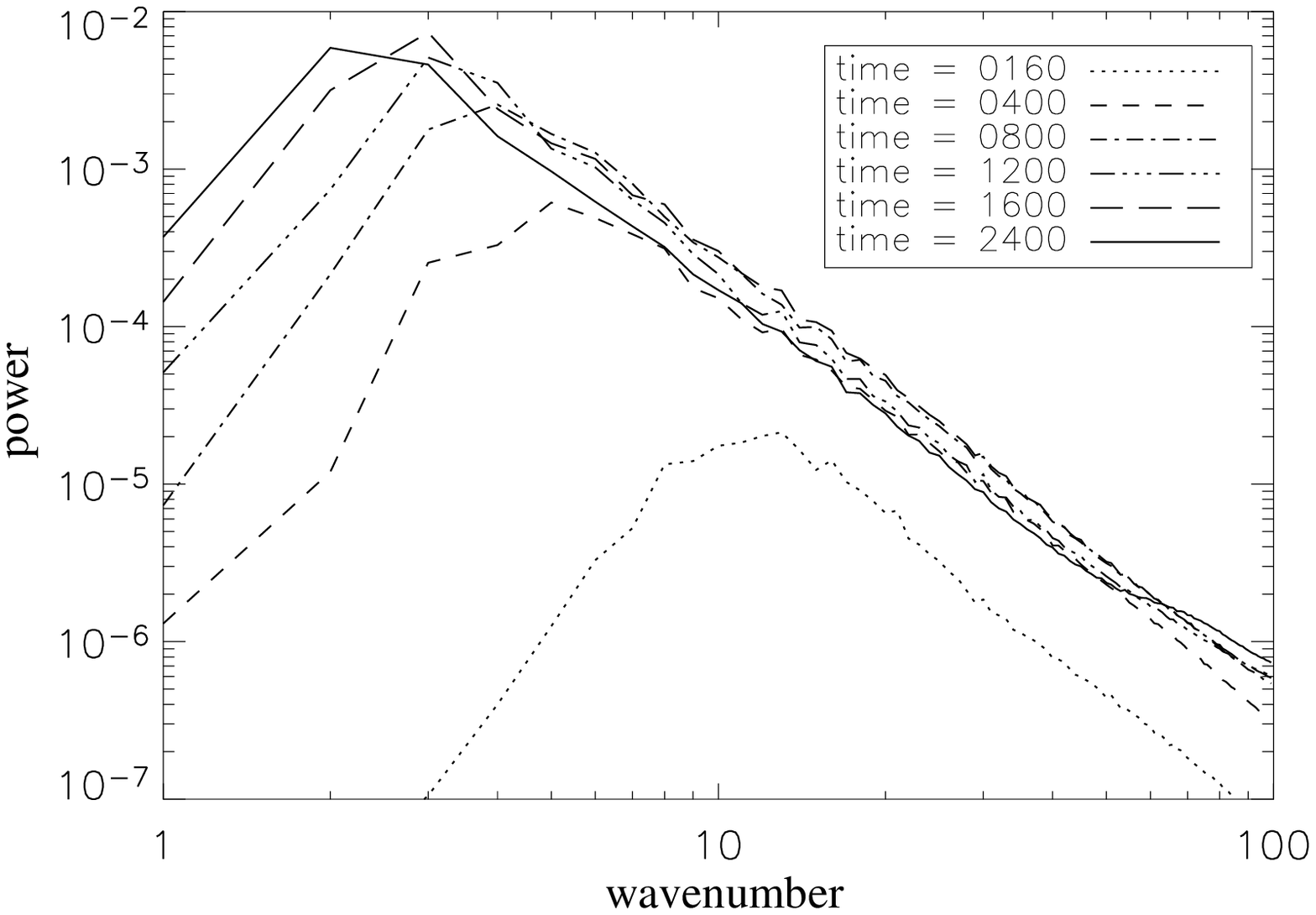,width=\textwidth}
\caption{Power spectrum of ${\mathbf B}_{\perp}$ for $z = 250$ at different times.}
\label{fig:power}
\end{center}
\end{figure}

The Debye shielding quenches the electron channels, while
at the same time supporting the ion-channels; the large
random velocities of the electron population allow the concentrated 
ion channels to keep sustaining strong magnetic fields.
Fig.~\ref{fig:Slice}, shows the highly concentrated ion currents, the more diffuse
-- and shielding -- electron currents, and the resulting magnetic field.
The electron and ion channels are further illustrated in Fig.~\ref{fig:jiz}.  
Note the limited $z$-extent of the electron
current channels, while the ion current channels extend throughout
the length of the box, merging to form larger scales downstream.
Because of the longitudinal current channels the magnetic field is 
predominantly transversal; we find $|B_z|/|B_{tot}| \sim 10^{-1} - 10^{-2}$.

\begin{figure}[!t]
\begin{center}
\epsfig{figure=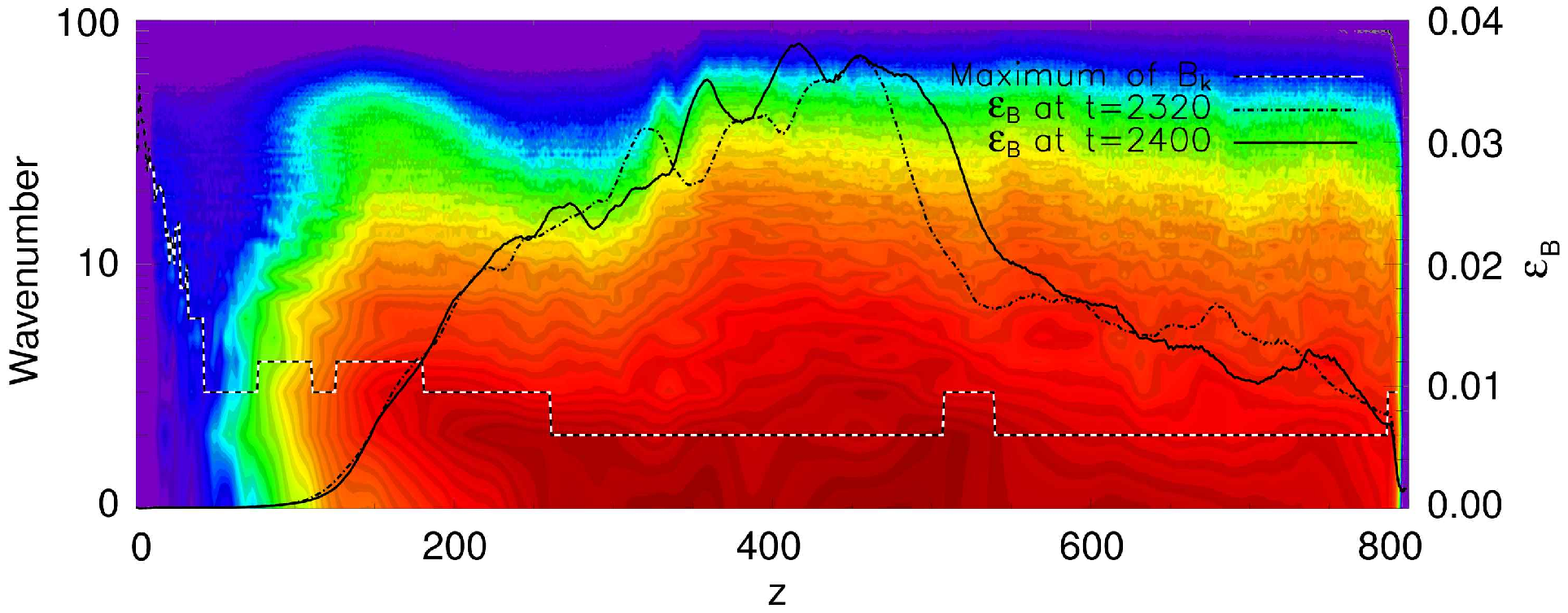,width=\textwidth}
\caption{Relative electromagnetic energy density $\epsilon_{B}$.
The contour colour plot shows the power in the transverse magnetic
field through the box distributed on spatial Fourier modes at $t=2400$,
with the dotted line marking the wavenumber with maximum power.
Superposed is the spatial distribution of $\epsilon_{B}$, averaged across the beam,
at $t=2320$ (dashed-dotted) and $t=2400$ (full drawn), highlighting how EM-fields 
are advected down through the box.
}
\label{fig:epsb}
\end{center}
\end{figure}

Figure \ref{fig:power} shows the temporal development of the 
transverse magnetic field scales around $z=250$.
The power spectra follow power-laws, 
with the largest scales growing with time.
The dominant scales at these $z$ are of the order $\delta_i$
at early times. Later they become comparable to $L_{x,y}$.
Figure \ref{fig:epsb} 
captures this scaling behaviour as a function of depth for $t=2400$.

\begin{figure}[!t]
\begin{center}
\epsfig{figure=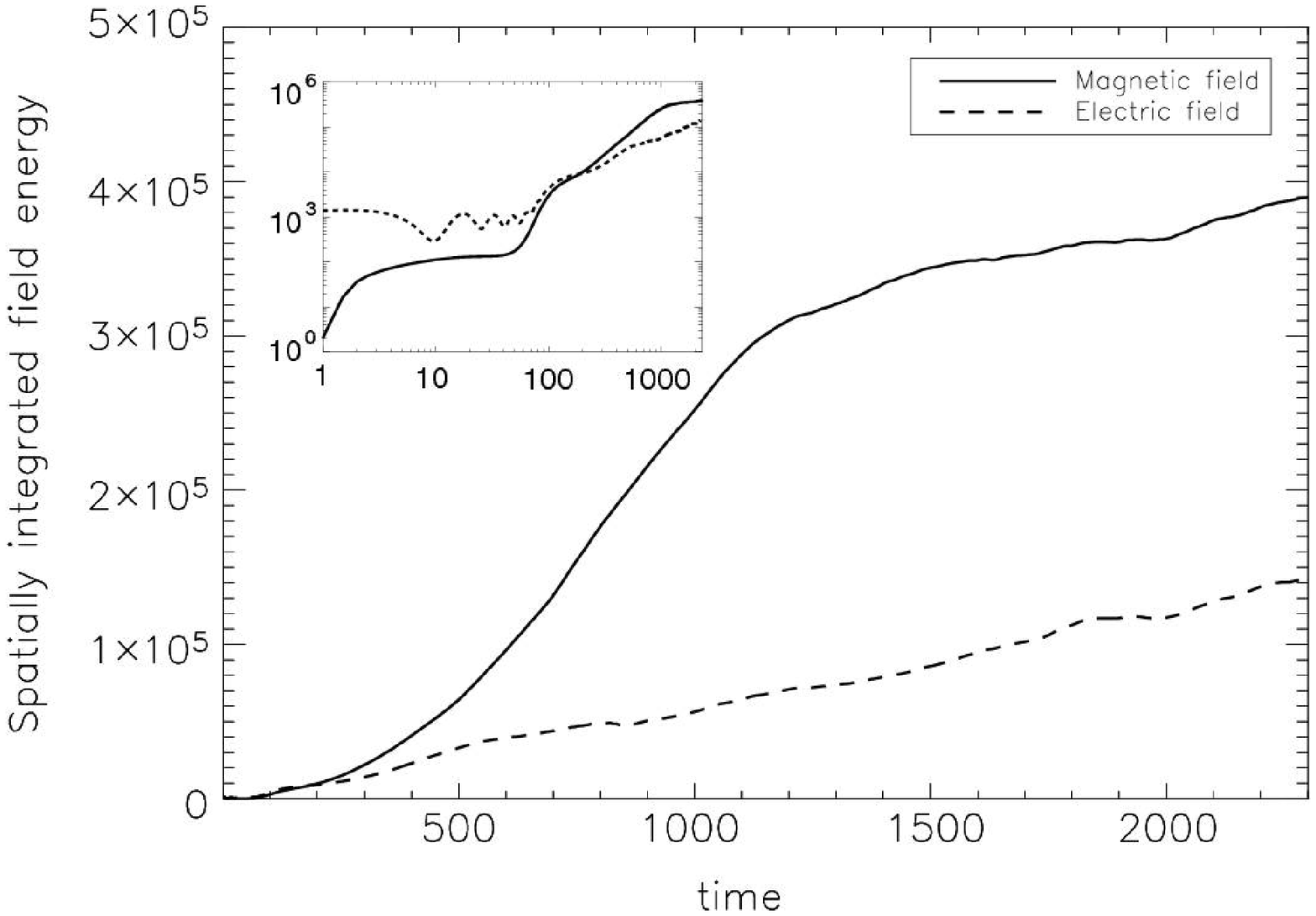,width=\textwidth}
\caption{
Total magnetic (full drawn) and electric (dashed) energy
in the box as a function of time. The inset shows a log-log 
plot of the same data.
}
\label{fig:B_energy}
\end{center}
\end{figure}

The time evolutions of the electric and magnetic field energies are shown in 
Fig.~\ref{fig:B_energy}. Seeded by fluctuations in the fields, mass 
and charge density,  the two-stream instability initially grows super-linearly 
($t=80-100$), reflecting approximate exponential growth in a small sub-volume. Subsequently the 
total magnetic energy grows more linearly, reflecting essentially the 
increasing volume filling factor as the non-linearly saturated magnetic 
field structures are advected downstream.

At $t\approx 1100$ the slope drops off, due to advection of the generated
fields out of  the box.  The continued slow growth, for $t > 1100$, reflects
the increase of the pattern size with time (cf.\ Fig.~\ref{fig:power}).  
A larger pattern size
corresponds, on the average, to a larger mean magnetic energy, since the
total electric current is split up into fewer but stronger ion current 
channels.  
The magnetic energy scales with
the square of the electric current, which in turn grows in inverse proportion
to the number of current channels.  The net
effect is that the mean magnetic energy increases accordingly.

The magnetic energy density keeps growing throughout our experiment,
even though the duration of the experiment (480 $\omega_{pe}^{-1}$) significantly 
exceeds the particle crossing time, and also exceeds the advection time of the 
magnetic field structures through the box.  This is in contrast to the results 
reported by Silva et al.~\cite{2003ApJ...596L.121S}, 
where the magnetic energy density drops back after about 10-30 $\omega_{pe}^{-1}$.
It is indeed obvious from the preceding discussion that the ion-electron 
asymmetry is essential for the survival of the current channels.

From the requirement that the total plasma momentum should be conserved,
the (electro)magnetic field produced by the two-stream instability
acquires part of the z-momentum lost by the two-stream population
in the shock; this introduces the possibility that magnetic field structures
created in the shock migrate downstream of the shock and thus
carry away some of the momentum impinging on the shock.

Our experiments show that this does indeed happen; 
the continuous injection of momentum transports the generated
field structures downstream at an accelerated advection speed.
The dragging of field structures through the dense plasma acts as to transfer momentum between 
the in-streaming and the shocked plasmas.

\subsection{Thermalisation and plasma heating}
\begin{figure}[!t]
\begin{center}
\epsfig{figure=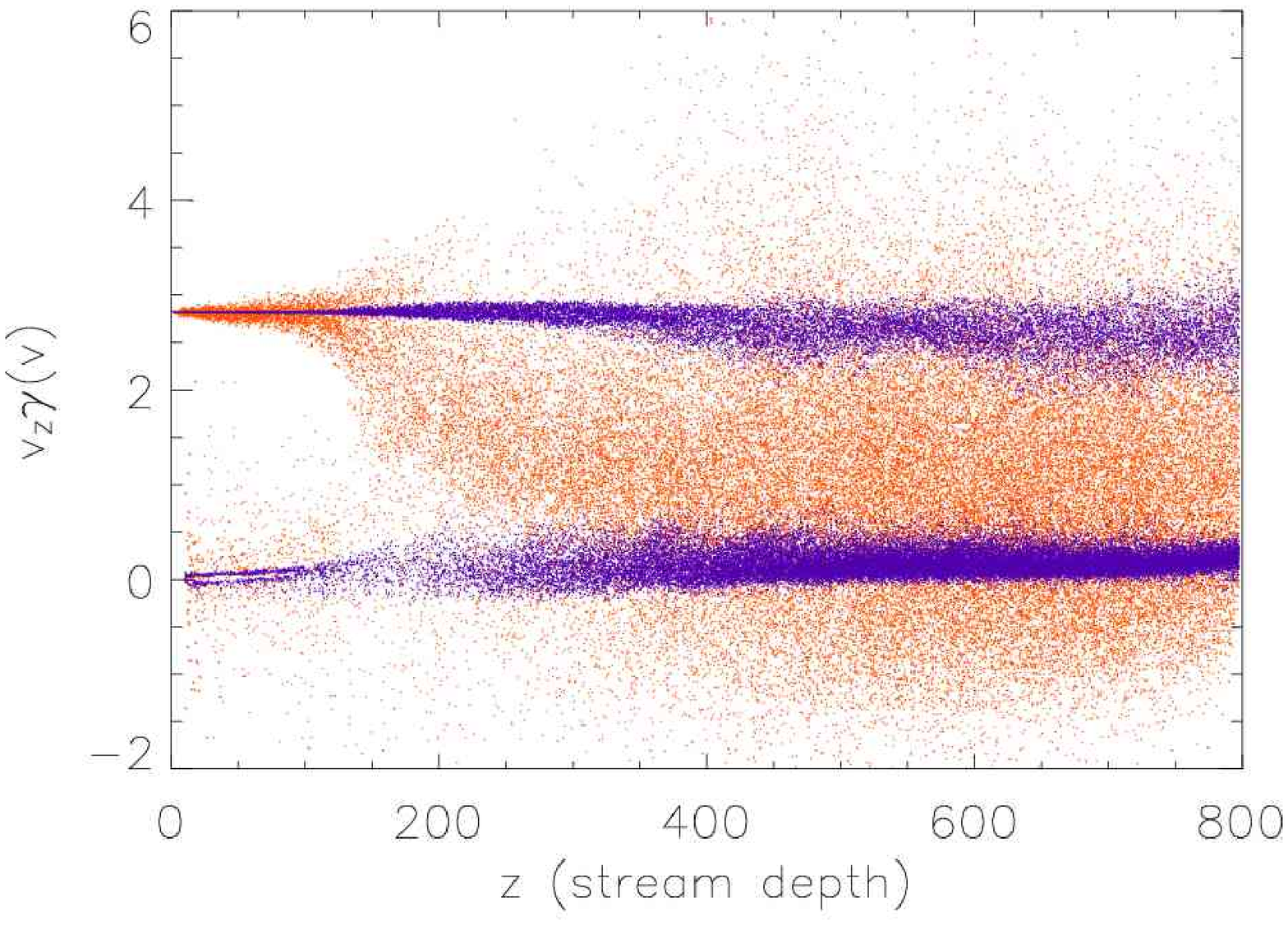,width=\textwidth}
\caption{Thermalisation and longitudinal acceleration,
illustrated by scatter plots of the electron (orange) and ion (blue) 
populations. 
Note the back-scattered electron population ($v_z\gamma(v) < 0$).
}
\label{fig:acc}
\end{center}
\end{figure}

At late times the entering electrons are effectively
scattered and thermalised: The magnetic field isotropises the velocity distribution 
whereas the electric field generated by the $e^{-}$--$p$ charge 
separation acts to thermalise the populations.
Figure \ref{fig:acc} shows that this happens over the $\sim$ 20 electron skin
depths from around $z=80$ -- $200$. 
The ions are expected to also thermalise, given sufficient space and time. This
fact leaves the massive ion bulk momentum constituting a vast energy reservoir for
further electron heating and acceleration. Also seen in Fig.~\ref{fig:acc}, the ions
beams stay clearly separated in phase space,  and are only slowly broadened (and
heated).

We do not see indications of a super-thermal tail in the heated electron
distributions, and there is thus no sign of second order Fermi-acceleration in 
the experiment presented in this Letter.
\cite{2003ApJ...595..555N} and \cite{2003ApJ...596L.121S} reported acceleration of particles 
in experiments similar to the current experiment, except for more limited
sizes and durations, and the use of an $e^-e^+$ plasma \citep{2003ApJ...596L.121S}.
On closer examination of the published results it appears that there is no
actual disagreement regarding the absence of accelerated particles. Whence,
\cite{2003ApJ...595..555N} refer to transversal velocities of the order of $0.2
c$ (their Fig.\ 3b), at a time where our experiment shows similar
transversal velocities (cf. \fig{fig:acc}) 
that later develop a purely thermal spectrum. \cite{2003ApJ...596L.121S} refer to 
transversal velocity amplitudes up to about $0.8 c$ (their Fig.\ 4), or $v\gamma\sim 2$,
with a shape of the distribution function that appears to be compatible with thermal.
In comparison, the electron distribution illustrated by the scatter plot in
Fig.~\ref{fig:acc} 
covers a similar interval of $v\gamma$, with distribution functions that are 
close to Lorentz boosted relativistic Maxwellians (see
App.~\ref{chap:maxwell} for a discussion of Lorentz boosted thermal profiles).
Thus, in the experiment reported on in this chapter there is
no compelling evidence for non-thermal particle acceleration.
Thermalisation is a more likely cause of the increases in transversal velocities.

Frederiksen et al.~\cite{bib:frederiksen2002} reported evidence for
particle acceleration, with electron gammas up to $\sim100$, in experiments 
with an external magnetic field present in the up-stream plasma.
This is indeed 
a more promising scenario for particle acceleration experiments 
(although in the experiments by \cite{2003ApJ...595..555N} results with an
external magnetic field were similar to those without).
Figure \ref{fig:acc} shows the presence of a population of
back-scattered electrons ($v_z\gamma < 0$).  In the presence of
an external magnetic field in the in-streaming plasma,
this possibly facilitates Fermi acceleration in the shock.

\section{Conclusions}

The experiment reported upon in this chapter illustrates a number of 
fundamental properties of relativistic, collisionless shocks:

1.
Even in the absence of a magnetic field in the up-stream plasma,
a small scale, fluctuating, and predominantly transversal magnetic field
is unavoidably generated by a two-stream instability reminiscent of the
Weibel-instability. In the current experiment the magnetic energy density 
reaches a few percent of the energy density of the in-coming beam.

2.
In the case of an $e^-p$ plasma the electrons are rapidly thermalised, while
the ions form current channels that are the sources of deeply
penetrating magnetic field structures.  The channels merge in the downstream 
direction, with a corresponding increase of the average
magnetic energy with shock depth.  This is expected
to continue as long as a surplus of bulk relative momentum remains in the 
counter-streaming plasmas.

3.
The generated magnetic field patterns are advected downstream at speeds 
intermediate of the streaming and rest frame plasmas.
The electromagnetic field structures
thus provide scattering centres that interact with both the fast, in-coming
plasma, and with the plasma that is initially at rest.  As a result the 
electron populations of both components quickly thermalise and form
a single, Lorentz-boosted thermal electron population.  The two ion populations 
merge much more slowly, with only gradually increasing ion temperatures.

4. The observed strong turbulence in the field structures at the shocked streaming interface 
provides a promising environment for particle acceleration.

We emphasise that quantification of the interdependence and development of
$\epsilon_U$ and $\epsilon_B$ is accessible by means of such experiments as reported
upon here. 

Rather than devising abstract scalar parameters $\epsilon_B$
and $\epsilon_U$, that may be expected to depend on shock depth, media densities
etc., a better approach is to compute synthetic radiation spectra directly from 
the models, and  then apply scaling laws to predict what would be observed from 
corresponding, real supernova remnants and Gamma-Ray Burst afterglow shocks.

\chapter{Non--Fermi Power law Acceleration in 
Astrophysical Plasma Shocks}\label{chap:acc}
Collisionless plasma shock theory, which applies for example 
to the afterglow of gamma ray
bursts, still contains key issues that are poorly understood.
In this chapter I discuss the results of
charged particle dynamics in a highly
relativistic collisionless 
shock numerically using $\sim 10^9$ particles first
published by Hededal et al.~\cite{bib:hededal2004}.  We find a power
law distribution of accelerated electrons, which upon detailed
investigation turns out to originate
from an acceleration mechanism that is decidedly
different from Fermi acceleration.
Electrons are accelerated by strong filamentation 
instabilities in the shocked interpenetrating plasmas and coincide
spatially with the 
powerlaw distributed current filamentary structures. These structures are an 
inevitable consequence 
of the now well established Weibel--like two--stream instability that operates 
in relativistic collisionless shocks.
The electrons are accelerated and decelerated instantaneously and
locally; a scenery that differs qualitatively from recursive
acceleration mechanisms such as Fermi acceleration.
The slopes of the electron distribution powerlaws are in concordance with the 
particle powerlaw spectra inferred from observed afterglow synchrotron radiation 
in gamma ray bursts, and the mechanism can possibly explain more generally the 
origin of non--thermal radiation from shocked inter-- and circum--stellar regions
and from relativistic jets.

\section{Introduction}
Given the highly relativistic conditions in the outflow from
gamma ray bursts (GRBs), the mean free path for particle Coulomb
collisions in the afterglow shock is several orders of magnitude
larger than the fireball itself.
In explaining the microphysical processes that work to define the
shock, MHD becomes inadequate and collisionless plasma shock theory
stands imperative.
In particular two key issues remain, namely 
the origin and nature of the magnetic field in the shocked region,
and the mechanism by which electrons are accelerated from a thermal
population to a powerlaw distribution $N(\gamma)d\gamma\propto\gamma^{-p}$.
Both ingredients are needed to explain the 
observed afterglow spectra
(e.g.~\cite{2000ApJ...538L.125K, 2001ApJ...560L..49P}).

Regarding the origin of the magnetic field in the shocked region, observations 
are not compatible with a compressed inter--stellar magnetic field, which would
be orders of magnitude smaller than needed \cite{1999ApJ...511..852G}. 
It has been suggested that a Weibel--like two--stream instability
can generate a magnetic field in the 
shocked region (see Chapter \ref{chap:field}, and
Medvedev \& Loeb \cite{1999ApJ...526..697M}; 
Frederiksen et al.~\cite{bib:frederiksen2002}; 
Nishikawa et al.~\cite{2003ApJ...595..555N}; 
Silva et al.~\cite{2003ApJ...596L.121S}).
Computer experiments presented in Chapter \ref{chap:field} and
\cite{bib:frederiksen2004} showed that the 
nonlinear stage of a two--stream instability induces a magnetic field
{\it in situ} with an energy content of a
few percent of the equipartition value, consistent with
that required by observations. 

Fermi acceleration \cite{1949PhRv...75.1169F} has, so far,
been widely accepted as the mechanism that provides the inferred electron 
acceleration. 
It has been employed extensively in Monte Carlo simulations 
(e.g.~\cite{bib:Niemiec} and references therein),
where it operates in conjunction with certain
assumptions about the scattering of particles
and the structure of the magnetic field.
The mechanism has, however,
not been conclusively demonstrated to occur in 
{\em ab initio} particle simulations.
As pointed out by Niemiec \& Ostrowski \cite{bib:Niemiec}, 
further significant advance in the study of relativistic shock 
particle acceleration is
unlikely without understanding the detailed microphysics of 
collisionless shocks. Also,
recently Baring and Braby \cite{bib:baring} found that
particle distribution functions (PDFs)
inferred from GRB observations are in contradistinction with standard
acceleration mechanisms such as diffusive Fermi acceleration. 

In this chapter we study {\em ab initio} the particle dynamics 
in a collisionless shock with bulk Lorentz factor $\Gamma=15$.
We find a new particle 
acceleration mechanism, which is presented in section \ref{sec:4.2}.  Detailed 
numerical results are presented 
and interpreted in section \ref{sec:4.3}, while section \ref{sec:4.4} contains the conclusions.
\begin{figure*}[!th]
\begin{center}
\epsfig{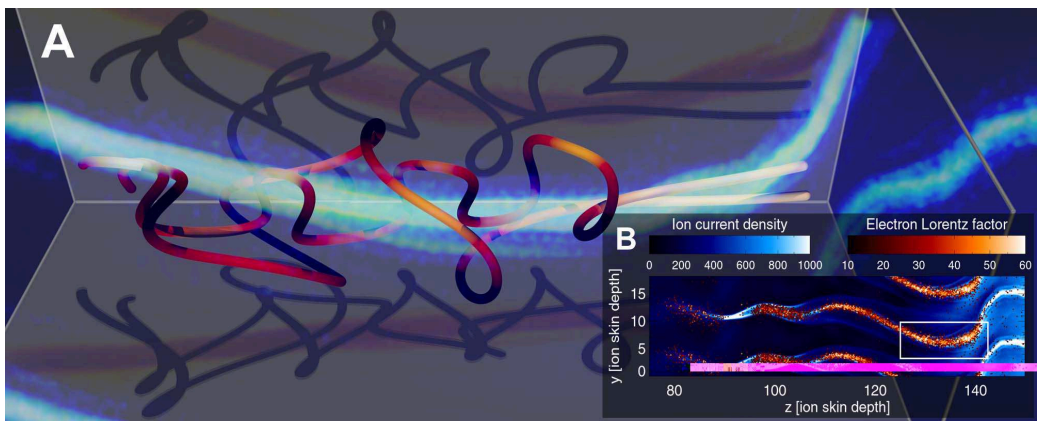}
\caption{(A) Ray traced electron paths (red) and current density (blue).
The colours of the electron paths reflect their four velocity according 
to the colour table in the inset (B). The shadows are
equivalent to the $x$ and $y$ projections of their paths. The ion
current density is
shown with blue colours according to the colour table in the inset. The inset also
shows the ion current density (blue) integrated along the
$x$ axis with the spatial distribution of fast 
moving electrons (red) over plotted.}
\label{fig:acceleration}
\end{center}
\end{figure*}

\section{A New Acceleration Mechanism}\label{sec:4.2}
A series of numerical experiments have been performed where collisionless 
shocks are created by two colliding plasma populations. These experiments 
are described in more detail below, but a common feature is 
that the electron PDF has a high energy tail which is powerlaw distributed.  By 
carefully examining the paths of representative accelerated electrons, 
tracing them backwards and forwards in time, it has been possible to 
identify the mechanism responsible for their acceleration.
The acceleration mechanism, which was presented for the first
time in \cite{bib:hededal2004}, works as follows:

When two non--magnetised collisionless plasma populations interpenetrate,
current channels are formed 
through a Weibel--like two--stream instability 
(see Chapter \ref{chap:field}; Medvedev \& Loeb \cite{1999ApJ...526..697M};
Frederiksen et al.~\cite{bib:frederiksen2002};
Nishikawa et al.~\cite{2003ApJ...595..555N};
Silva et al.~\cite{2003ApJ...596L.121S}).
In the nonlinear stage of evolution of this instability, ion current 
channels merge 
into increasingly stronger patterns, while electrons act to 
Debye shield these channels, as shown in Chapter \ref{chap:field}. 
Further it was showed
that a Fourier decomposition of the transverse structure of the
ion current filaments 
exhibits powerlaw behaviour which has been recently confirmed by
Medvedev et al.~\cite{2005ApJ...618L..75M}.

At distances less than the Debye length, the ion current channels are surrounded 
by transverse electric fields that accelerate the electrons toward the current 
channels. However, the magnetic fields that are induced around 
the current channels act to deflect the path of the accelerated electrons, 
boosting them instead in the direction of the ion flow.
Since the forces working are due to quasi--stationary fields the acceleration is a
simple consequence of potential energy being converted into 
kinetic energy. Therefore the electrons are decelerated again 
when leaving the current channel, and reach their maximal velocities 
at the centres of the current channels. Hence, as illustrated by
Fig.~\ref{fig:acceleration}B, the spatial distribution
of the high energy electrons is a direct match to the ion current channels and 
the properties of the accelerated electrons depend
primarily on the local conditions in the plasma.

One might argue that the near--potential behaviour of the electrons, 
where they essentially must loose most of their energy to escape from 
the current channels, would make the mechanism uninteresting as an 
acceleration mechanism since fast electrons cannot easily escape.  
However, this feature may instead be a major advantage, since it means that
energy losses due to escape are small, and that the electrons 
remain trapped long enough to have time to loose their energy 
via a combination of bremsstrahlung and synchrotron or jitter radiation.
We observe that only a very small fraction of the electrons manage to escape,
while still retaining most of their kinetic energy. 
This happens mainly at sudden bends or mergers of
the ion channels, where the electron orbits cannot be described in terms of 
a particle moving in a static electromagnetic field.

\begin{figure}[!t]
\begin{center}
\epsfig{figure=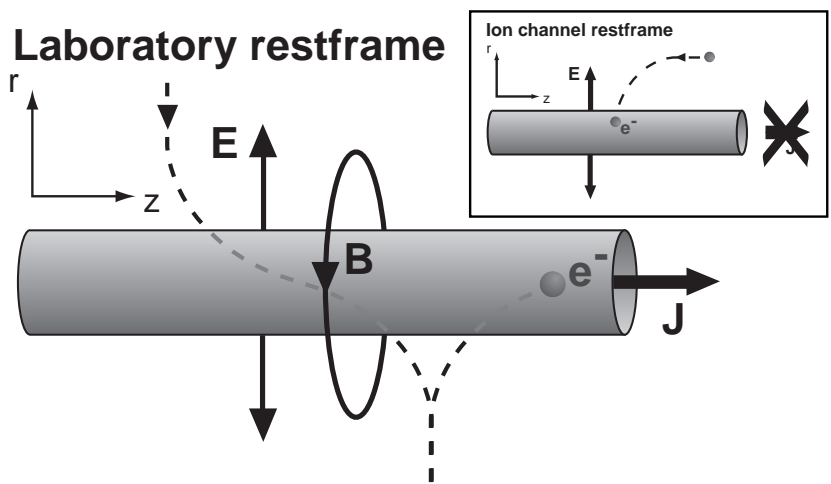,width=0.8\textwidth}
\caption{An ion current channel surrounded by an electric -- and a magnetic field.
Electrons in the vicinity of the current channels are thus subject to 
a Lorentz force with both an electric and magnetic component, 
working together to accelerate the
electrons along the ion flow. Crossing the centre of the channel the process
reverses leading to an oscillating movement along the channel.}
\label{fig:current_acc}
\end{center}
\end{figure}
To analyse the acceleration scenario quantitatively 
we construct a toy model. It has been sketched in Fig.~\ref{fig:current_acc}.
We assume that the ion current channel has radius $R$, that 
the total charge inside
the cylinder per unit length is $\lambda$ and the ions all stream with velocity
$u$ and Lorentz factor $\Gamma$ in the laboratory rest frame
(see Fig.~\ref{fig:current_acc} and inset for definition of rest frames).

Consider an electron
with charge $-q$ and mass $m$ at a distance $r$ from the centre of the channel,
initially having no velocity components perpendicular to the cylinder,
and four velocity $\gamma_0 v_{z,0}$ parallel to the cylinder, and disregard
for the moment any other shielding electrons.

By analysing everything in the ion channel rest frame the problem reduces
to electrostatics and it is possible to analytically calculate the change in 
four velocity of the electron when it reaches the surface of the cylinder.
In the ion channel rest frame the electron has the
Lorentz factor and four velocity
\begin{align}
\label{eq:initgamma}
\gamma'_0          &= \Gamma \gamma_0(1 - u v_{z,0})\,, \\
\label{eq:initvz}
\gamma'_0 v'_{z,0} &= \Gamma \gamma_0 (v_{z,0}-u)\,,
\end{align}
where quantities in the ion channel rest frame are denoted with a prime.
The ions were before moving with velocity $u$, and hence subject 
to a Lorentz contraction, but are now in their rest frame. The line charge 
density is reduced by a factor of $\Gamma$: $\lambda' = \lambda/\Gamma$.
The electron will be attracted to the cylinder and will gain downward momentum in
the $r$--direction. This is simply a free fall in the electric potential and
the final velocity, when the electron reaches the outer edge of the cylinder,
can be found by calculating the change in potential energy
\begin{equation}\label{eq:potenergy}
        \Delta {E'}_{pot}^{r\rightarrow R} = 
                  -\int^{r}_{R} q\vec{E}' \cdot \vec{dr}
        = -\frac{q\lambda'}{2\pi\epsilon_0} \ln(r/R)\,.
\end{equation}
The change in the Lorentz boost $\gamma'$ is then
$m c^2 \Delta\gamma' = \Delta E'_{kin} = - \Delta E'_{pot}$.
The electric force only works along the $r$-axis and the four velocity
along the $z$--axis of the electron is conserved in the ion channel rest frame.
Exploiting this we can calculate not only the total change in energy
but also the change in the different velocity components.
Returning to the laboratory rest frame we find
\begin{align}\label{eq:acc}
\Delta\gamma_{electron} &= \frac{q \lambda}{2 \pi m c^2 \epsilon_0} 
             \ln \frac{r}{R}\,, \\
\Delta(\gamma v_z )_{electron} &= u \Delta\gamma_{electron}\,.
\end{align}
The change in the Lorentz boost is directly proportional to
the total charge inside the channel and inversely
proportional to the electron mass. In reality the Debye shielding 
reduces the electric field further away from the ion channel, so the estimate
above is only valid for distances smaller than a Debye length. 
Inside the ion channel the electron is accelerated too, but the amount
depends on the detailed charge distribution of the ions, and one should
remember, that in general the electrons do indeed have velocity
components perpendicular. The above estimate then can be understood as
an upper limit to the observed acceleration.

\section{Computer Experiments}\label{sec:4.3} 
The experiments were performed with the
three-dimensional relativistic kinetic and electromagnetic particle--in--cell code
described briefly in Chapter \ref{chap:field} and more thoroughly in
\cite{bib:trier}. The code works from first principles, 
by solving Maxwell's equations for 
the electromagnetic fields and solving the Lorentz force equation of motion 
for the particles.

Two colliding plasma populations are set up in the rest frame of one 
of the populations (downstream, e.g. a jet). A less dense population (upstream,
e.g. the ISM) is continuously injected at the left boundary with a 
relativistic velocity corresponding
to a Lorentz factor $\Gamma=15$. The two populations initially differ in 
density by a factor of 3.
We use a computational box with $125\times125\times2000$ grid points and a
total of $8\times10^8$ particles. The ion rest frame plasma frequency in 
the downstream
medium is $\omega_{pi}=0.075$, rendering the box 150 ion skin depths long.
The electron rest frame plasma frequency is $\omega_{pe}=0.3$ in order to resolve
also the microphysics of the electrons.
Hence the ion-to-electron mass ratio is $m_i/m_e = 16$. Other mass ratios and
plasma frequencies were used in complementary experiments. 
Initially, both plasma populations are unmagnetised.

The maximum experiment duration has $t_{max} =$ 340 $\omega_{pi}^{-1}$, which 
is sufficient for the continuously injected upstream plasma 
($\Gamma = 15$, $v\sim c$)
to travel 2.3 times the length of the box.
The extended size and duration of these experiments enable
observations of the streaming instabilities and concurrent particle 
acceleration through several stages of development \citep{bib:frederiksen2004}. 
Momentum losses to radiation (cooling) are presently not included in the 
model. We have, however, verified that none of the accelerated particles in the
experiment would be subject to significant synchrotron
cooling.   The emitted radiation may thus be
expected to accurately reflect the distribution of accelerated electrons.

When comparing numerical data with \Eq{eq:acc}
we take $r$ to be the radius where Debye shielding starts to be
important. Using a cross section
approximately in the middle of \fig{fig:acceleration}
we find $\Delta(\gamma v_z)_{electron} = 58 \ln (r/R)$. It is hard to determine
exactly when Debye shielding becomes effective, but looking at electron paths 
and the profile of the electric field we estimate that
$\ln (r/R) \approx 1.3$. Consequently, according to \Eq{eq:acc}, the 
maximally attainable four velocity in this experiment is in the neighbourhood of 
$(\gamma v_z)_{max}=75$. This is in good agreement with the results from our 
experiments, where the maximum four velocity is $(\gamma v_z)_{max}\simeq80$.

\begin{figure}[!t]
\begin{center}
\epsfig{figure=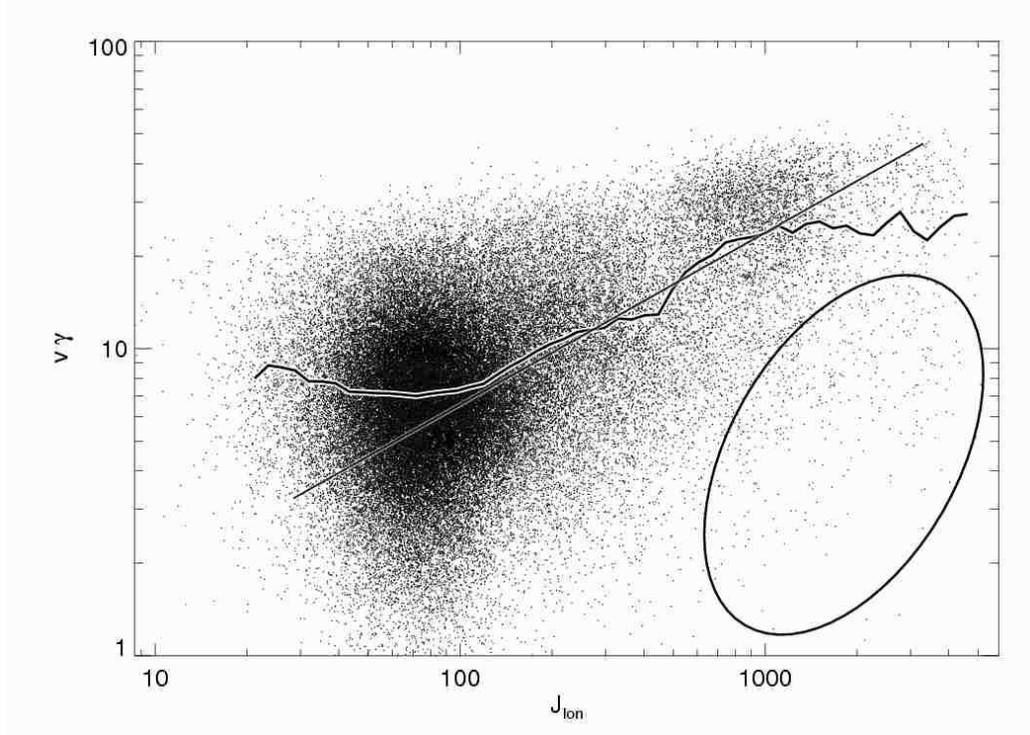,width=\textwidth}
\caption{A scatter plot of the local ion current density $J_{Ion}$ versus the 
four velocity of the electrons in a region downstream of the shock. Over 
plotted is a line (thin) showing the average four velocity as a function 
of $J_{Ion}$, and a line (thick) showing a straight line fit.
Because 'cold' trapped thermal electrons (indicated
with the ellipse) exist inside the ion current channel they count towards 
lowering the average four velocity at high $J_{Ion}$. If the scatter plot
was cleaned, statistically removing all thermal electrons, we would see 
a much tighter relation. Such cleaning, though, is rather delicate and 
could introduce biases by itself. The trend is clearly there though even for the
'raw' data.}
\label{fig:jvg}
\end{center}
\end{figure}

The theoretical model does of course not cover all details of the experiment. 
For example, in general the
electrons also have velocity components parallel to the magnetic field; instead of
making one dimensional harmonic oscillations in the plane perpendicular 
to the current
channel the electrons will describe complicated ellipsoidal paths.
Fig.~\ref{fig:acceleration}A shows the path of two electrons in the vicinity of
an ion channel. But, overall, the electrons behave as expected from the model 
considerations. Consequently, high speed electrons are tightly coupled 
to the ion channels, as clearly illustrated by Fig.~\ref{fig:acceleration}B.

\begin{figure}[!t]
\begin{center}
\epsfig{figure=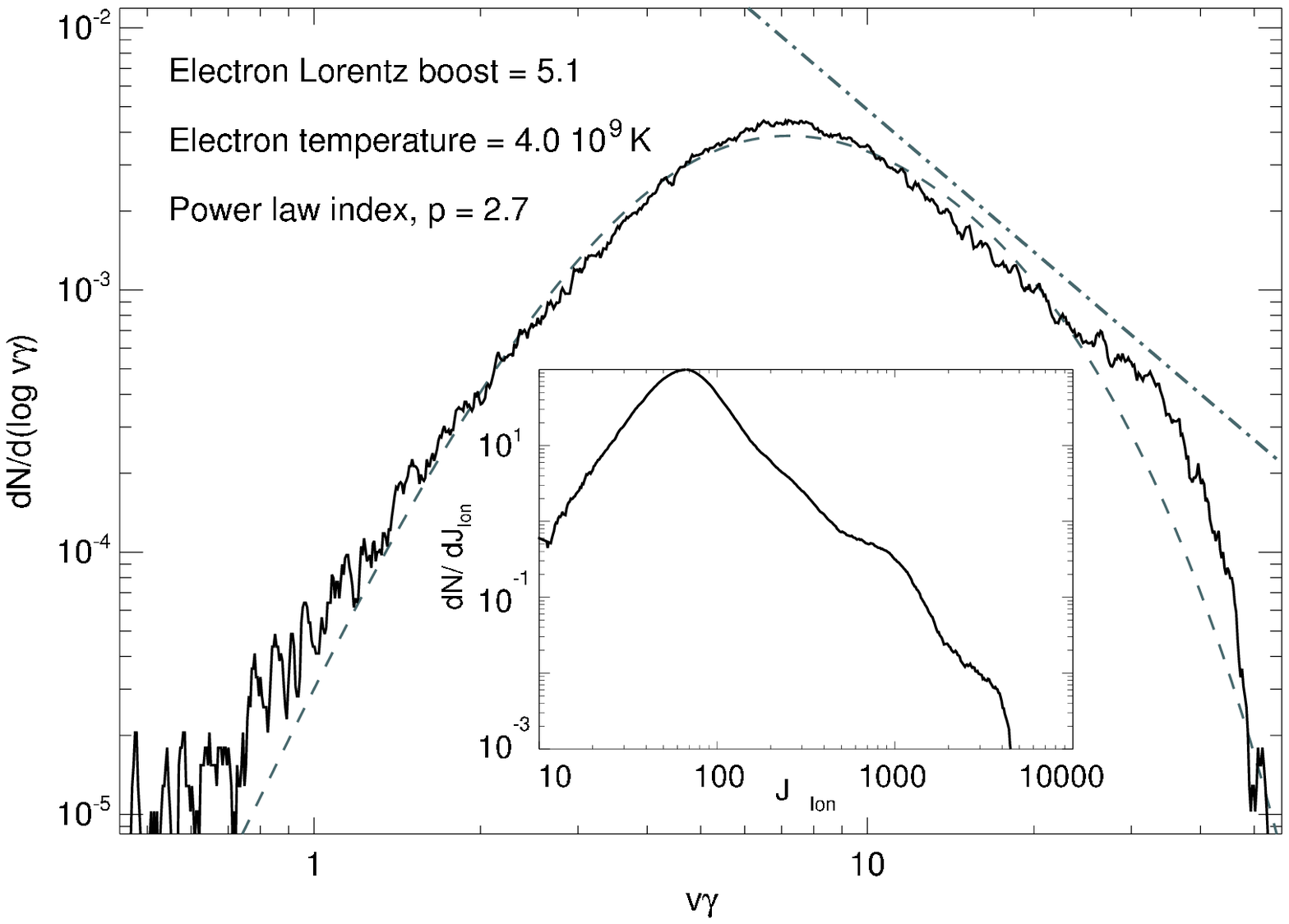,width=\textwidth}
\caption{The normalised electron particle distribution function downstream of
the shock. The dot--dashed line is a powerlaw fit to the non--thermal high
energy tail, while the dashed curve is a Lorentz boosted thermal electron
population. The histogram is made from the four velocities of 
electrons in a thin slice in the $z$--direction
of the computational box. The inset shows a similar
histogram for ion current density
sampled in each grid point in the same slice. The bump
in the inset is a statistical fluctuation due to a single
ion channel.
}
\label{fig:pdfpower}
\end{center}
\end{figure}

Figure \ref{fig:pdfpower} shows that the electrons are powerlaw distributed at 
high energies, with index $p=2.7$. 
The electrons at the high gamma cut-off are found where the ion current peaks, 
as may be seen from Fig.~\ref{fig:jvg}. The maximum ion current is limited 
by the size of our box; larger values would probably be found if the 
merging of current channels could be followed further down stream.
The PDF is not isotropic in any frame of reference due to the high 
anisotropy of the Weibel generated electromagnetic field.
The powerlaw in the electron PDF is dominant for $10<\gamma<30$. 
Likewise, a powerlaw dominates the ion current channel strength, $J_{Ion}$, for 
$100<J_{Ion}<1000$ (inset). 
A relation between the powerlaw distributions of these two quantities 
on their respective 
intervals is provided with Fig.\ref{fig:jvg}: We see that 
the average four velocity is proportional 
(straight line fit) to a power of the local ion current 
density on their respective relevant intervals, 
$10<\gamma<30$ and $100<J_{Ion}<1000$. Their kinship stems 
from the fact that acceleration is
local. $J_{Ion}$ has a powerlaw tail and its potential 
drives the high energy distribution of
the electrons according to Eq.~(\ref{eq:acc}), thus 
forming a powerlaw distributed electron PDF.

Measuring the rate at which the in--streaming ions transfer momentum to the 
ion population initially at rest allows us to 
make a crude estimate of the length scales
over which the two--stream instability in the current 
experiment would saturate due to ion
thermalisation. A reasonable estimate appears to be approximately 10 times
the length of the current computational box, or 
about 1500 ion skin depths. Assuming that
the shock propagates in an interstellar environment with a plasma density of 
$\sim 10^6$ m$^{-3}$ we may calculate a typical
ion skin depth. Comparing this value with the upstream 
ion skin depth from our experiments, 
we find that the computational box corresponds to 
a scale of the order  of $10^7$ m,
or equivalently that the collisionless shock transition 
region of the current experiment corresponds to about $10^8$ m.
For an ion with a Lorentz factor $\gamma=15$ this length corresponds roughly
to 40 ion gyro radii in the average strength of the generated magnetic field. 
But it should be stressed that the in--streaming ions actually do not
really gyrate 
since they mainly travel inside the ion current channels where the magnetic 
field, by symmetry, is close to zero. Also, the strong electromagnetic fields 
generated by the Weibel instability and the non-thermal electron acceleration, 
which is crucial from the interpretation of GRB afterglow observations,
emphasise the shortcoming of MHD in the context of collisionless shocks.

In the computer experiments presented here we have used a mass ratio 
$m_i/m_e=16$ in order to resolve the dynamics of 
both species. \Eq{eq:acc} suggests that
reducing the electron mass to $1/1836\,m_i$ will increase the acceleration of 
the electrons, but the gained  energy is independent of the mass (see
\Eq{eq:potenergy}). In this experiment we observe electrons with energies of
approximately 5 GeV. 
Even further acceleration may occur as ion channels keep growing down stream,
outside of our computational box.

The scaling estimates above depend, among other things, on plasma 
densities, the bulk Lorentz factor, and the mass ratio ($m_i/m_e$).
A parameter study is necessary to explore these dependencies,
but this is beyond the scope of the present chapter. We thus stress that the
 extrapolations performed here are speculative and that 
unresolved physics could influence the late stages
of the instability in new and interesting ways as discussed in the following
chapter. 

When the in--streaming ions are fully thermalised they can no longer support the 
magnetic field structures. Thus one might speculate that the radiating region of 
the GRB afterglow is actually very thin, as
suggested by Rossi \& Rees \cite{2003MNRAS.339..881R}. 
Further, traditional synchrotron radiation theory does not apply to intermittent 
magnetic field generated by the two--stream instability, since the 
electron gyro radii often are larger than the scales of
the magnetic field structures. 
We emphasise the importance of the theory of 
jitter radiation for understanding the generated radiation
\cite{2000ApJ...540..704M}.

\section{Conclusions}\label{sec:4.4}
In this chapter we have proposed an acceleration mechanism for electrons in
collisionless shocks. The theoretical considerations were suggested by
particle--in--cell computer experiments, which also allowed quantitative 
comparisons with the theoretical predictions. We have shown that the 
non--thermal acceleration of electrons is directly related to the 
ion current channels in the shock transition zone.
The results are applicable to
interactions between relativistic outflows and the interstellar medium.
Such relativistic outflows occur in GRB afterglows and in jets from
compact objects \cite{2004Natur.427..222F}. The suggested acceleration
scenario might overcome some of the problems pointed out by Baring \&
Braby \cite{bib:baring} regarding the apparent contradiction between
standard Fermi acceleration and spectral observations of GRBs.

The mechanism has important implications for the way 
we understand and interpret observations of collisionless shocks:

1. The acceleration mechanism is capable of creating a powerlaw 
electron distribution in a collisionless shocked region. 
In the computer experiment presented here a bulk flow with
$\Gamma=15$ results in a powerlaw slope $p=2.7$ for the electron PDF.
Additional experiments will be needed to disentangle what determines
the exact value of the slope.

2. The acceleration is local; electrons
are accelerated to a powerlaw in situ. 
Therefore the observed radiation field may 
be tied directly to the local conditions of the plasma and could be a strong 
handle on the physical processes.

3. Our results strengthen the point already made in Chapter \ref{chap:field};
that the fractions of the bulk kinetic energy that go into in the electrons 
and the magnetic field, $\epsilon_e$ and $\epsilon_B$ respectively, are not 
free and independent parameters
of collisionless shock theory. Most likely they represent interconnected parts
of the same process. 

4. In the case of a weak or no upstream magnetic field, 
the Weibel--like two--stream
instability is able to provide the necessary electromagnetic 
fields. We have shown here
that the collisionless shocked region is relatively thin, and we suggest that the
non--thermal radiation observed from GRB afterglows 
and relativistic jets in general is emitted from such a relatively thin shell. 

It is clear that the non-thermal electron acceleration, the ion current
filamentation, the magnetic field amplification/generation, and hence the
strong non-thermal radiation from the shock, is beyond the explanatory
capacity  of MHD. Whether or not the relativistic MHD jump conditions
become valid on any
larger scale is not possible to decide from the simulations presented in
this chapter.

\newpage
$\phantom{.}$
\newpage
\chapter{The Global Structure of Collisionless Shocks}\label{chap:global}
In collisionless shocks the mean free path of a particle is greater
than the extent of the shock interface. Hence the particle distribution
functions are highly anisotropic and one cannot study them using fluid
methods. Rather, the dominant means of collision is
indirect, mediated by the collective electromagnetic field.
In Chapter \ref{chap:field} it was demonstrated that in collisionless shocks,
where no large scale magnetic field exists beforehand, the resulting
electromagnetic field is largely dictated by the evolution of two-stream
instabilities.
In this chapter I study global charged particle dynamics in relativistic
collisionless $e^+e^-$ pair--plasma shocks numerically, using three dimensional
computer experiments with up to $\sim 10^9$ particles, and present
results on the application and limitations of two dimensional
simulations for the study of the global structure in ion-electron plasmas.
The pair plasma simulations are advanced to a quasi-steady state,
and compared to a fluid model. There is good agreement with the fluid model
for physically relevant quantities, such as bulk density and momentum, which
shows that the evolution can be extrapolated to larger timescales. We
find that the two-stream instability decays over 50-100 electron skin depths
and hence for a pair plasma shock remains firmly in the microphysical
domain.
This type of microphysical experiment may be used to determine
empirically an effective equation of state in a collisionless shock,
and the necessary sink and source terms that describe the conversion
of kinetic to magnetic energy due to the two-stream instability, which
could then be implemented in global fluid models,
leading to more accurate large scale simulations of phenomena such as gamma
ray bursts and relativistic jets from AGN's, where collisionless shocks
are of importance.
The technique would be similar in spirit to the role played by subgrid models
in understanding large scale turbulence, where models
of the small scale behaviour are integrated into the fluid simulations to
extend the dynamical range.

\section{Introduction}
Three dimensional particle-in-cell experiments of
ion-electron collisionless shocks with open boundaries in the
streaming direction have been considered in 
Chapters \ref{chap:field} \& \ref{chap:acc}, and by
Fredriksen et al.~\cite{bib:frederiksen2002,bib:frederiksen2004},
Hededal et al,~\cite{bib:hededal2004,bib:hededal2005} and
Nishikawa et al.~\cite{2003ApJ...595..555N}, 
but as shown in Chapter \ref{chap:acc} the estimated true extent of
a collisionless ion-electron shock is much larger than
the computational domains that have been considered up to now. 
In Chapter \ref{chap:acc} I found that for
a mass ratio of $m_i/m_e=16$ the shock extends at least
1500 ion skin depths. Unfortunately, with current
computer technology there is no hope of performing
three dimensional experiments that resolve scales all
the way from sub \emph{electron}-skin depths to 1500 ion skin
depths.

In Chapter \ref{chap:field} a qualitative difference
between ion-electron shocks and pair plasma shocks 
was noted: In the case of an ion-electron plasma the
heavier ions disrupt the electron channels and the
electrons form a Debye shielding cloud around
the ion channels. This cloud of electrons stabilises the ion
channels; indeed this is why the channels survive significantly
longer and ions from the
upstream and downstream mediums interact less.
The consequence is that thermalisation of the ions and decay of
the two-stream instability in an
ion-electron dominated shock interface happens
on a fundamentally longer time scale than in a shock interface
dominated by a pair plasma.

Even though full three dimensional ab initio experiments of
ion-electron shocks are out of reach, that is not
so for pair plasma shocks. In a pair plasma the
electrons and positrons generate channels on the same time scale,
and with no shielding they are quickly disrupted. In terms
of the electron skin depth time scale $\omega_p/c$, thermalisation is
faster than in the ion-electron case. Furthermore, the electrons
and positrons have the same mass, and therefore many more
skin depths can be resolved in a single box.

\section{Collisionless Pair Plasma Shocks}
The two-stream instability deflects particles in the transverse direction
to the flow, and to correctly describe a collisionless shock the model has to
be fully multi-dimensional. It was shown in Chapter \ref{chap:field},
that the current channels merge in a self similar process generating
a powerlaw distributed magnetic field. Medvedev et
al.~\cite{2005ApJ...618L..75M} investigated the problem theoretically
and demonstrated that the magnetic field correlation length in the
shock interface grows with the speed of light for relativistic shocks.

It is therefore necessary that our computational domain perpendicular
to the shock is comparable in size to the longest two-stream unstable
regions in the box. Otherwise the process may be limited numerically, and
with periodic boundaries perpendicular to the shock interface the experiment
reaches a state containing just a
few self interacting current channels; then it cannot
be entirely clear if the saturation of the magnetic field, decay of the
current channels, and thermalisation of the particles happen due to
numerical or physical effects.

The basic setup of the numerical experiment has already been described in
the previous chapters. In this section I consider three variants of the
same experiment. All are pure pair plasmas. Initially the dense downstream
medium is at rest. The density jump is a factor 3, and the inflow Lorentz factor
of the upstream medium is 3.
The only differences between the setups are in the box sizes and the
plasma frequencies considered. They all contain initially 8 particles per
species per cell in the medium at rest. In the main experiment, $A$,
the box size is $nx\times ny\times nz=80\times80\times2800$ and
the plasma frequency
is $\omega_p = 0.42$. In the two complementary experiments, $B$ and $C$, the box
sizes are $160\times160\times1400$ and $80\times80\times2800$, with
plasma frequencies of $\omega_p=0.21$ and $\omega_p=0.105$ respectively.

\begin{figure*}[!t]
\begin{center}
\epsfig{figure=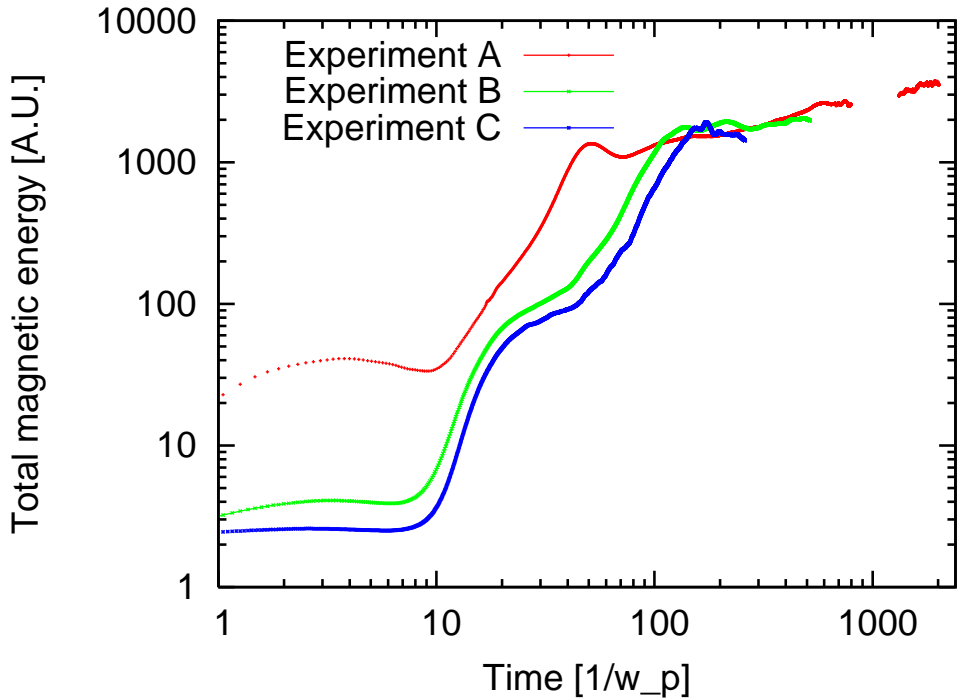,width=0.7\textwidth,angle=270}
\caption{The evolution of the total magnetic energy in the box as a function
of time for the three runs.}
\label{fig:maghist}
\end{center}
\end{figure*}
In experiment $A$ plasma oscillations are resolved with only
2.4 cells, which is close to the Nyquist frequency of the grid, and
indeed there is a higher level of small scale numerical noise than in
experiments $B$ and $C$. While it would have been preferable with an
experiment with a
smaller plasma oscillation frequency, the presented runs are at the limit
of current computer capacities, containing up to a billion particles, and
only experiment $A$ settles to a steady state with selfsimilar evolution.
Experiment $B$ and $C$
are used to validate the behaviour for early times. In fact the first
stages of a thermalised shock
is observed in experiment $B$, but it does not separate into different states
before reaching the edge of the box.
Specifically, the evolutions in averaged current and mass
densities are equal for early times in the different experiments.
Figure \ref{fig:maghist} shows
the evolution of the total magnetic energy in the box. There is a clear
difference in the initial level of fluctuations between the experiments,
due to the difference in plasma oscillation frequencies, but the growth
rate is the same for the three cases, and the experiments show the
same late time behaviour. We can separate the evolution in
three phases: First an initial inflow phase, where the particles
have not yet undergone the two-stream instability. At around
$t=10\,\omega_p^{-1}$ the two-stream instability commences, and it
is saturated at around $t=40-100\,\omega_p^{-1}$. From then
on the region containing shocked material and a diffuse turbulent
magnetic field expands, while a reverse and a forward shock
are formed, and a slow rise in the total magnetic field is seen.
Both at the forward and the reverse shock interface a permanent
two-stream instability is observed.

\begin{figure*}[!th]
\begin{center}
\epsfig{figure=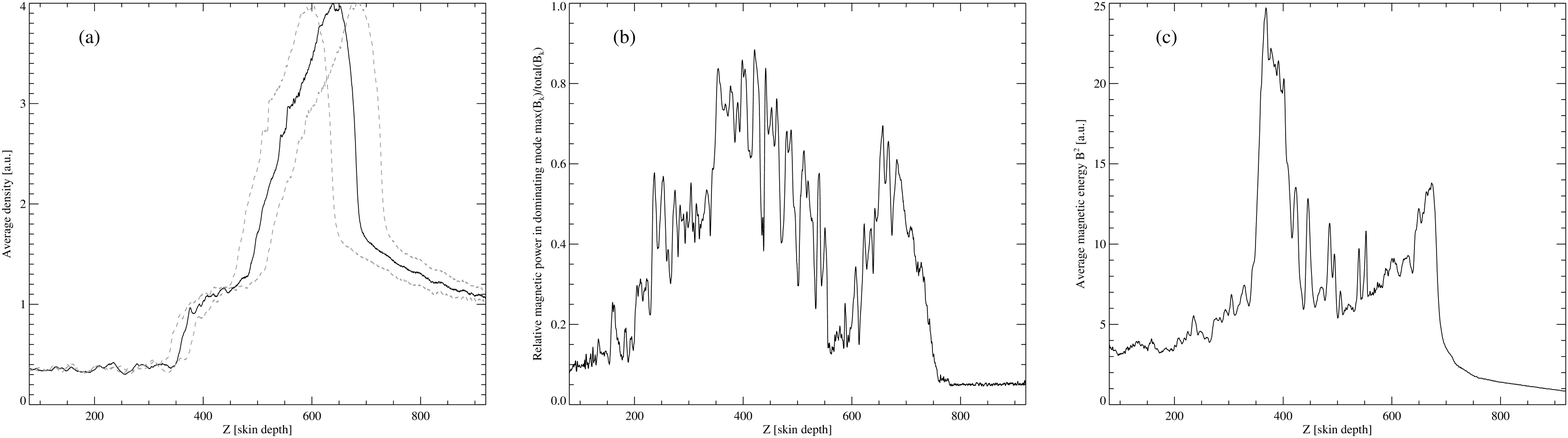,width=\textwidth}
\epsfig{figure=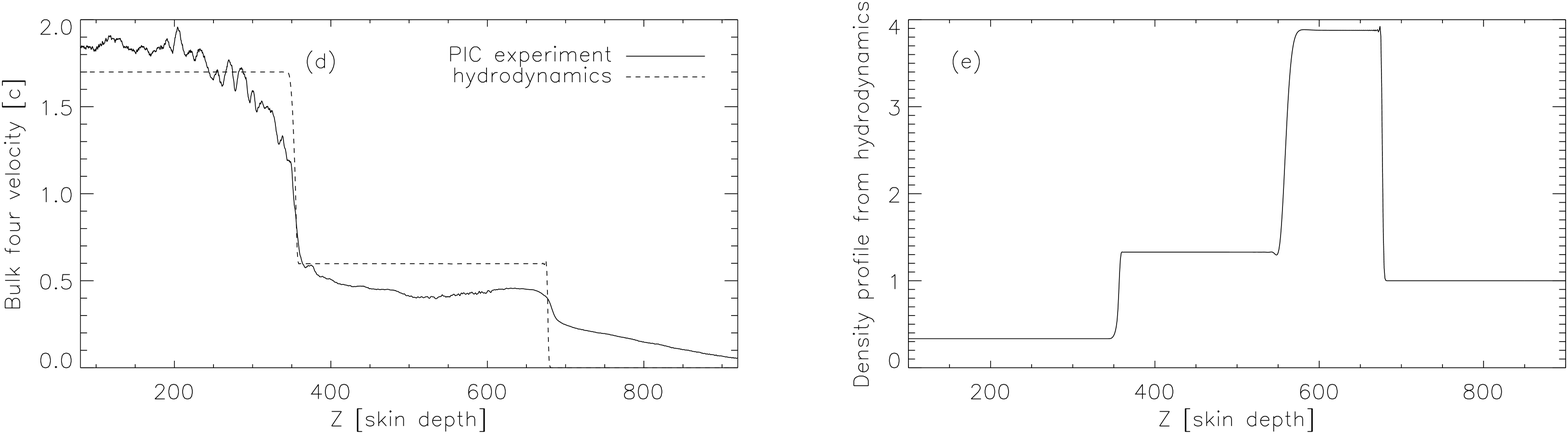,width=\textwidth}
\caption{The large scale structure at $t=1100\,\omega_p^{-1}$.
(a) The average density of electrons in an $x-y$ slice, with
similar plots for $t=1016\,\omega_p^{-1}$ and $t=1184\,\omega_p^{-1}$
as dashed lines. (b) The relative amount of power in the dominating
mode compared to the total power in a two dimensional Fourier
transform of the transverse magnetic field. (c) Average energy
density in the magnetic field. (d) Average four velocity in the $z$-direction
and a similar profile from an HD run. (e) Density from an HD run.}
\label{fig:lscale}
\end{center}
\end{figure*}
The large scale structure at $t=1100\,\omega_p^{-1}$ for experiment $A$ is
shown in Fig.~\ref{fig:lscale}. The average density profile has a
pile-up of matter from $Z=350-700$, this is the shocked area.
In Fig.~\ref{fig:lscale} (a) profiles for earlier
and later times have been overplotted to illustrate how the shocked
area expands with time, with the
forward shock to the right moving faster than the reverse shock to
the left.
Panel (b) in Fig.~\ref{fig:lscale} shows the relative amount of power in the
strongest mode of a Fourier transform of the transverse
magnetic field compared to the integrated power. It has been calculated
by taking the Fourier transform of the magnetic field in the $x-y$ plane,
finding the largest amplitude mode, and then dividing that with the
integrated power.
In the two-streaming shock interface, the largest transverse scale
of the box is dominant, and the ratio is close to 1, while in the centre of the
shocked medium the magnetic field has decayed and power is distributed
over a range of scales. In (c) the magnetic energy in the field is plotted
and the two shock interfaces are clearly seen to be separated.
From Fig.\ref{fig:lscale} (c) it can be estimated that the two-stream unstable
regions have a width of between 50 and 100 \emph{electron} skin depths.

In order to validate that the jump conditions are satisfied I have used the
fluid code described in Chapter \ref{chap:GrMHD} to setup a similar shock
problem, but in fluid dynamics. I have chosen a relativistic equation of
state with $\Gamma=4/3$ and a density jump of 3, an inflow four velocity of
1.7 and a uniform pressure of $P=0.2$. The velocities and densities are
taken in accordance with the unperturbed states seen in experiment $A$ around
$t=1100\,\omega_p^{-1}$. A priori it is not clear how to measure the pressure,
since there are contributions from the random magnetic field, the two-stream
generated magnetic field, heating
from backscattered particles to the left of the shock, and heating from
particles that were not scattered to the right of the shock.
Instead of trying to measure an ill defined pressure, I have chosen
the pressure $P$ such that the maximum density in the right shock wave is in
accordance with the particle data.
This is a reasonable approach, because by fixing the pressure according
to one single parameter, we see good
agreement between the two models for the velocity profile
and density profile in the other parts of the shock wave.
Furthermore the shock profile
is seen to move with a velocity of 0.6 (Approximately 50 skin depths in 84
skin depth times, see Fig.~\ref{fig:lscale} (a)),
while the corresponding fluid shock, in good agreement, is moving with a
velocity of 0.65.
Naturally, the profiles and bulk velocities do not correspond exactly,
since not all facets of the particle experiment are reflected
in the fluid shock. The hydrodynamical experiment does not include magnetic
fields in any way, since no magnetic fields are present in the initial
state. In contrast to that, in experiment $A$, strong magnetic field
generation at the discontinuities in the velocity profile is seen. Moreover, the
discontinuities are still rather smooth in experiment $A$, due to the
collisionless nature of the shock. To take an
example, some of the upstream particles, coming from the left,
do not scatter in the shocked region. As a result there is a smooth transition
in velocity and density at the forward shock front to the right.
If the box size was larger, and the shock could be followed
for longer times, making the extent of the shocked medium larger compared
to the two-stream interfaces and effective mean free path, this smooth transition
would stay constant and the solution would converge even better to that of a
fluid shock. A detailed analysis between larger experiments
running for longer timescales and MHD shock tubes with a range of magnetic
field configurations has to be done to fully understand
the implications for the jump conditions of the two-stream
instability; but this first experiment has shown that indeed it is possible
to recreate the fluid representation ab initio of a pair plasma collisionless
shock using a particle code and working from first principles.

Recently similar experiments were presented by
Spitkovsky~\cite{bib:spitkovsky}, and his results are in agreement
with our findings.

The relatively short thermalisation length observed in this experiment implies
that in the case of a pair plasma shock for astrophysical applications,
the two-stream instability remains a purely microphysical phenomenon, which
probably has little impact on any observed quantities in astrophysical
shocks, simply due to the small volume in which it takes place. 

\section{Collisionless Ion-Electron Shocks and Limits to Current Experiments}
The computer experiments presented in last section are only a few of 
a series of large scale experiments that I have performed during
the last year. The aim is to understand the global structure
in ion-electron dominated collisionless shocks, by performing a series
of three-dimensional experiments with lower and lower mass ratios
in order to finally obtain a thermalised profile, and furthermore with this body
of experiments to be able to rescale the results to realistic mass ratios.
But even using mass ratios down to $m_i/m_e=4$,
with current computer limitations of around a billion particles, it
is not possible to reach a state in the experiments, where both the
ions and electrons fully thermalise and two shock interfaces emerge.

A related but possible project is instead to understand collisionless
shocks in two dimensions. Two-dimensional collisionless shocks have
been considered for some time in the literature (e.g.~Califano
et al.~\cite{1998PhRvE..57.7048C} and Kazimura et al.~\cite{bib:Kazimura}),
but until now no large scale 2D simulations have
been made with open boundaries (though see Medvedev et
al.~\cite{2005ApJ...618L..75M} for an experiment with periodic
boundaries in the streaming direction, and the work
by Hededal \cite{bib:hededalthesis}).

To compare 2D and 3D simulations, in this section I present two
experiments with \emph{exactly} the same initial conditions as
the 3D experiment considered in 
Chapter \ref{chap:acc} (from now on experiment $C$).
They both have an inflow Lorentz boost of $\Gamma=15$, an electron
plasma frequency
of $\omega_{p,e}=0.3$ and a mass ratio of $m_i/m_e=16$. I use 8 particles
per species per cell.
Experiment $A$ has the same box size transverse to the
flow as experiment $C$ with $nx\times nz=125 \times 4000$, while
experiment $B$ is much wider with $nx\times nz=1600\times 4000$.
 
\begin{figure*}[!th]
\begin{center}
\epsfig{figure=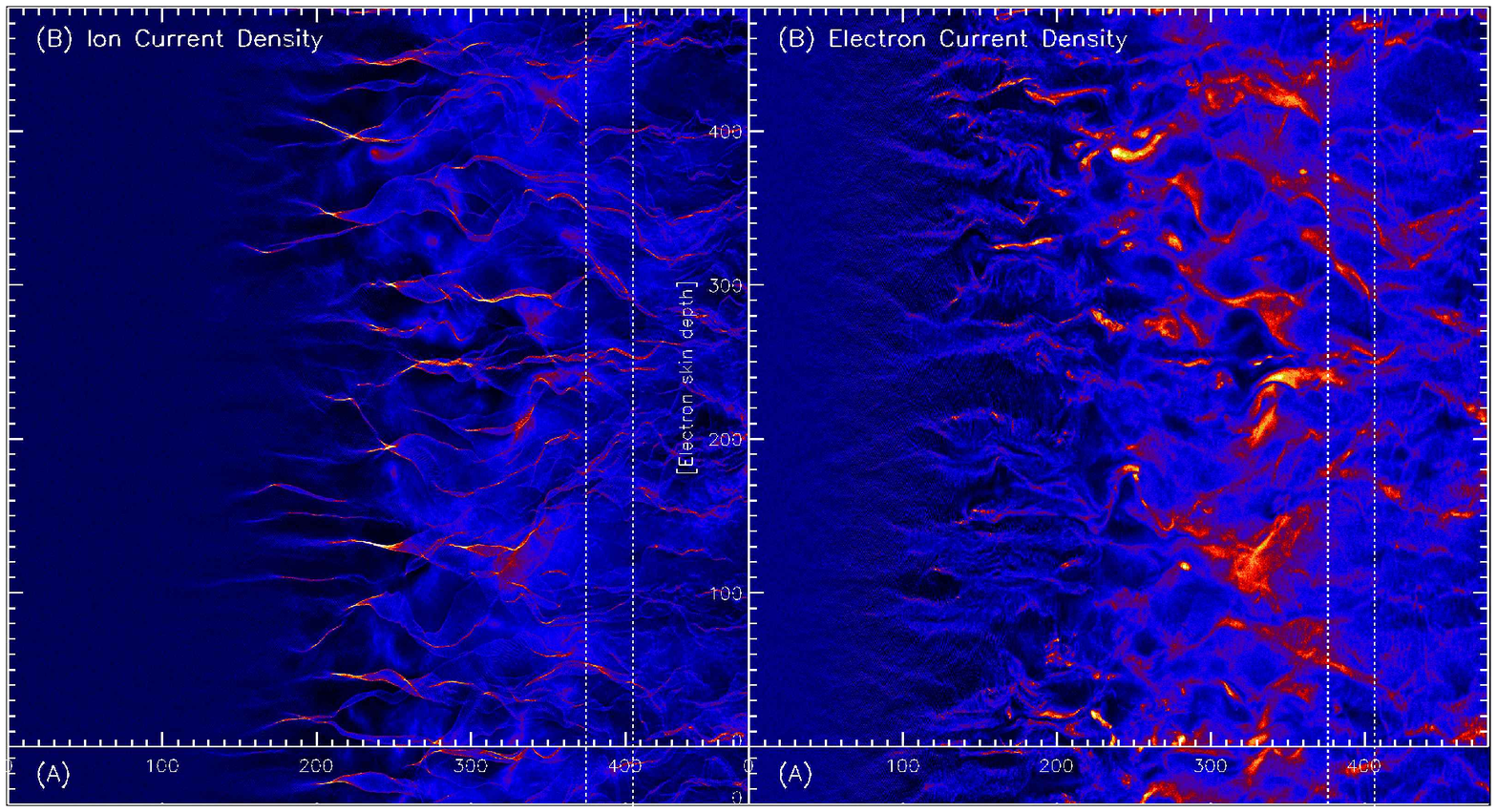,width=\textwidth}
\epsfig{figure=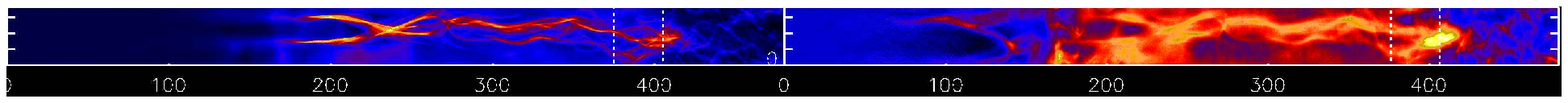,width=\textwidth}
\caption{From top to bottom: The current density of the ions and electrons
of experiment $B$, experiment $A$ and the averaged current density in the
$y$-direction
of experiment $C$, reported on in Chapter \ref{chap:acc}.
The dashed lines indicate the region used for constructing particle
distribution functions. The figure is shown with the correct aspect ratio
and the snapshots are taken at $t=125\,\omega_{p,e}^{-1}$. Length units
are given in electron skin depths.}
\label{fig:lscale2d}
\end{center}
\end{figure*}
I have selected the two experiments to address the following
questions: 1) How much do 2D and 3D experiments differ, both quantitatively
but also in the underlying physical process. 2) What is the impact
of the narrow boxes that have been considered until now. In nature
a collisionless shock is much wider than the shock interface is long,
and during the instability, the different regions, by causality, can
only interact with a finite area of the shock front. But in some of the
3D experiments, to grasp the streaming nature of the shock properly,
the boxes have been far too small in the transverse direction of the shock.

In Fig.~\ref{fig:lscale2d} the size of the current densities for the
three experiments are shown at time $t=125\,\omega_{p,e}^{-1}$. We see basic
agreement between experiment $A$ and experiment $C$ for the morphology,
though the ion current channels in experiment $C$ are thicker and more
well structured.

Comparing experiment $A$ and $B$ we see that the larger box size leads to a
much more complex picture than the simple idea of current channels
streaming strictly in the direction of the flow. There is a dazzling
array of interactions going on, with nontrivial interactions between
the channels. In the lower right of experiment $B$ there is an almost
square formed complex of current channels, which in itself is wider
than both experiments $A$ and $C$. The filamentary structure is sustained,
but in contrast to the simple toy model presented in
Chapter \ref{chap:field} and by Medvedev et al.~\cite{2005ApJ...618L..75M},
one cannot speak of a merging hierarchy of ion channels, and the lifetime
of an individual channel is quite small.

The process is initially ignited by the two-stream instability, and in
experiment $B$ abundant examples of forming channels may be found, not
only at the initial interface at $Z=200$, but also downstream
of the first generation, due to two-stream like configurations
of the magnetic field. Moreover we observe direct merging and head on collisions
between the individual channels.
Counteracting this process and partly responsible for the decay of ion
channels is the electric potential. Near the centre of the channel
there is a strong over density of positive charges and even after taking
into account a relativistic
time dilation factor -- or equivalently -- the self generated magnetic fields
of the channels, they will ``explode''. In two dimensions they leave a
cone-like
structure containing two trails of ions, with some symmetry, since
everything happens at the speed of light, while in three dimensions a
ring-like structure is created.
A three-dimensional version of this explosion may be seen in
Fig.~\ref{fig:acceleration}B, where, except for the helix structure,
everything is stabilised and symmetric
due to selfinteraction of the channel. This electrostatic explosion makes
the evolution of collisionless shocks even more dynamic and intermittent.
It is important to point out that the effect depends 
on the effectiveness by which the electrons Debye shield the ion channels,
and therefore on the mass ratio of the experiment. At higher, more realistic
mass ratios, the shielding is more effective and the timescale for the
breakup of the ion channel longer.

The relatively short time scale of experiment $B$ does not make it
possible to assess what the long time implications are of this richer
structure, but does show
that it is important to have an adequate resolution transverse to
the flow, and not only in the streaming direction, if a full 
understanding of collisionless ion-electron shocks is to be obtained.
The highly dynamic nature and evolution along the streaming direction
also show that streaming experiments with open boundaries are essential
to understand the state of the plasma far downstream of a collisionless
shock interface.

\begin{figure*}[!t]
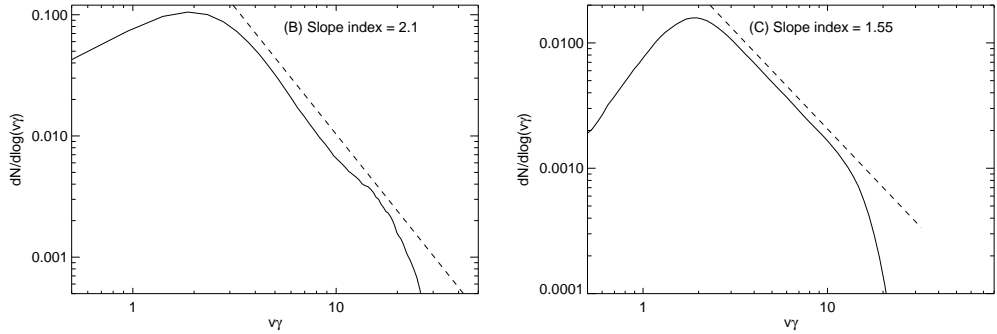

\begin{center}
\epsfig{figure=hist_2000.eps,width=0.49\textwidth}
\epsfig{figure=hist_2050.eps,width=0.49\textwidth}
\caption{Particle distribution function for the electrons in a slice around
$Z=400\,\omega_{p,e}^{-1}$. To the left is shown the PDF for experiment $B$
and to the right the PDF for experiment $C$. The PDF for experiment $A$
is identical to $B$.}
\label{fig:slope}
\end{center}
\end{figure*}
It is important to understand if there are differences in the morphology
of the currents observed in experiments with two- and 
three-dimensional shocks. But in order to make a more formal and
quantitative investigation we have to look at the particle statistics.
I have measured the particle distribution functions (PDFs) for the electrons in
a slice of the domain delimited in Fig.~\ref{fig:lscale2d} by dashed lines for
the three experiments. The slice has been selected to lie at the point in
the shock
where the electrons are on the brink of merging to one single continuous
population, but still the form of the PDF is dominated by remnants of the
initial upstream and downstream populations. The slope of the PDF
indicated in the figure depends on the amount of heating in the populations.
A warmer upstream population will be broader in phase space, and consequently
the maximum is lower, giving rise to a steeper slope. It should be
emphasised that the perceived power law seen in the figure is but
a consequence of the merging populations. It can easily be verified
by noting that the ``powerlaw'' breaks at around $v\gamma=15$, the
velocity of the instreaming population.
We find perfect agreement
between the PDFs of the two-dimensional experiment $A$ and $B$, and they both
have a slope index of $2.1$, while experiment $C$ has a slope index of $1.55$
(see Fig.~\ref{fig:slope}). The reason for this significant difference
in the heating rates of the electrons in two-dimensional shocks compared
to three dimensional is the same as the reason for the lower lifetime
of the ion channels in the two-dimensional experiments.

Essentially we can understand the differences between two- and
three-dimensional simulations by considering the \emph{real}
space available in the two cases. 1) There is a difference in the dynamics of
the ion channels: If we make a transverse cut through the flow
 in the three-dimensional
case, the channels may be likened to 1D particles in a plane 
(see Fig.~\ref{fig:Slice} in Chapter 
\ref{chap:field} for an illustration of this).
We have carefully studied the time evolution of this 1D gas of current channels
in the experiment considered in Chapter \ref{chap:field}. It is found
that, if channels collide head on, they will generally not coalesce, and
instead in
many cases destabilise, because
of the inertia of each channel, while in cases where the impact parameter is
very low, or the two channels slightly miss each other, a much smoother
collision process is initiated by the in-spiralling of
the two channels, ultimately leading to the formation of a single channel.
In a two-dimensional simulation, though, the two-dimensional transverse plane
reduces to a one dimensional line, and consequently two merging channels will
always collide head on, making coalescence more difficult, the transverse
velocities, i.e.~the temperature, of the ions higher, and the lifetime of
the channels lower. 2) There is a difference in the dynamics of the electrons:
The electrons Debye shield the ion channels, and move generally in accordance
with the toy model described in Chapter \ref{chap:acc}. In the three-dimensional
case the paths of two such
electrons have been depicted in Fig.~\ref{fig:acceleration} in
Chapter \ref{chap:acc}.
In general the electrons do not move exactly through the centre of the current
channel, but instead traverse some kind of complicated ellipsoidal path.
In the two-dimensional case though, because there is no ``third dimension''
to miss the centre in, the electrons have to go right through
the centre of the ion channel, and gain the maximum amount of acceleration.
The acceleration is to a large extend potential, and the electron looses
most of the energy climbing out of the potential well, but because of the
time dependence of the fields, statistically there will be some momentum
transfer. In the two-dimensional case the electrons have to pass through
the local minimum of the electrostatic potential, and hence the heating
is more effective than in the three-dimensional case.

The experiments presented in this section have shown that, while the
basic morphology and dynamics do carry over from three to two dimensions,
there are some quantitative differences. The heating of electrons and ions
is more effective and the two-stream generated ion channels are not as
stable in two as in three-dimensional experiments. Nonetheless, if we take these
differences into consideration when interpreting two-dimensional experiments,
these experiments are still the most promising tools for understanding
the global structure
of ion-electron collisionless shocks. From the above discussion it is clear that
the extent of the two-streaming region in
an ion-electron collisionless shock, as inferred from future two-dimensional
experiments of the global shock structure,
will most likely 
be smaller than in the case of a three-dimensional shock. This conclusion
may be drawn directly from the higher heating rate alone, as observed in
Fig.~\ref{fig:slope}.

\section{Conclusions}
Collisionless shocks arise in many astrophysical objects and the correct
understanding of relativistic collisionless shocks have implications for the
observations of outflows from compact objects, such as gamma ray burst afterglows
and relativistic jets. We have seen in the preceding chapters that magnetic field
generation and particle acceleration are integral parts of collisionless shocks
in the case of weak or absent large scale magnetic fields.
To understand the impact on observations it is essential to investigate how far
down stream of the initial shock the two-stream unstable region
extends. With this in mind I have, in the current chapter, 
discussed the global structure of collisionless shocks.

In the first part of the chapter I have presented a fully three-dimensional
model of colliding pair plasmas using
a particle-in-cell code, and observed the thermalisation of the plasma due to
the collective electromagnetic field, and the formation of a macro physical
shock structure. Comparing the results to a fluid simulation, with the same
initial conditions, good agreement is found, implying that the full structure
of the shock has been resolved.

Crucially, I have estimated that the decay of the two-streaming region and
subsequent thermalisation happens over 50-100 \emph{electron} skin depths.
Thus, for the specific case we have considered it renders the two-stream
instability for pair plasmas completely microphysical. Hence,
the two-stream instability in collisionless shocks comprised
purely of leptonic matter may have few direct observational consequences.

In the second part of the chapter I have considered the global structure
of ion-electron dominated collisionless shocks. With current computer capacities
it is impossible to correctly model the global structure of an ion-electron
shock in three dimensions. Two-dimensional collisionless shocks remain a
promising alternative, and I have investigated their applicability
in understanding
three-dimensional models. It has been
shown that while indeed two-dimensional shocks, for the time being, are
our best hope to grasp numerically the global structure of ion-electron
collisionless shocks, there are some differences, and caution should be
voiced in directly generalising results from two-dimensional experiments to
three dimensions. The ion channels that form
due to the two-stream instability are less stable, and the heating
rate of the electrons is higher. Both factors contribute to a faster
thermalisation than what can be expected from three-dimensional experiments
in the future, and hence give rise to an underestimation of the extent
of the two-stream
unstable region. Nonetheless, the overall physical picture is the same, and these
differences can be taken into account.

\chapter{A Next Generation PIC Code}\label{chap:photonplasma}
\section{Introduction}
Over the last couple of years the Copenhagen group has been using PIC
models that include electromagnetic fields and charged particles to
understand the plasma microphysics of collisionless shocks
\cite{bib:frederiksen2002,bib:frederiksen2004,bib:hededal2004,bib:hededal2005}. It has
turned out to be a very successful tool, but it is still limited in the scope of
phenomena that may be addressed.
Even though a large class of astrophysical
environments are indeed collisionless, scattering and collision processes
do play an important role in several key scenarios.
Examples are given below. Another key ingredient, which has been
missing in charged particle simulations, is a full treatment of
photon propagation. It can be argued that photons are represented
directly on the mesh by electromagnetic waves, which certainly is
correct. But the mesh can only represent waves with frequencies
smaller than the Nyquist frequency. The physical length
of a typical cell has in our applications typically been $10^5-10^6\
\textrm{cm}$ and hence it is clear that only low frequency radio waves can be
represented. High frequency photons have to be implemented as
particles that propagate through the box and interact, either
indirectly through messenger fields on the mesh, or directly with
other particles. A valuable consequence of modeling the detailed
photon transport is that extraction of electromagnetic spectra is
trivial. Even in cases where the photon field is only a passive
participant, this fact should not be underestimated as
it enables direct comparison with observations.

There exists Monte Carlo based particle codes (see e.g.~\cite{bib:stern95}
and references therein) that address various particle interactions, but one
of their main shortcomings is the poor spatial resolution. This makes it
impossible to couple the particle aspects to a self consistent evolution
of the plasma.

Our goal has been to develop a framework where both
electromagnetic fields and scattering processes are included in a consistent way. We can then
correctly model the plasma physics and the radiative dynamics. The
scattering processes include, but are not limited to, simple
particle-particle scattering, decay and annihilation/creation
processes. Our new code is not limited in any way to charged
particles, but can also include neutrals such as photons and neutrons.

In the next section we describe some of the physics that can be addressed
with this new code. In section \ref{sec:NGPimplementation} we discuss how
the code has been implemented, the general framework and in detail
which physical processes that are currently implemented. In section
\ref{sec:NGPresults} we present the first results in the form of a toy
experiment that we have performed to validate the code. In the last
section \ref{sec:NGPdiscussion} we summarize.

\subsection{Motivation}
Before we continue and describe in detail the methods, physics and test
problems we have implemented and used, it is important to consider
the general class of scenarios we have had in mind as motivation for
developing the code. There are several key objects, where only the
bulk dynamics is understood, and we are lacking detailed
understanding of the microphysics.

\subsubsection{Internal shocks in gamma ray bursts}
In the internal/external GRB shock model, the burst of gamma-rays is
believed to be generated when relativistic shells collide and dissipate
their relative bulk energy \cite{bib:rees1992,bib:meszaros1993}.
The nature of the radiation is presumably inverse Compton scattering and
synchrotron radiation. Particle/photon
interactions might also play an important role in the very early
afterglow as suggested by
\cite{bib:thompson2000,bib:beloborodov2002}: Even though the medium
that surrounds the burst (ISM or wind) is optically very thin to
gamma-rays, a tiny fraction of the gamma-rays will Compton
scatter on the surrounding plasma particles. This opens up for the
possibility of pair-creation between back scattered and outgoing
gamma-rays. The creation of pairs may increase the rate of back
scattered photons in a run-away process \cite{bib:stern2003}.
The Compton scattering may accelerate the pair-plasma through the surrounding medium with many
complicated and non-linear effects, including streaming plasma
instabilities and electromagnetic field generation. Hence, it is
crucial that plasma simulations of internal GRB plasma shocks
include lepton-photon interactions.

\subsubsection{Solar corona and the solar wind}
Space weather (defined as the interaction of the solar wind on
the Earth) is in high focus for several reasons. Not only is the Sun
our closest star, providing us with invaluable data for stellar
modeling, but coronal mass ejections from the Sun potentially have
impact on our every day life. The strong plasma outflows from the
sun can induce large electrical discharges in the Earths ionosphere.
This may disrupt the complex power grids on Earth, causing rolling
blakcouts such as the one in Canada and North America in 1989. Also
high-energy particles can be hazardous to astronauts and airline
passengers. Computer simulations have provided a successful way of
obtaining insight in these complex plasma physical processes.
However, in the solar coronal and in the solar wind plasma out to
distances beyond the earth orbit, difficulties arise in finding the
right formalism to describe the plasma. Neither a
collisionless model based on the Vlasov equation nor an MHD fluid
model provides a adequate framework for investigation. The problem
has already been studied using three dimensional PIC simulations
but without taking collisions into account (e.g.
\cite{bib:buneman1992,bib:hesse2001}).

\subsubsection{The corona of compact objects}
The bulk dynamics of accreting compact objects have been modeled for many
years using fluid based simulations (e.g. \cite{bib:balbus2003} and references therein). Nevertheless,
it has been a persistent problem
to extract information about the radiating processes. Furthermore in
the corona the MHD approximation becomes dubious, just as in the
solar corona. The environment around a compact object
is much more energetic than the solar corona, and therefore
radiative scattering processes play an important role. Pair
production is also believed to be abundant. Using our new code it would
be possible to model a small sub box of the corona.
The main problem here -- as in most numerical implementations --  is to
come up with realistic
boundaries for the local model. A shearing box approach may be
appropriate, but in fact we can do even better.

The size of a stellar mass black hole is around $10^6\ \textrm{cm}$.
In a fluid simulation we want to model the accretion disk--compact
object system out to
hundreds of radii of the compact object. The normal approach is
to use a non uniform mesh. Nonetheless, the Courant criterion,
which determines the time step, is still limited by the sound
crossing time of the compact object. I.e.~the time step is limited
by the size of the innermost (and smallest) cells in the mesh. The very small
time step corresponds to those found in a typical particle
simulation, where the strict time step arises from the need to
resolve plasma oscillations. Hence data from an MHD simulation
could provide temporally well resolved fluxes on the boundaries of
the much smaller sub box containing the particle simulation.

In this sense the particle simulation will act as a probe or
thermometer of the fluid model. The particle model includes the
full microphysics in a realistic manner and most importantly
includes photon transport. Realistic spectra
could be obtained indirectly from the fluid model, testing
fluid theory against observations. We have already
worked on interfacing fluid models with the old PIC code.

\subsubsection{Pre-acceleration in cosmic ray acceleration}
Accepting Fermi acceleration as a viable mechanism for accelerating
electrons and creating the
the non-thermal cosmic ray spectrum is still left with some big
unanswered questions. One is that the Fermi
mechanism requires injection of high energy electrons while still keeping a large, low-energy population to sustain
the magnetic turbulence. Hence, a pre-acceleration mechanism needs
to be explained.

The shocks in supernova remnants are believed to be cosmic ray
accelerators. However, the Fermi acceleration process in shocks is
still not understood from first principles but rely on assumptions on
the electromagnetic scattering mechanism. PIC codes would seem ideal
in exploring the mechanism from first principles, since they include
field generation mechanisms and the back-reaction that the high
energy particles have on this scattering agent. In Supernova
remnants however, the mean free path for Coulomb collisions are
comparable to the system and particle-particle interactions cannot
be fully neglected.

\section{Implementation}\label{sec:NGPimplementation}
Implementing any state-of-the-art large scale numerical code is a big
undertaking, and can
easily end up taking several man years. We estimate that the final version of the next generation
code will contain more than 50.000 lines of code. Starting in
February this year, it has taken us three man months to implement
the current incarnation of the code which has already grown to
approximately 10.000 lines. Besides T.~Haugb{\o}lle and
C.~B.~Hededal, the development is done together with
{\AA}.~Nordlund and J.~T.~Frederiksen. Fortunately we have a good
tradition and expertise for numerical astrophysics in Copenhagen
and we have been able to port different technical concepts and
solutions from our suite of fluid codes and to a lesser extent
from the old PIC code. The aim is to build an extremely scalable
code that is able to run on thousands of CPUs on modern cluster
architectures and
utilize MPI as the inter node communication protocol. In this
chapter we will not go further into technical details. Instead
we will put emphasis on the important concepts and physics and how we have
implemented these.

\subsection{Concepts}
The two fundamental objects in a particle-in-cell code are the
mesh and the particles. We have adopted the solver and interpolation routines from the old PIC code
to solve the Maxwell equations and find fluxes and densities on the mesh.
The mesh is used to distribute messenger fields
-- such as the electromagnetic fields -- and to calculate volume averaged
fluxes and densities of the particles The latter are used as source terms in the evolution of the messenger
fields.
The particles really represent an ensemble of particles and are often
referred to as \emph{pseudoparticles} \cite{bib:birdsall} or {\em large particles}. A so called smoothing kernel describes the
density distribution of a single pseudoparticle on the mesh. In our
implementation the volume of a particle is comparable to a cell in the mesh.

\subsubsection{Pseudoparticles with variable weights}
The concept of
pseudoparticles is introduced since the ``real space'' particle
density easily exceeds any number that is computationally reasonable
(i.e. of the order of a billion particles). The pseudoparticle charge to mass ratio
is kept the same as the ratio for a single particle.

In ordinary PIC codes the weight of each pseudoparticle of a given
species is kept constant throughout the simulation. The benefit is a
simple code and a unique identity for each particle. The first is a
convenience in the practical implementation, the second important
when understanding the detailed dynamics and history of a single
particle.

Notwithstanding possible conveniences, as detailed below in section
\ref{scat}, we have decided to improve this concept to a more
dynamical implementation where each pseudoparticle carries a
individual weight. Particles are then allowed to merge and split up
when a cell contains too many/few particles, or when particles are
scattered. The concept is sometimes used in smooth particle
hydrodynamics (SPH), where different
techniques have been proposed for the splitting and merging of
particles. It is both used to adjust the density of individual
particles \cite{bib:trulsen2001} and in the conversion of gas--
to star particles in galaxy formation models \cite{bib:governato2004}.
An important quality of SPH is its adaptive resolution
capabilities. These are important in the description of collapsing
self gravitating systems, ranging from core collapse supernovae to
the formation of galaxy clusters, scenarios where matter is collapsing
many orders of magnitude, and therefore the smoothing length or
volume of the individual particles is readjusted accordingly.
Consequently, when splitting particles or adjusting the weights in
an SPH code, it is important to match precisely the spatial density
distribution of the parent particle to the spatial distribution
of the child particles. In PIC codes, though, the spatial size or
smoothing parameter of an individual particle is
determined beforehand by the mesh spacing. This is reasonable since
we are not interested in adaptive resolution but rather a kinetic
description of the plasma dynamics. Splitting a {\it
parent} particle with weight $w_p$ into {\it child} particles
with weights $w^i_c$ is therefor trivial.
The requirements of conservation of mass and
four velocity together with conservation of the density and flux
distribution in the box, can all be satisfied by setting
\begin{align}
w_p &= \sum^n_{i=1} w^i_c\,, & e_p &= e^i_c\,, &
\gamma_p\vec{v}_p &= \gamma^i_c\vec{v}^{\, i}_c\,,
\end{align}
since the smoothing kernel is determined by the mesh spacing, not the
mass of the individual particle.

The merging or renormalization of pseudoparticles requires a much more thorough
analysis. Up to now we have investigated two schemes, one that respects conservation
of mass, energy and four velocity by merging three particles into two at a time,
and one where only mass, energy and average direction is conserved by merging two particles into one
particle. While these schemes probably
work well for approximately thermal distributions, it will easily give
rise to a large numerical heating when considering head on beam collisions.
We believe it can be improved by first selecting a random ``merger particle'' and
then find other particles in the local cell, that are close to the merger
particle in momentum space. A more radical approach is to resample
the full phase distribution in a cell every time the number density becomes
above a certain threshold. Nevertheless, it requires testing of different extreme
situations to find the optimal method to merge particles, and it is still a work in progress.

To obtain the results that we present in section \ref{sec:NGPresults}, we
ran the code without merging of the pseudoparticles activated.

\subsubsection{Scattering processes and splitting of particles}\label{scat}
In Monte Carlo based particle codes the generic way to compute an
interaction is first to calculate the probability for the interaction
$P_S$, then compute a random number $\alpha$. If $\alpha \le P_S$ then
the full pseudoparticle is scattered otherwise nothing happens. This
probabilistic approach is numerically rather efficient and simple to implement, but
it can be noisy, especially when very few particles are present in a cell.
In large particle Monte Carlo codes the typical cell contains up to $10^4$
particles per species per cell (hence ``large particle''). In our
PIC code typical numbers are $10^1-10^2$ particles per species per cell, since
we need many cells to resolve the plasma dynamics. For our requirements the
probabilistic approach would result in an unacceptable level of noise. For example, in a beam
experiment the spectra of the first generation of scattered particles may come
out relatively precise, but the spectra of higher generation scattered particles
(i.e.~particles that are scattered more than once) will come out with
poor resolution or require an excessive amount of particles. Another well known
consequence of the probabilistic approach is that for a given experiment
the precision goes in the best case inversely proportional to
\emph{the square root} of the number of particles used in the experiment.
\begin{figure}[t]
\begin{center}
\epsfig{figure=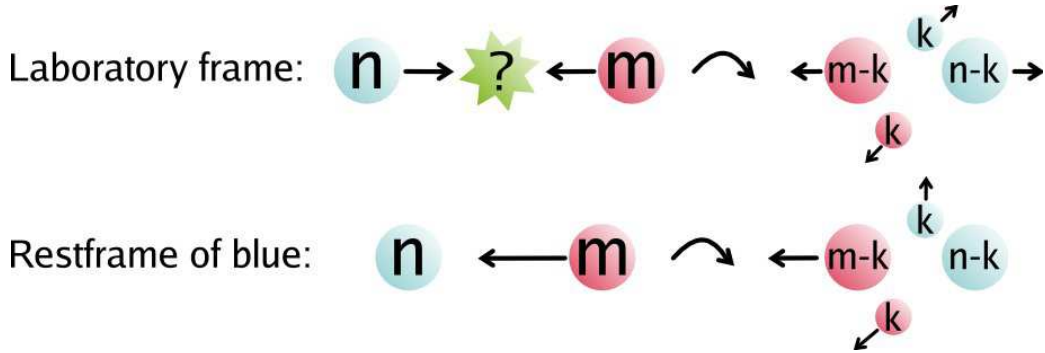,width=\textwidth}
\caption[Schematics of a generic scattering process]{To implement the
scattering of
two pseudoparticles we transform to the rest frame of the target particle
(shown as red/light gray) and computes the probability $P(n)$ that a single
incident particle (shown as blue/dark gray) during a time step $\Delta t$ is
scattered on the $n$ target particles. If the incident particle has weight
$m$, then $k=P(n) m$ particles will interact and two new pseudoparticles are
created.}
\label{fig:splitting_schematic}
\end{center}
\end{figure}
To increase effective spectral resolution we have instead decided to take a
more direct approach. For simplicity we will here describe the method for a
two-particle interaction, and disregard
all factors converting code units to physical units. For example, the weight
of a pseudoparticle is proportional to the number of physical particles in
the
pseudoparticle. Although, these prefactors all represent trivial conversions of units,
they must be taken into account in the real code.

Consider a single cell containing a single pseudoparticle (red) with weight $w_t=n$ and a single pseudoparticle (purple) with weight
$w_i=m$, where
$n>m$ (see Fig.~\ref{fig:splitting_schematic}). We first select the red particle
as the \emph{target}, since $n>m$, and the purple as the \emph{incident}
particle. We then transform the four velocity of the incident particle to the rest
frame of the target particle, and calculate the total cross section
$\sigma_t$ of the interaction. Conceptually we consider the process as a single incident particle
approaching a slab of the target particle. The number density of target
particles in the slab can be calculated from the weight $w_t$ as
$\rho_t = w_t/\Delta V$, where
$\Delta V = \Delta x \Delta y \Delta z$ is the volume of a single cell. Given the number density
the probability that a single incident particle is scattered
\emph{per unit length} is
\begin{equation}
P_l = \rho_t \sigma_t = \frac{w_t \sigma_t}{\Delta V}\,.
\end{equation}
During a time step $\Delta t$ the incident particle travels
$\Delta l =v_{i} \Delta t$,
and the probability that a single incident particle is scattered then becomes
\begin{align}\nonumber
P_S &= 1 - \exp\left[ - P_l \Delta l\right] \\ \label{eq:scat}
    &= 1 - \exp\left[ - \frac{w_t \sigma_t v_i \Delta t}{\Delta V}\right]\,.
\end{align}
The weight of the incident pseudoparticle is $w_i=m$. Pseudoparticles
represent an ensemble of particles. Therefore
$P_S$ is the fraction of incident particles that are scattered on the
target. To model the process we create two new particles with weight
$w_{new} = w_i P_S = k$. Given the detailed interaction, we can calculate
the theoretical angular distribution of scattered particles in accordance with the
differential scattering cross section. Drawing from this distribution we
find the momentum and energy of the new scattered particles. The
weights of the target and incident particles are decreased to $w_t=n-k$ and
$w_i=m-k$ respectively (see Fig.~\ref{fig:splitting_schematic}).

Our method faithfully represents the actual physics even for small cross
sections. However, if all the particles are allowed to interact, the number of
particles in the box will increase at least proportionally to the total number of
particles squared. This is potentially
a computational run away. Normally we will have on the order of up to $100$
particles per species per cell, but to be computationally efficient we only calculate interactions for
a subset of the particles in a cell. This subset is chosen at random according
to an arbitrary distribution we are free to select. If the probability that
two particles are selected for scattering in a given time step is $Q$ then
the traveling length $\Delta l$ simply has to be adjusted as $\Delta l/Q$.
If this arbitrary distribution is chosen cleverly, the particles with the
largest cross section are actually the ones selected most often for
scattering, and everything ends up as a balanced manner: We only
calculate the full cross section and scattering as often as needed, and the
computational load that is given to a certain particle is proportional to the
probability of that particle to scatter.
We rely on the merging of particles as described above to avoid the copious
production of pseudoparticles. Every time the number of pseudoparticles in a
given cell crosses a threshold, pseudoparticles are merged and this way the
computational load per cell is kept within a given range.

\subsection{Neutron decay}
Free neutrons not bound in a nucleus will decay with a
half-life a little longer than ten minutes. The neutron
decays into an electron and a proton and an electron antineutrino
to satisfy lepton number conservation
\begin{equation}
n \to p + e^{-} + \bar{\nu}_e
\end{equation}
The rest mass difference of the process (0.78 MeV) goes into kinetic energy
of the proton, electron and neutrino. Let the neutron lifetime be $\tau$ in
code units. If $\tau$ is comparable to or less than a typical time step, then
practically all neutrons decay in one iteration, and it is irrelevant to
include them. If $\tau$ is much larger than the total runtime, the
neutron can be considered a stable particle (unless the
neutron density in the box is much larger than the proton-- or electron density). If instead $\tau \simeq \alpha
\Delta t$ where $\alpha \sim 100$, then we can select a fraction $f$ of the
pseudoparticle neutrons in each cell and let them decay. This is done in an
analogous manner to the generic scattering process described above in
section \ref{scat}. The weight of the selected neutron
is decreased with a factor
\begin{equation}
\exp\left[-\frac{f\Delta t}{\gamma \tau}\right]\,,
\end{equation}
where $\gamma$ is the Lorentz boost of the neutron pseudoparticle and $f$
is chosen to give reasonable values for the decrease in the weight. At the
same time a pair of electron
and proton pseudoparticles is created with the same weight.
The generated particles share the excess mass of the process (where the
neutrino is neglected for now, but could be included in the future).
The momenta are selected to give an isotropic distribution in the rest frame of the
decaying neutron.

\subsection{Compton scattering}
Here we briefly describe a specific physical scattering mechanism which
has already been implemented in the code, namely Compton scattering.

\begin{figure}[t]
\begin{center}
\epsfig{figure=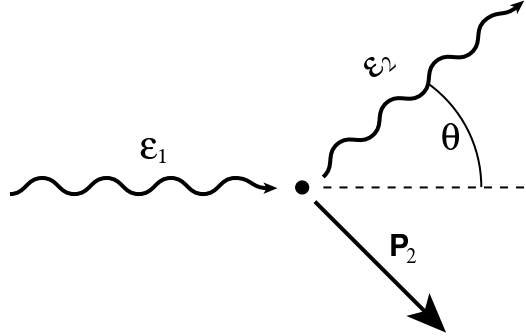,width=.5\textwidth}
\caption[Schematic picture of Compton scattering] {Schematic view
of the Compton scattering process.
Impinging on the electron, an incoming photon with energy
$\epsilon_1$ is scattered into the angle $\theta$ with
energy $\epsilon_2$. In the initial rest-frame of the electron, the
electron will be recoiled to conserve energy and momentum.}
\label{fig:compton_schematic}
\end{center}
\end{figure}

Compton scattering is the relativistic generalization of the classical
Thompson scattering process, where a low energy photon scatters of on
a free electron. In the rest frame of the electron, the photon
changes direction and loses energy to the electron which is set in
motion. The cross section for Thompson scattering is
\cite{bib:rybicki}
\begin{equation}
\sigma_T=\frac{8\pi}{3}r_0^2\,,
\end{equation}
where $r_0\equiv e^2/(m c^2)$ is called the {\it classical electron
radius}. The Thompson scattering approximation is valid as long as
the photon energy is much lower than the electron rest mass $h\nu\ll
m_ec^2$ and the scattering can be regarded as elastic. For photon
energies comparable to, or larger than, the electron rest mass,
recoil effects must be taken into account.
Measured in the electron rest frame we define $\epsilon_1$ as the
photon energy before the scattering,
$\epsilon_2$ as the photon energy after the scattering and $\theta$
the photon scattering angle (\ref{fig:compton_schematic}).
By conservation of energy and momentum one can show
(e.g. \cite{bib:rybicki}) that
\begin{equation}\label{eq:comptonenergy}
\epsilon_2=\frac{\epsilon_1}{1+\frac{\epsilon_1}{m_e c^2}
               (1-\cos\theta)}\,.
\end{equation}
The differential cross section as a function of scattering angle is
given by the Klein-Nishina formula
\cite{bib:klein1929,bib:heitler1954}
\begin{equation}\label{eq:kn}
\frac{d\sigma_C}{d\Omega}=\frac{r_0^2}{2}
                          \frac{\epsilon_2^2}{\epsilon_1^2}
              \left(\frac{\epsilon_1}{\epsilon_2}
+\frac{\epsilon_2}{\epsilon_1}-\sin^2\theta\right).
\end{equation}

The Klein-Nishina formula takes into account the relative intensity
of scattered radiation, it incorporates the recoil factor, (or
radiation pressure) and corrects for relativistic quantum mechanics.
The total cross section is then
\begin{equation}
\sigma_C=\sigma_T\frac{3}{4}\left[\frac{1+x}{x^3}
                 \left\{\frac{2x(1+x)}{1+2x}-\mathrm{ln}(1+2x)
\right\}+ \frac{1}{2x}\mathrm{ln}(1+2x)-\frac{1+3x}{(1+2x)^2}\right]\,,
\end{equation}
where $x\equiv h\nu/(mc^2)$.

\section{Results}\label{sec:NGPresults}
To test the new code and it capabilities in regard to the inclusion
of collisions, we have implemented and tested a simple scenario
involving Compton scattering.
\begin{figure}[t]
\begin{center}
\epsfig{figure=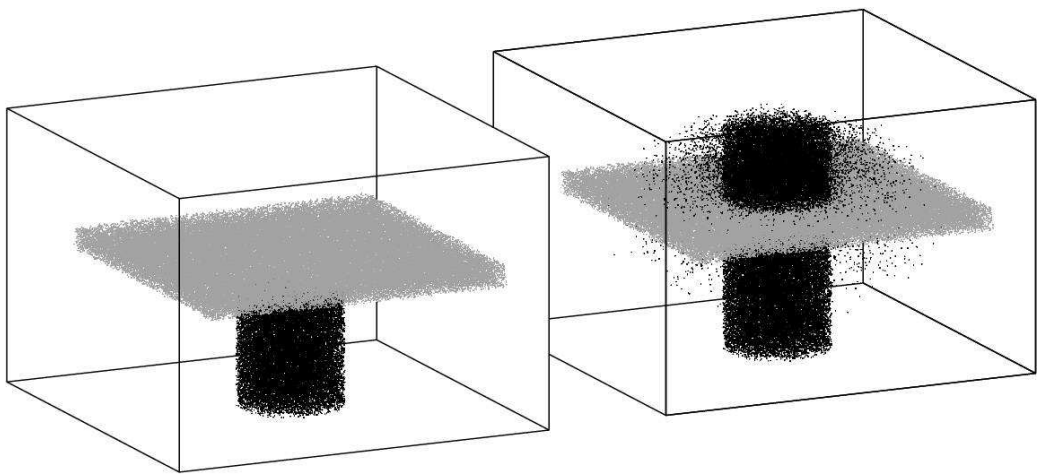,width=1.\textwidth}
\caption[3D
scatter plot of a photon beam passing through a pair plasma] {3D
scatter plot of a photon beam ({\it black}) passing through a cold
pair plasma ({\it gray}). Left panel show initial setup where a
photon beam is injected in the upward direction. Right panel shows
how photons are scattered on the electron-positron pairs}
\label{fig:compt_scatterplot}
\end{center}
\end{figure}

In the test setup, we place a thin layer of cold electron-positron pair plasma
in the computational box. From the boundary, we inject a monochromatic beam of photons
all traveling perpendicular to the pair-layer (Fig. \ref{fig:compt_scatterplot} left panel).
As the beam passes through the plasma layer, photons are scattered (Fig. \ref{fig:compt_scatterplot} left panel).

\begin{figure}[htb]
\begin{center}
\epsfig{figure=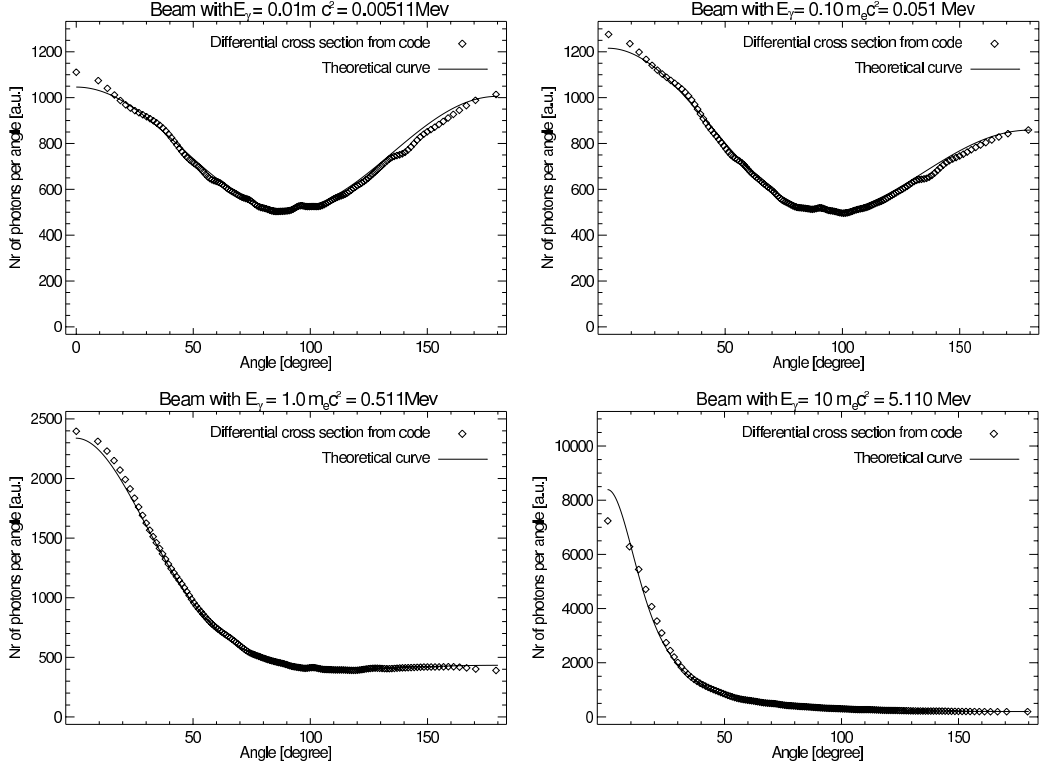,width=1.\textwidth} \caption[The
theoretical Compton scattering cross section compared to data] {The
theoretical Compton scattering differential cross section. We have
performed a test experiment with an incoming laser beam on a very
cold electron population. Over plotted the differential distribution
is the theoretical curve according to Eqs.~(\ref{eq:kn}) and
(\ref{eq:comptonenergy}). } \label{fig:compt_theory}
\end{center}
\end{figure}

For each scattered photon we sample the weight of the photon
and its direction (remembering that all particles are pseudoparticles
that represent whole groups of particles).
Fig. \ref{fig:compt_theory} shows the theoretical cross section as
function of scattering angle compared with the result from the simulations.
Four plots for different energies of the incoming photon
beam are shown. We find excellent agreement between the simulation results and
the theoretical predictions.

\section{Discussion}\label{sec:NGPdiscussion}
A next generation PIC code that includes many different kinds of
scattering process is under development.
It will enable us to target problems that resides in the grey zone
between the MHD and collisionless plasma domains.
This domain covers many astrophysical scenarios of great interest
counting internal shocks in gamma-ray bursts, solar flares
and magnetic substorms, compact relativistic objects, supernova
remnants and many more.

The concept of splitting/merging particles and keeping individual weights of
each particle carry many important features. Variable weights represent the true statistics of a scattering process
in an optimal way compared to the Monte Carlo approach.
Also, for MPI-parallelization it is crucial that the
number of particles per cell is kept more or less constant to ensure an
optimal CPU load-balancing. To localize calculations we are employing a
sorting algorithm that maintains neighboring particles on the mesh as
neighbors in memory. This is not only good for parallelization, but also
makes all computations very cache efficient; a crucial requirement on modern
computer architectures.

To test the infrastructure of the new code we have implemented
Compton scattering as a simple scattering mechanism.
The first results are very promising in form of excellent
agreement with the theoretical prediction.
We note that a recent paper by \cite{bib:moderski2005}
provide an interesting test suite for various kind of particle-photon interactions
that can be tested in the future.
Merging particles has not been satisfactorily implemented yet.
Parallelization of code is still not there yet, and is necessary to obtain the
capability of performing truly large scale experiments.
In summary: Work has still to be done before we can start to investigate
non trivial astrophysical scenarios, nevertheless solid progress has already been made

\vspace{2ex}
This chapter has been written jointly by Christian Hededal and
Troels Haugb{\o}lle, reflecting the fact that the development
process of the next generation PIC code
has been highly team based. Essentially everybody have contributed
time and effort to every single source file of the code. It would
not make sense to write the chapter separately, essentially
repeating each other and reusing the same figures.

\newpage
$\phantom{.}$
\newpage
\chapter{Summary \& Conclusions}
In the past chapters of this thesis I have presented
different numerical methods, as well as applications of the methods to a
number of current problems in relativistic astrophysics.
The thesis
is logically structured into three parts, and below I would like to
summarise the most important points of the work presented.\\[1.5ex]

In the first part (Chapter \ref{chap:GrMHD})
I have presented the theoretical foundation and numerical
implementation of a new general relativistic
magnetohydrodynamics code. I have derived a new form of
the equations of motion, with global coordinates evolving the dynamical
variables from the point of view of a local observer.
When deriving the equations of motion, I have not made any assumptions
about the background metric, so the design is ready to be coupled with
methods solving the full Einstein equations.
The code has been tested on a variety of demanding problems,
and it has been demonstrated that it is able to deal with huge pressure
and density gradients.
The computer code is fully three-dimensional and parallelised and
shows a spectacular
performance on modern computer architectures exploiting up to
30\% of the theoretical peak performance.
It has been tested and verified to scale to hundreds of CPUs, making it possible
to exploit massive supercomputers at national centres
to the full extent.\\[1.5ex]

In the second part of the thesis (Chapters \ref{chap:field}--\ref{chap:global})
I have presented important results in the understanding of collisionless
shocks using a charged relativistic particle-in-cell code.
Together with Jacob Trier Frederiksen, Christian Hededal and {\AA}ke
Nordlund I have investigated the fundamental consequences of the two-stream
instability for observations of collisionless shocks in general, and the
implications for gamma ray afterglows in particular. In Chapter \ref{chap:global}
I extended our analysis and presented results on the global structure and
transition of collisionless shocks to fluid shocks.

In Chapter \ref{chap:field} we have shown that
even in the absence of a magnetic field in the up-stream plasma,
a small scale, fluctuating, and predominantly transversal magnetic field
is unavoidably generated by a two-stream instability reminiscent of the
Weibel-instability. In the current experiments the magnetic energy density 
reaches a few percent of the energy density of the in-coming beam.

In Chapter \ref{chap:acc} we proposed an acceleration mechanism for electrons in
ion-electron collisionless shocks.
The acceleration mechanism is capable of creating a powerlaw 
electron distribution in a collisionless shocked region. 
The theoretical considerations were suggested by
particle--in--cell computer experiments, which also allowed quantitative 
comparisons with the theoretical predictions. We have shown that the 
non--thermal acceleration of electrons is directly related to the 
ion current channels in the shock transition zone and is local in nature.
The electrons are accelerated to a powerlaw in situ. 
Therefore the observed radiation field may 
be tied directly to the local conditions of the plasma and could be a strong 
handle on the physical processes.

To understand the impact on observations it is essential to investigate how far
down stream of the initial shock the two-stream unstable region
extends. With this in mind I have analysed, in Chapter \ref{chap:global}
the global structure of collisionless shocks.
I have presented three-dimensional experiments of colliding pair plasmas using
the particle-in-cell code, and observed the thermalisation of the plasma, due to
the collective electromagnetic field, and the formation of a macrophysical
shock structure. Comparing the results to a fluid simulation, made using
the code presented in Chapter \ref{chap:GrMHD}, with the same
initial conditions, good agreement is found, implying that the full structure
of the shock has been resolved.
I have estimated that the decay of the two-streaming region and
subsequent thermalisation happens over 50-100 \emph{electron} skin depths.
Hence, the two-stream instability in collisionless shocks comprised
purely of leptonic matter may have few direct observational consequences.

In the second part of Chapter \ref{chap:global}
I have considered the global structure of ion-electron
dominated collisionless shocks.
I have investigated the applicability of global models using
two-dimensional shocks -- just possible with current computer technology --
in the understanding of the complete three-dimensional shock structure
It is demonstrated that caution should be
observed in generalising results from two-dimensional experiments to
three dimensions. In two dimensions the ion channels that form
due to the two-stream instability are less stable, and the heating
rate of the electrons is higher. Both factors contribute to a faster
thermalisation than what may be expected from three-dimensional experiments
in the future, and hence cause an underestimation of the extent
of the two-stream
unstable region. Nonetheless, the overall physical picture is the same, and these
differences may be taken into account.\\[1.5ex]

In the third part of the thesis (Chapter \ref{chap:photonplasma}) together
with Christian Hededal I have presented a new code under development by our
group, which will enable us to study not only charged particle dynamics, but
also the propagation of neutral particles, such as photons and neutrons,
as well as interactions between these.

The code is an extension of the current particle-in-cell code, and
also solves the full Maxwell equations, but
furthermore considers particle-particle interactions and
microphysical processes, such as scattering, pair production,
decay, and annihilation of particles.

Especially the inclusion of photons and related radiative processes is
important. In the future we will be able to extract self consistent spectra
from our numerical experiments, thereby gaining the ability to directly
compare our models with observations.\\[1.5ex]

Even though the different tools presented in this thesis
\emph{per se} are not connected,
they all revolve around the same physical problems.
In Chapter \ref{chap:global} we saw
the first example of connecting the codes, to obtain different points of view
on the same physical situation.
In conclusion, and with a look to the future, I believe that the coupling of
the GrMHD code with the new photon plasma
code yields a great potential for obtaining
realistic synthetic light curves from fluid
simulations, connecting them directly with observations.

\newpage
$\phantom{.}$
\newpage
\appendix
\chapter{The Relativistic Maxwell Distribution}
\label{chap:maxwell}
In this appendix I briefly consider the relativistic Maxwell
distribution. When working with data from the
particle code, we have often needed to 
assess how thermal or non thermal a given
particle distribution function (PDF) for a subset of
the particles is, and evaluate the
temperature and the overall Lorentz boost of
the population. Even if the particles are in fact
thermally distributed, they can still be moving with
an overall velocity $u$. To find the temperature
and the boost factor we need to compare our
data, not to the standard Maxwell distribution, but rather
a Lorentz boosted Maxwell distribution.
 
In principle this is a straight forward
exercise, but it becomes complicated because the different
components of the velocity couple through
the Lorentz factor. Then, the Maxwell distribution
of a Lorentz boosted thermal population is not
merely the Lorentz boost of the Maxwell distribution
of the population at rest.

Below in \Eq{eq:bvmaxwell} and \Eq{eq:bvgmaxwell} I
present the Maxwell distribution functions as function
of the boost factor $\Gamma$, boost velocity $u$ and temperature $T$.

\section{The standard relativistic distribution}
The standard Maxwell distribution for a population at rest
in its most basic form can be written
\begin{equation}
dN = N(T)\exp\left(-\frac{\gamma-1}{T}\right)dv_xdv_ydv_z\,,
\end{equation}
where $dN$ is the number of particles per $dv_x dv_y dv_z$
and $N(T)$ is an overall normalisation factor.
Going to spherical coordinates and integrating out the
angle dependence it changes to
\begin{equation}\label{eq:spherical}
dN = 4\pi N(T)\exp\left(-\frac{\gamma-1}{T}\right)v^2dv\,,
\end{equation}
while the most convenient system for boosting the distribution is cylindrical
coordinates, where it can be written
\begin{equation}\label{eq:cylindrical}
dN = 2\pi N(T)\exp\left(-\frac{\gamma-1}{T}\right)v_\perp dv_\perp dv_z\,.
\end{equation}

When considering PDFs, from a numerical
point of view, the most natural variable to work in is not
the normal velocity. The three velocity is bounded by the speed of light
and the PDFs are squeezed towards $c$ at high temperatures.
Instead normally the four velocity $v \gamma$ is used,
which is linear all the way from non relativistic to ultra relativistic
velocities. The Maxwell distribution in terms of $v\gamma$ and $\gamma$ is
\begin{equation}
dN =4\pi N(T)\frac{\sqrt{\gamma^2-1}}{\gamma^4}
    \exp\left(-\frac{\gamma-1}{T}\right)d\gamma
\end{equation}
and
\begin{equation}\label{eq:vg}
dN =4\pi N(T)\frac{(v\gamma)^2}{(1+(v\gamma)^2)^{5/2}}
    \exp\left(-\frac{\gamma-1}{T}\right)d(v\gamma)\,.
\end{equation}

\section{Boosting the Maxwell distribution}
To generalise the above distributions to those seen by
observers moving with four velocity $u\Gamma$ along the
$z$-axis we need to Lorentz transform the variables.
The Lorentz transformation together with the inverse
transformation between the two rest frames are
\begin{equation}\label{eq:tgamma}
\gamma' = \Gamma\gamma(1 - u v_z) \quad \Leftrightarrow \quad
\gamma  = \Gamma\gamma'(1 + u v'_z)
\end{equation}
\begin{equation}\label{eq:tvz}
v'_z = \frac{v_z-u}{1 - u v_z} \quad \Leftrightarrow \quad
v_z  = \frac{v'_z+u}{1 + u v'_z}
\end{equation}
\begin{equation}\label{eq:tvperp}
v_\perp  = \frac{v'_\perp}{\Gamma(1 + u v'_z)}
\quad \Leftrightarrow \quad
v'_\perp  = \frac{v_\perp}{\Gamma(1 - u v_z)}\,,
\end{equation}
where $v_\perp$ is a velocity component perpendicular to the
boost direction and prime denots the boosted reference frame.
To derive the Maxwell distribution, as seen by
an observer moving in the $z$--direction, we
have to transform either \Eq{eq:spherical} or \Eq{eq:cylindrical}
and reexpress it in terms of the new coordinates.
The Maxwell distribution in cylindrical coordinates is best suited,
since from \Eq{eq:tgamma} we see that the transformation of $\gamma$
will pick up a dependence on $v'_z$. Using Eqs.~(\ref{eq:tvz})
and (\ref{eq:tvperp}) to evaluate the Jacobian of the differentials
and substituting the new variables into \Eq{eq:cylindrical}, the boosted Maxwell
distribution in cylindrical coordinates may be found to be
\begin{equation}\label{eq:bmaxwell}
dN = 2\pi N(T) \exp \left( - \frac{\Gamma\gamma'(1 + u v'_z) - 1}{T} \right)
     \frac{v'_\perp}{\left[ \Gamma(1 + u v'_z) \right]^4} dv'_\perp dv'_z\,.
\end{equation}
In this form the distribution function cannot be compared directly
with PDFs obtained from numerical data,
since it is still two dimensional. We need to marginalise one of the
two dimensions, to reduce it to a one dimensional PDF.

\section{The boosted Maxwell velocity distribution}
To find the velocity distribution we shift to spherical coordinates, setting
\begin{align}
v'_z     & = v' \cos(\theta) & v'_\perp & = v' \sin(\theta)\,,
\end{align}
where $\theta \in [0,\pi], v' \in [0,1]$. Inserting the new coordinates
in \Eq{eq:bmaxwell} and integrating over angles after some algebra
 the boosted Maxwell distribution binned linearly in the velocity is
\begin{equation}\label{eq:bvmaxwell}
dN = 2\pi N(T)T \frac{\gamma'^3v'dv'}{\Gamma u} \int^{\alpha_+}_{\alpha_-} 
                \frac{e^{-\beta} d\beta}{(1+T\beta)^4}\, ,
\end{equation}
where the temperature dependent integral has the limits
$\alpha_\pm = \frac{\Gamma\gamma'(1\pm uv')-1}{T}$.

As mentioned above, when analysing particle data it is important to compute
the PDFs with a linear behaviour from sub-- to ultra relativistic velocities.
Changing from $dv'$ to $d\gamma'v'$ we find the final result
\begin{equation}\label{eq:bvgmaxwell}
dN = 2\pi N(T)T \frac{(v'\gamma')d(v'\gamma')}{\Gamma u\sqrt{1+(v'\gamma')^2}} 
                \int^{\alpha_+}_{\alpha_-} 
                \frac{e^{-\beta} d\beta}{(1+T\beta)^4}\, .
\end{equation}
The integral in \Eq{eq:bvgmaxwell} may be simplified by
repeated partial integration
\begin{equation}
\int \frac{e^{-\beta} d\beta}{(1+T\beta)^4} =
\frac{-(1 + T\beta)^2 + T(1+T\beta) - 2T^2}{6T^3{\left( 1 + T\beta  \right) }^3}e^{-\beta}
- \frac{1}{6T^3}
\int \frac{e^{-\beta} d\beta}{(1+T\beta)}
\end{equation}
and everything reduces to an exponential integral, that depends on $T$.

When analysing data I use \verb|IDL|. It already contains a function
to evaluate the exponential integral, and it is rather trivial
to implement \Eq{eq:bvgmaxwell} into a computer program that, given
a set of particles, evaluates the PDF, fits a boosted Maxwell distribution
and finds the corresponding temperature and velocity.

\chapter{Transformation of tensors between different metrics}\label{chap:appa}
In Chapter \ref{chap:GrMHD} it was argued that calculating variables in a local frame,
retaining at the same time global coordinates is the best approach for our numerical method.
Methods used for special relativity may then be employed with minimal changes in arbitrary
space times.

In this appendix, I give the detailed transformation rules for
vectors and two tensors. I consider the transformation between three different
frames. The global star fixed coordinate system (SFCS) has the metric
\begin{equation}
ds^2 = -\alpha^2 dt^2 + \gamma_{ij} \left(dx^i + \beta^i dt\right)
                                    \left(dx^j + \beta^j dt\right)\,.
\end{equation}
The local laboratory (LOLA) frame has the metric
\begin{equation}
ds^2 = - d\hat{t}^2 + \gamma_{ij} d\hat{x}^i d\hat{x}^j\,,
\end{equation}
while the pseudo fiducial observer (PFIDO) frame has the metric
\begin{equation}
ds^2 = -d\t{t}^2 + \sum_{i,j} \frac{\gamma_{ij}}{\sqrt{\gamma_{ii}\gamma_{jj}}}
            d\t{x}^id\t{x}^j\,.
\end{equation}
In the case that the metric contains no off diagonal spatial components,
the PFIDO frame is, in fact, the frame of a fiducial observer.
In the worst case the PFIDO
metric contains three non-trivial components.

\noindent The three metrics are related by the relations
\begin{align}
\left(dx^i + \beta^i dt\right) &= d\hat{x}^i & \alpha dt &= d\hat{t} \\
\sqrt{\gamma_{ii}}d\hat{x}^i   &= d\t{x}^i   &    d\t{t} &= d\hat{t}\,.
\end{align}
The differentials transform as contravariant vectors. The transformation laws for
contravariant vectors may be found by multiplying with metrics and doing a bit of
linear algebra. Tensors by definition transform as the product of the corresponding vectors
and it is a straight forward, though tedious, exercise to find all the combinations.
I have written them down here, since they are essential for the implementation of any
physics; deriving how the local variables are related to the global one.
The following relations have been of interest  when transforming to and from
different frames:\\
\textbf{{\SFCS} $\leftrightarrow$ {\LOLA} frame:}\\
(vectors)
\begin{align}
\h{U}^t &= \alpha U^t              & \h{U}^i &= U^i + \beta^i U^t \\
\h{U}_t &= \frac{1}{\alpha}(U_t-\beta^iU_i) & \h{U}_i &= U_i \\
U^t     &= \frac{1}{\alpha}\h{U}^t & U^i     &= \h{U}^i - \frac{\beta^i}{\alpha}\h{U}^t \\
U_t     &= \alpha\h{U}_t + \beta^i\h{U}_i & U_i &= \h{U}_i \\
\end{align}
\textbf{{\SFCS} $\rightarrow$ {\LOLA} frame:}\\
(contravariant two--tensors)
\begin{align}
T^{tt} &= \frac{1}{\alpha^2} \h{T}^{tt} \\
T^{ti} &= \frac{1}{\alpha}\left(\h{T}^{ti} - \frac{\beta^i}{\alpha}\h{T}^{tt}\right) \\
T^{ij} &= \h{T}^{ij} - \frac{\beta^i}{\alpha}\h{T}^{tj}
          - \frac{\beta^j}{\alpha}\left( \h{T}^{it} - \frac{\beta^i}{\alpha} \h{T}^{tt}\right)
\end{align}
(mixed type two--tensors)
\begin{align}
T^t_t  &= \h{T}^t_t + \frac{\beta^i}{\alpha}\h{T}^t_i \\
T^t_i  &= \frac{1}{\alpha}\h{T}^t_i \\
T^i_t  &= \alpha \left(\h{T}^i_t - \frac{\beta^i}{\alpha} \h{T}^t_t \right)
          + \beta^j \left(\h{T}^i_j - \frac{\beta^i}{\alpha} \h{T}^t_j \right) \\
T^i_j  &= \h{T}^i_j - \frac{\beta^i}{\alpha}\h{T}^t_j
\end{align}
(covariant two--tensors)
\begin{align}
T_{tt} &= \alpha^2\left(\h{T}_{tt} + \frac{\beta^j}{\alpha}\h{T}_{tj}\right)
          + \alpha\beta^i\left(\h{T}_{it} + \frac{\beta^j}{\alpha} \h{T}_{ij}\right) \\
T_{ti} &= \alpha\left(\h{T}_{ti} + \frac{\beta^j}{\alpha}\h{T}_{ji}\right) \\
T_{ij} &= \h{T}_{ij}
\end{align}
\textbf{{\LOLA} $\rightarrow$ {\SFCS} frame:}\\
(contravariant two--tensors)
\begin{align}
\h{T}^{tt} &= \alpha^2 T^{tt} \\
\h{T}^{ti} &= \alpha\left(T^{ti} + \beta^i T^{tt}\right) \\
\h{T}^{ij} &= T^{ij} + \beta^i T^{tj}
          + \beta^j \left( T^{it} + \beta^i T^{tt}\right)
\end{align}
(mixed type two--tensors)
\begin{align}
\h{T}^t_t  &= T^t_t - \beta^i T^t_i \\
\h{T}^t_i  &= \alpha T^t_i \\
\h{T}^i_t  &= \frac{1}{\alpha} \left[ T^i_t + \beta^i T^t_t
          - \beta^j \left(T^i_j + \beta^i T^t_j \right) \right]\\
\h{T}^i_j  &= T^i_j + \beta^i T^t_j
\end{align}
(covariant two--tensors)
\begin{align}
\h{T}_{tt} &= \frac{1}{\alpha^2}\left(T_{tt} - \beta^j T_{tj}
	     - \beta^i T_{it} + \beta^i\beta^j T_{ij}\right) \\
\h{T}_{ti} &= \frac{1}{\alpha}\left(T_{ti} - \beta^j T_{ji}\right) \\
\h{T}_{ij} &= T_{ij}
\end{align}
\textbf{{\LOLA} frame $\leftrightarrow$ PFIDO frame:}\\
(vectors)
\begin{align}
\h{U}^t &= \t{U}^t & \h{U}^i &= \frac{1}{\sqrt{\gamma_{ii}}} \t{U}^i \\
\t{U}^t &= \h{U}^t & \t{U}^i &= \sqrt{\gamma_{ii}} \h{U}^i
\end{align}
(contravariant two--tensors)
\begin{align}
\h{T}^{tt} &= \t{T}^{tt} & \h{T}^{ti} &= \frac{1}{\sqrt{\gamma_{ii}}} \t{T}^{ti} \\
\h{T}^{ij} &= \frac{1}{\sqrt{\gamma_{ii}\gamma_{jj}}} \t{T}^{ij}
\end{align}
(mixed type two--tensors)
\begin{align}
\h{T}^t_t &= \t{T}^t_t & 
\h{T}^t_i &= \sqrt{\gamma_{ii}} \t{T}^t_i \\
\h{T}^i_j &= \sqrt{\frac{\gamma_{jj}}{\gamma_{ii}}} \t{T}^i_j &
\h{T}^i_t &= \frac{1}{\sqrt{\gamma_{ii}}} \t{T}^i_t
\end{align}
(covariant two--tensors)
\begin{align}
\h{T}_{tt} &= \t{T}_{tt} & \h{T}_{ti} &= \sqrt{\gamma_{ii}} \t{T}_{ti} \\
\h{T}_{ij} &= \sqrt{\gamma_{ii}\gamma_{jj}} \t{T}_{ij}
\end{align}

\backmatter
\bibliographystyle{plain}
\bibliography{main}
\end{document}